\UseRawInputEncoding
\pdfoutput=1

\documentclass[twocolumn]{revtex4}
\newlength{\widthfigcolumn}
\newlength{\widthfigpage}
\usepackage{graphicx}
\usepackage[caption=false, subrefformat=parens,labelformat=parens]{subfig}

\usepackage[]{amsmath}
\usepackage[]{algorithm}
\usepackage[noend]{algpseudocode}
\alglanguage{pseudocode}
\usepackage{algorithmicx,eqparbox,array}
\usepackage{mathtools}

\usepackage{amsthm}

\usepackage{placeins}

\newcommand{\powerset}{\raisebox{.15\baselineskip}{\Large\ensuremath{\wp}}}

\algnewcommand{\LongComment}[1]{\hfill// \begin{minipage}[t]{\eqboxwidth{COMMENT\thealgorithm}}#1\strut\end{minipage}}

\begin{document}

	\title[Topological transitions during grain growth on a finite element mesh]{Topological transitions during grain growth on a finite element mesh}
	\author{Erdem Eren}
	\email[]{ereren@ucdavis.edu}
	\affiliation{Department of Materials Science and Engineering, University of California at Davis, Davis, CA 95616, USA}

	\author{Jeremy K. Mason}
	\email[]{jkylemason@ucdavis.edu}

	\affiliation{Department of Materials Science and Engineering, University of California at Davis, Davis, CA 95616, USA}

	\begin{abstract}
		The topological transitions that occur to the grain boundary network during grain growth in a material with uniform grain boundary energies are believed to be known. The same is not true for more realistic materials, since more general grain boundary energies in principle allow many more viable grain boundary configurations. A simulation of grain growth in such a material therefore requires a procedure to enumerate all possible topological transitions and select the most energetically favorable one. Such a procedure is developed and implemented here for a microstructure represented by a volumetric finite element mesh. As a specific example, all possible transitions for a typical configuration with five grains around a junction point are enumerated, and some exceptional transitions are found to be energetically similar to the conventional ones even for a uniform boundary energy. A general discrete formulation to calculate grain boundary velocities is used to simulate grain growth for an example microstructure. The method is implemented as a C++ library based on SCOREC, an open source massively parallelizable library for finite element simulations with adaptive meshing.
	\end{abstract}

	\maketitle

	\section{Introduction}
	One of the overarching goals of integrated computational materials engineering (ICME) \cite{2008ICME} is to accurately predict microstructure and property evolution during thermomechanical processing. At a minimum this would require a simulation incorporating crystal plasticity and grain boundary motion, and ideally interactions involving multiple phases and additional material physics. Such simulations would benefit from recent advances in three-dimensional microscopy \cite{2005MatSciFormZaefferer}, and specifically three-dimensional X-ray diffraction microscopy (3DXRD) that enables non-destructive three-dimensional imaging of millimeter-sized samples \cite{2013ApplCrysLi, 2014ActaMateLi}. These could both provide initial conditions for and allow verification of the output of predictive simulations of microstructure evolution. 
	
	Historically, one major difficulty with simulations of microstructure evolution has been the use of unrealistic grain boundary energy (GBE) functions. Such functions are difficult to determine experimentally due to the number of independent variables, but Morawiec recently suggested a procedure to estimate the GBE from distributions of grain boundary angles around triple junctions \cite{2000ActaMateMorawiec}. Saylor et al.\ subsequently used a related technique to estimate the GBE from EBSD analysis of the surface of aluminum samples \cite{2003Saylorexp, 2003Saylortheo}. While explicit functions for the grain boundary energy are not yet widely available (with a few exceptions \cite{2014ActMateBulatov,2016JourMechPhySolidRunnels}), this will likely change in the near future. When that happens, a code for microstructure evolution that is able to make full use of them would ideally already be available.
	
	Existing simulations of microstructure evolution include Monte Carlo (MC) Potts, cellular automata (CA), phase field (PF) and front tracking models. The Monte Carlo Potts \cite{1986ActaMetalMateSrolovitzGrestAndersonvol1, 1991PhysRevAHolm} and cellular automata \cite{2002AnnuRevMatRsRaabe, 2006JourCrysGrowDing, 2010MathemCompSimJanssens} methods are popular partly because of their low computational complexity and ease of implementation, but suffer from two relevant shortcomings. First, the underlying voxel lattice introduces an artificial anisotropy that can be difficult to eliminate \cite{2015ActaMateMason82, 2015ActaMateMason94}, and a predictive model requires kinetics relatively independent of any underlying grid. The second limiting property of MC Potts and CA models is the difficulty of connecting the model with physical units of measure. Zhang et al.\ scaled quantities defining characteristic time, length and energy but observed that the grid size affected the bulk energy driving force \cite{2012ActaMateZhang}. Mason established spatial and temporal dimensions in a CA model using the Turnbull relation and a uniform grain boundary energy, but the technique is not easily generalized to other situations \cite{2015ActaMateMason94}. 

	The phase field method is an implicit boundary approach that was initially developed to study phase transitions \cite{1999PhysicaDSteinbach}, and can be modified to include small deformations and mildly anisotropic interface energies \cite{2008CalphadMoelans}. One drawback is the high memory and computational demand associated with representing grains by continuous fields, since numerical instabilities associated with steep gradients limit the time step. Modern implementations often use sparse data structures \cite{2006ModSimMatSciEngGruber, 2006PhysRevEVedantam, 2017NPJMiyoshi} and adaptive meshing \cite{2010DorrJourCompPhy} to address this issue. Still, finite deformations and arbitrary boundary energies that can depend on the grain boundary plane pose difficulties. Moreover, the use of diffuse boundaries can complicate the study of topological aspects of the grain boundary network and can introduce subtle numerical errors. Jin et al.\ compared the accuracy of level set and phase field methods coupled with the Finite Element Method (FEM) in representing the motion of triple lines during isotropic and anisotropic grain growth \cite{2015CompMatSciJin}. They observed that under proper grid and time refinement, both methods performed similarly for the isotropic case. For anisotropic grain growth though they observed 14.2\% error in triple junction velocity for the level set method and as much as 68.7\% error for the PF method. Some recent functional methods allow for anisotropic grain boundary properties \cite{2019ModSimMatSciRibot}, but modeling of finite mechanical deformation is still not addressed.
	
	Early front tracking methods had the advantage of concentrating computational resources just on the boundaries, and were often used to study mean curvature flow \cite{1989PhilMagBKawasaki, 1990PhaseTransNagai}. FEM-based approaches are a natural extension of these that can support additional physics, e.g.,\ boundary energies can be explicitly defined, and volumetric meshes allow for crystal plasticity \cite{2010ActaMateRoters}. However, FEM-based methods introduce additional challenges with scalability and require explicit handling of the topology and mesh. The complexity of the latter has encouraged use of an MC Potts, CA or PF method in conjunction with a FEM solver. These hybrid schemes use an implicit boundary representation to model grain growth, and transfer the resulting microstructure to the FEM to model deformation. Sequential evolution is achieved by transferring the microstructure back and forth \cite{2008PhilMagLoge, 2012CompMatSciTonks, 2000ModellSimulMatSciRaabe}. This does not resolve accuracy concerns though, since transferring the solution potentially introduces information loss and increases computational complexity. 
	
	Of the purely FEM-based approaches, Kuprat developed a three-dimensional adaptation of the gradient weighted moving finite element (GWFE) method and implemented GRAIN3D, a serial finite element framework for microstructure modeling of grain growth \cite{2000SIAMJourSciCompKuprat}. The code had an element regularization scheme to improve the quality of low-quality elements, handled changes in the microstructure as boundaries evolved, and supported volumetric physics. While the initial implementation only supported constant grain boundary energies, more general energies were investigated by Gruber et al.\ \cite{2005ScripMateGruber}. There are two main concerns with using this for general purpose simulations of microstructure evolution though. First, Kuprat implemented the topological transitions by switching the last remaining set of elements of a collapsing boundary segment or volume to the appropriate neighboring volumes \cite{2000SIAMJourSciCompKuprat}. This is not necessarily physical, and the relabeling can cause a substantial and artificial perturbation of the boundaries. Although the likely changes to the overall evolution are limited for an isotropic grain boundary energy, this could substantially affect microstructure trajectory for the anisotropic case. Second, the existing implementation of the implicit finite element solver is serial. This prohibits simulating microstructures on physically relevant scales, such as the $ 1 $ mm$^3 $ cylindrical copper sample imaged using 3DXRD by Li et al.\ \cite{2014ActaMateLi}.
	
	Using a surface mesh representation, Syha and Weygand studied the effects of an anisotropic grain boundary energy \cite{2010ModelSimuMaterSciEngSyha}. They proposed to decompose topological transitions into simpler sequential operations and used a force-based criteria to select changes to the grain boundary network. While this could accommodate anisotropic grain boundary energies, decomposing a topological transition into a sequence of simpler ones could alter the eventual trajectory of microstructure evolution. Moreover, the implementation is not volumetric and therefore cannot support volumetric physics.
	
	Lazar et al.\ studied ideal grain growth by using a surface mesh representation, a fixed set of topological transitions applicable for uniform grain boundary energy, and evolving the microstructure with a discretized formulation satisfying the MacPherson-Srolovitz relation \cite{2011ActaMateLazar, 2007NatureMacPhersonSrolovitz}. Although this approach provided insight into ideal grain growth, it is not applicable to general microstructure evolution for two reasons. First, the boundary evolution formulation assumes that the microstructure is composed of quadruple points and triple junctions at all times except for the moments where transitions occur. While this is generally applicable for ideal grain growth, it does not hold for experimental microstructures. For instance, highly twinned microstructures often contain junction lines joining four grain boundaries, and accommodating such configurations would require implementing more general topological transitions. Second, the implementation doesn't support volumetric physics, and is only intended to model ideal grain growth. 
	
	A FEM code to be used for ICME would ideally be able to handle substantial volumes of material since many grains are required to accurately reflect variations in the local deformation response and to model stochastic processes like recrystallization. Tucker et al.\ studied convergence of large scale crack propagation simulations as a function of the number of grains and mesh refinement in  microstructures with abnormal grains \cite{2015ModSimMatSciTucker}. They observed that the overall damage response was not significantly affected by mesh resolution, but that more than 200 grains were required in the sample microstructure for the convergence of the local response. This shows that a scalable framework is necessary to accurately capture the local response during deformation. 
	
	To summarize, existing implementations of FEM-based grain growth codes are limited in several respects. First, they are generally serial, prohibiting large scale simulations \cite{2000SIAMJourSciCompKuprat, 2010ModelSimuMaterSciEngSyha, 2011ActaMateLazar}. Second, topological transitions are achieved by merging mesh entities with one of the neighboring grains \cite{2000SIAMJourSciCompKuprat}, by sequentially splitting points \cite{2010ModelSimuMaterSciEngSyha}, or selecting from a restricted set of operations \cite{2011ActaMateLazar}, all of which could substantially change the trajectory of the microstructure evolution. That is, a general FEM framework to study grain growth and deformation at physically relevant scales does not appear to exist.
	
	The main contributions of this paper are four-fold. First, a method for finding all possible topological transitions that can occur around junction points during grain growth is proposed. Second, operations on the simplicial mesh have been developed to modify the mesh corresponding to these topological transitions. Third, a criteria based on the rate of energy dissipation is used to compare different topological transitions, providing an unambiguous selection criterion. Fourth, a discrete formulation to simulate grain boundary motion has been implemented that allows for effectively arbitrary grain boundary properties \cite{2017ActMateMason}. The formulation is explicit and solves for the motion of each vertex individually, reducing the computational load greatly compared to the weak formulation of the FEM at the cost of increased error. A C++ library called VDLIB implements all these operations. VDLIB interfaces with SCOREC, a massively parallel mesh management library with local adaptive re-meshing \cite{scorecweb, 2012SCCompanionSeolShephardPUMI}. The intention is to provide the foundations for large scale simulations of microstructure evolution within the framework of ICME.

	\section{Microstructure representation} \label{sec:MSrepresentation}
	
	\begin{figure}
		\centering
		\subfloat[][]{		\label{subfig:MSFEM}
			\includegraphics[width=0.5\widthfigcolumn]{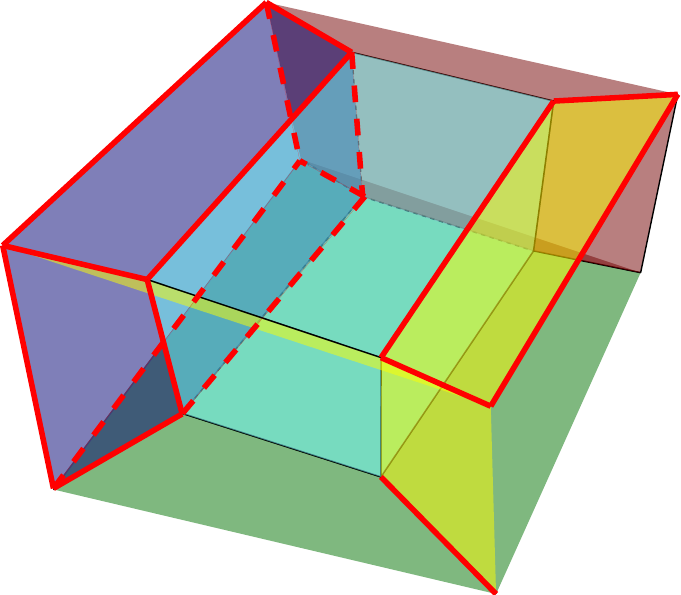}
		}%
		\subfloat[][]{		\label{subfig:FEMFEM}
			\includegraphics[width=102pt]{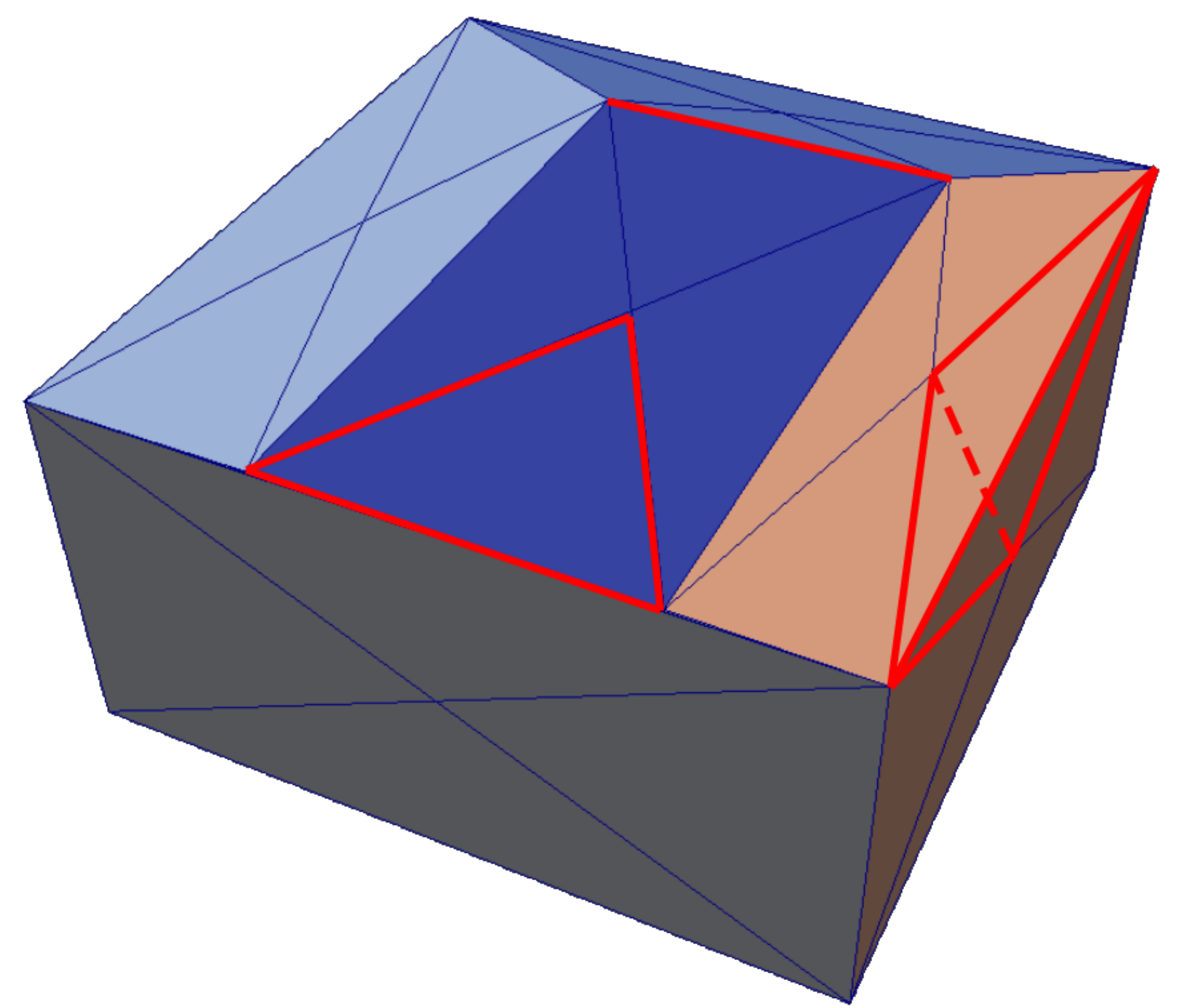}
		}%
		\caption{(a) The rectangular prism example is comprised of seven grains. A central rectangular grain is surrounded by six grains, with examples of a volume, surface, and line outlined in red. (b) A finite element representation of this microstructure where tetrahedra, triangles, edges and vertices are used to discretize volumes, surfaces, lines and points. Examples of a tetrahedron, triangle and edge are outlined in red.}
		\label{fig:FEM}
	\end{figure}
	
	Our intention is to simulate the evolution of a microstructure at a scale that resolves the grain structure. It will be useful in the following to introduce specific terminology to identify the various microstructure components. A grain will be called a volume, a boundary a surface, a boundary junction line a line, and a boundary junction point a point. A microstructure where each of these components is outlined in red is shown in Figure \ref{subfig:MSFEM}. The volumes, surfaces, lines, and points composing the microstructure formally comprise a stratified space, and for that reason the microstructure components will occasionally be referred to as $ d $-strata where $ d $ is the dimension of the stratum. The connectivity of the topological components of the microstructure is defined by the adjacencies of $ d $-strata and $ (d-1) $-strata; that is, a volume is bounded by surfaces, surfaces by lines, and lines by points.
	
	\begin{figure}%
		\centering
		
		\subfloat[][]{		\label{subfig:upreg_0c}
			\includegraphics[height=41pt]{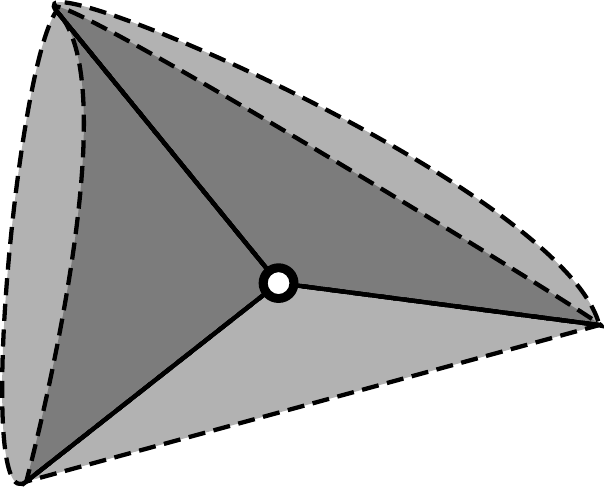}
		}%
		\subfloat[][]{		\label{subfig:upreg_1c}
			\includegraphics[height=41pt]{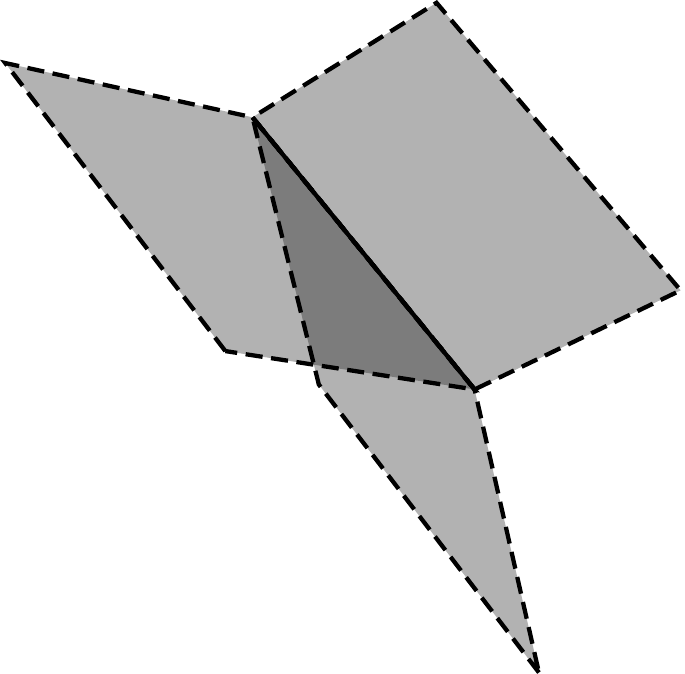}
		}%
		\subfloat[][]{		\label{subfig:upreg_emb1}
			\includegraphics[height=41pt]{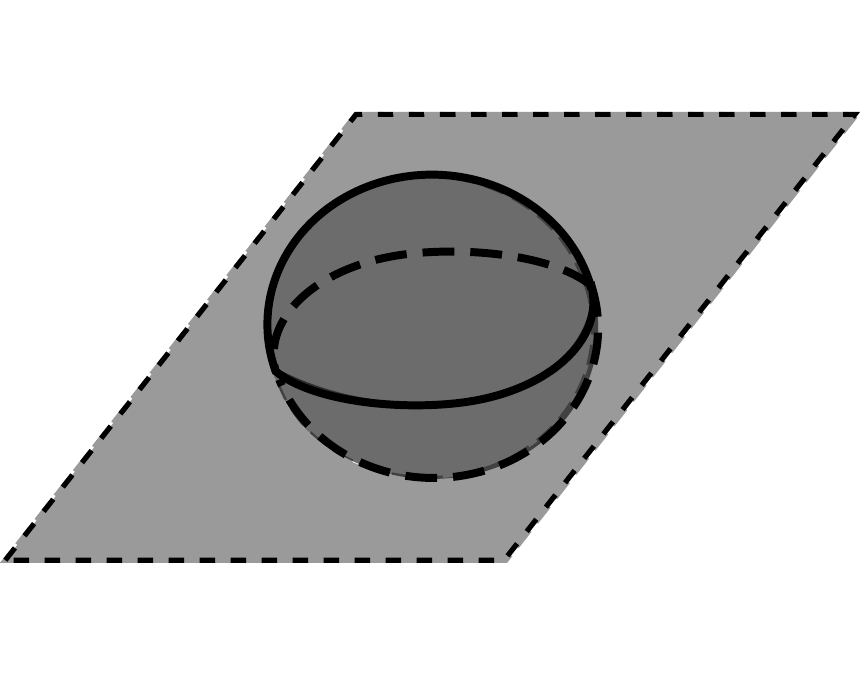}
		}%
		\subfloat[][]{		\label{subfig:upreg_emb2}
			\includegraphics[height=41pt]{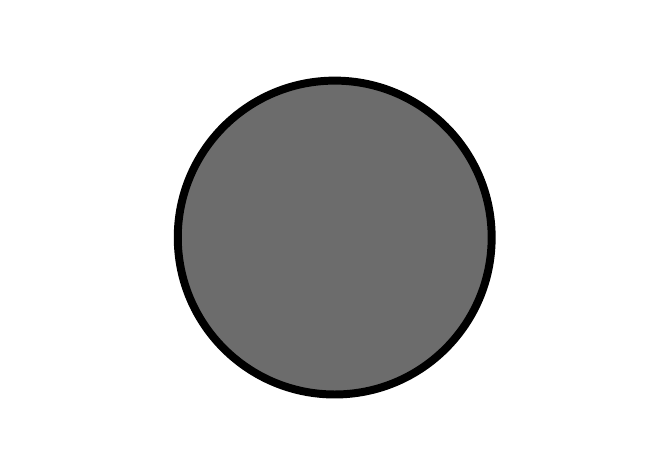}
		}%
		\caption{Examples indicating adjacency rules. (a) A point should bound at least three lines. This point bounds three lines, two conical volumes on the left and right, and two volumes above and below the page. (b) A line should bound at least three surfaces. (c) A surface separating a top and a bottom volume and ball embedded in the surface. The line of intersection has no bounding points. (d) A sphere inside another volume, with a surface that has no bounding lines.}
		\label{fig:Valreg}
	\end{figure}
	
	A point is required to bound at least three lines (Figure \ref{subfig:upreg_0c}), a line at least three surfaces (Figure \ref{subfig:upreg_1c}), and a surface at least two volumes. One can show that any topological component not satisfying these relationships is spurious in the sense that it can be removed by merging the adjacent components of the next higher dimension. There are no constraints imposed on the number of adjacent components of the next lower dimension; this allows e.g., a small spherical volume to be embedded in the middle of a surface (Figure \ref{subfig:upreg_emb1}), or a ball to be embedded in the interior of a volume (Figure \ref{subfig:upreg_emb2}).

	\section{Operations on the microstructure} \label{sec:microstructure}
	
	During the course of grain growth, grain boundaries move to reduce the energy of the microstructure. Occasionally a surface or volume will shrink to a point or will expand from a point to participate in the subsequent evolution; such events are called topological transitions. From the standpoint of the finite element mesh the corresponding operations are either collapses, where disappearing boundary segments or volumes are removed, or insertions, where new boundary segments are introduced to allow the microstructure evolution to continue.
	
	\subsection{Stratum collapses} \label{subsec:collapse}
	
	\begin{figure}
		\centering
		\subfloat[][]{		\label{subfig:col_initial}
			\includegraphics[width=73.8pt]{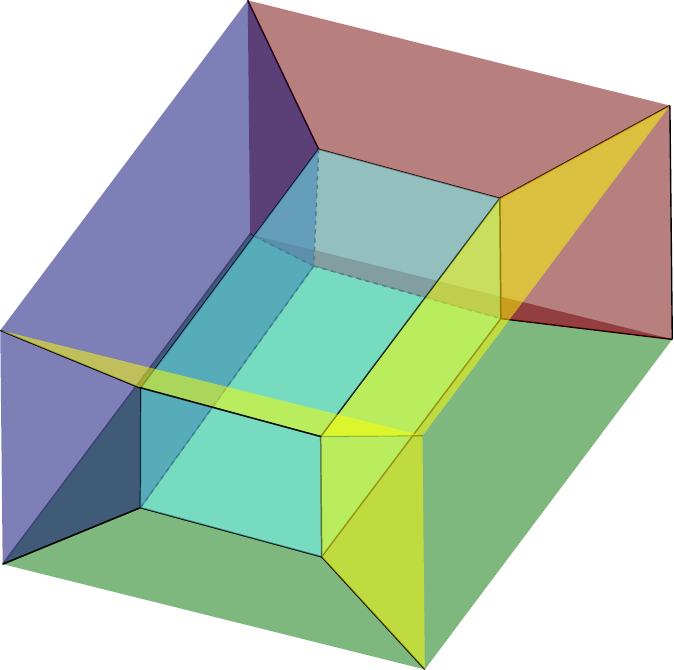}	
		}%
		\subfloat[][]{		\label{subfig:col_1stratum}
			\includegraphics[width=73.8pt]{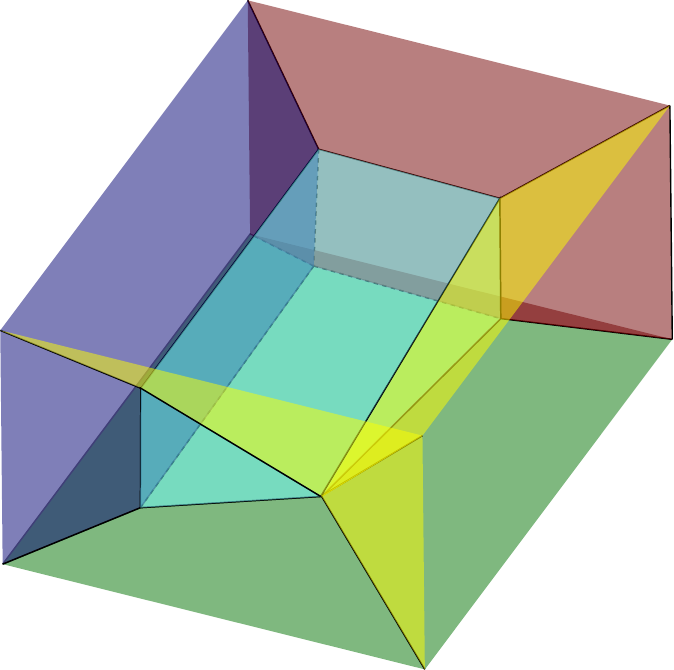}	
		}%
		\qquad
		\subfloat[][]{		\label{subfig:col_2stratum}
			\includegraphics[width=73.8pt]{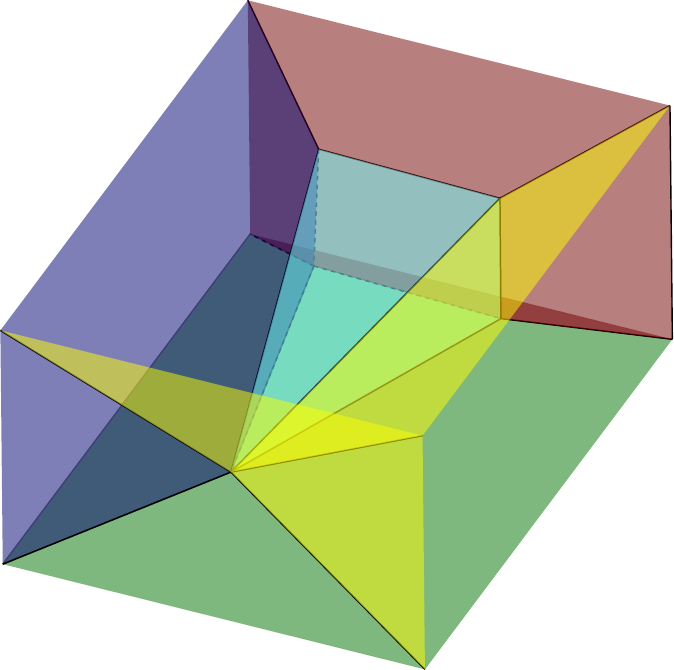}	
		}%
		\subfloat[][]{		\label{subfig:col_3stratum}
			\includegraphics[width=73.8pt]{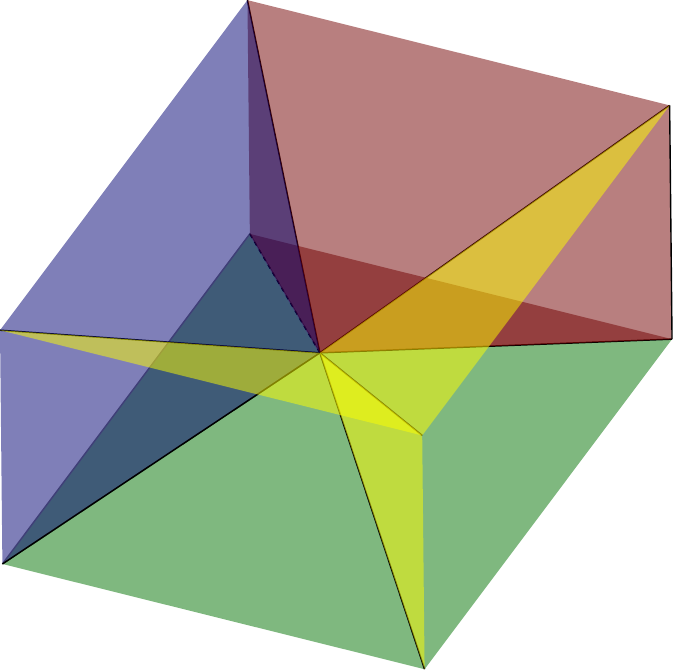}
		}%
		\caption{The cases of collapse shown on the rectangular prism example. 
			(a) The initial microstructure. 
			(b) Line collapse. 
			(c) Surface collapse. 
			(d) Volume collapse.}
		\label{fig:collapse}
	\end{figure}

	The average grain size increases during grain growth, meaning components of the grain boundary network should generally vanish. The criterion for this topological transition in practice is that the length of a line, area of a surface, or volume of a grain is shrinking and passes below a threshold tied to the overall scale of the microstructure. The collapse is effected by merging all of the bounding points and adjusting the adjacency lists of the surrounding components as appropriate. Examples of this operation are shown in Figure \ref{fig:collapse}, with several specifics of the algorithm given in Section \ref{app:gen_col} of the supplementary material (SM). 
	
	\subsection{Stratum insertions} \label{subsec:stratum_ins}
	\begin{figure}
		\centering
		\subfloat[][]{		\label{subfig:adj1}
			\includegraphics[width=54.5pt]{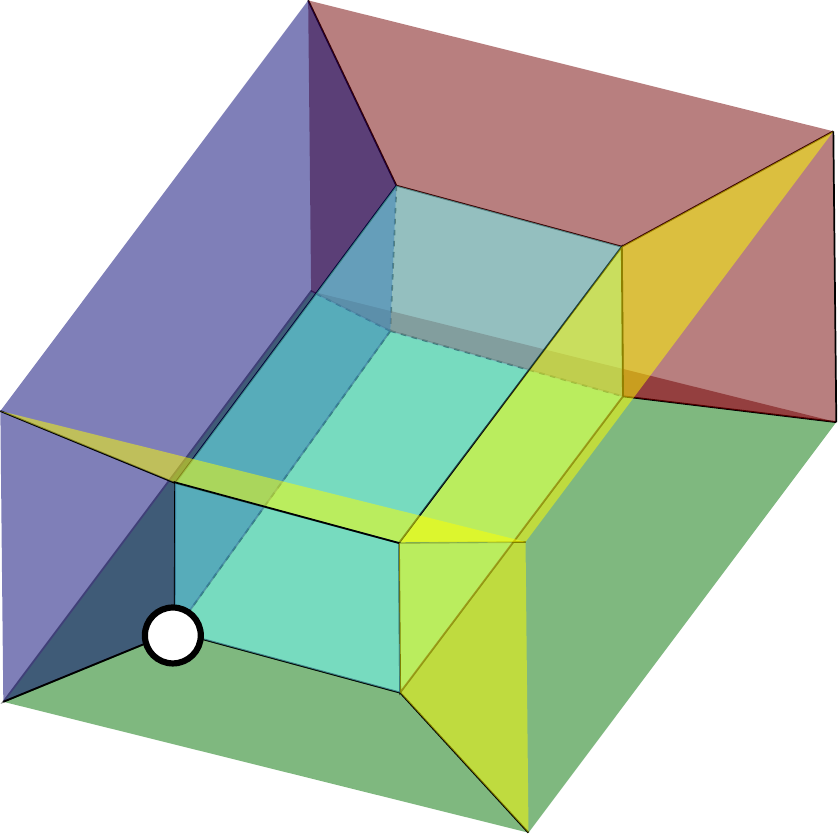}
		}%
		\subfloat[][]{		\label{subfig:adj2}
			\includegraphics[width=54.5pt]{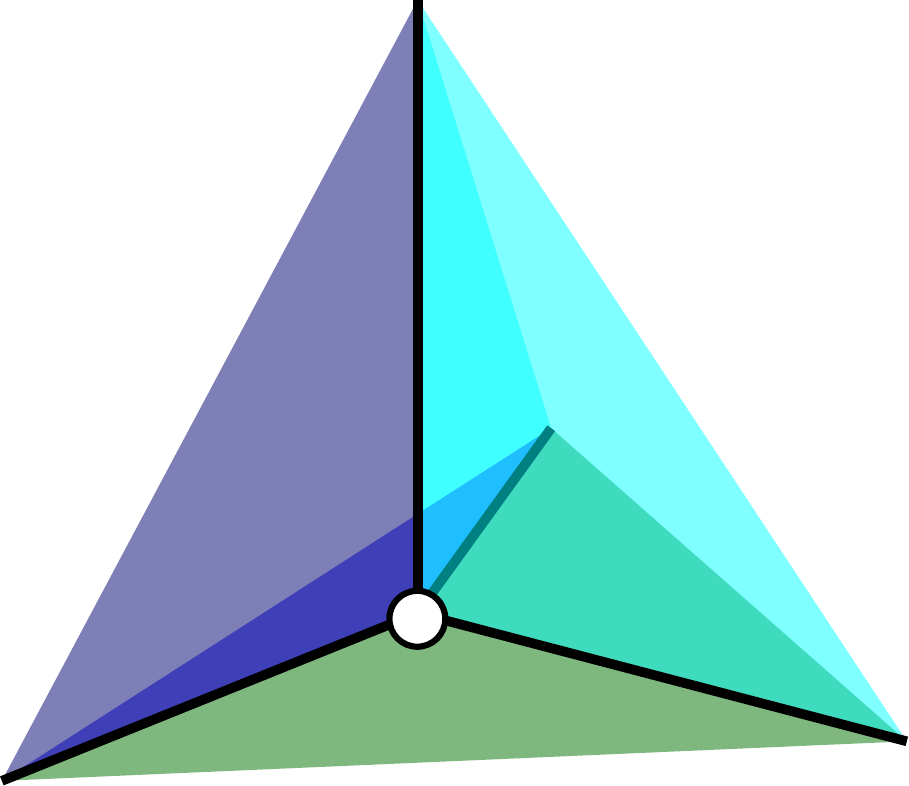}
		}%
		\subfloat[][]{		\label{subfig:adj3}
			\includegraphics[width=54.5pt]{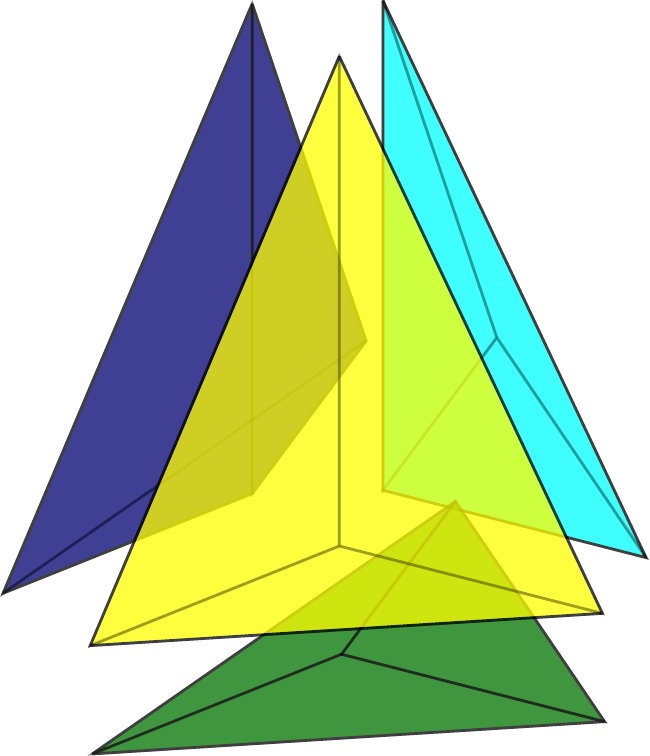}
		}%
		\subfloat[][]{		\label{subfig:adj4}
			\includegraphics[width=54.5pt]{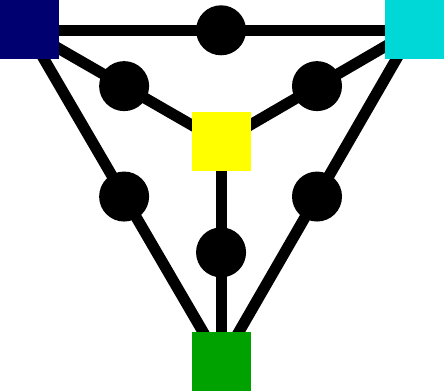}
		}%
		\qquad
		\caption{(a) Consider the point at the bottom left corner of the central volume. (b) The neighborhood of the point shows the relationships with the surrounding surfaces and volumes. (c) The volumes in an exploded view. (d) The adjacency graph showing the volumes as squares and the surfaces as disks. In this figure, volumes and squares are the same color.}
		\label{fig:adjacency}
	\end{figure}
	
	Often the configuration resulting from a stratum collapse is unstable and the energy could be lowered by splitting the point to insert a line or a surface. There are usually many such possible insertions, and the identification of the most likely one necessarily involves enumerating these possibilities. This analysis can be performed using the adjacency graph of surfaces and volumes. The adjacency graph is constructed by placing a node for each volume and surface and an edge between adjacent volumes and surfaces. The steps involved are shown in Figure \ref{fig:adjacency} for a particular point. Formally, for non-periodic microstructures, there is a volume surrounding the simulation cell that is connected to the surfaces bounding the simulation cell. For the purpose of enumerating the possible insertions, this is treated similarly to the volumes within the simulation cell, with the specifics given in Section \ref{app:shell} of the SM. 
	
	\begin{figure}
		\centering
		\subfloat[][]{		\label{subfig:circuit_stratum}
			\includegraphics[height=135.3pt]{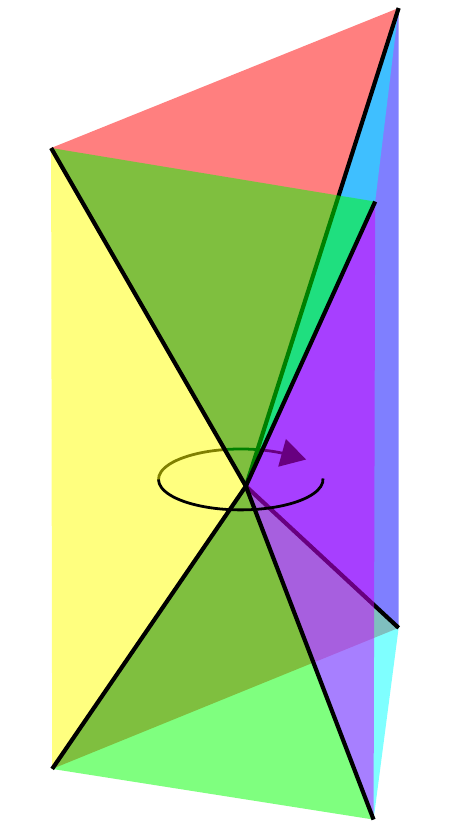}
		}%
		\subfloat[][]{		\label{subfig:1stratum_ins}
			\includegraphics[height=135.3pt]{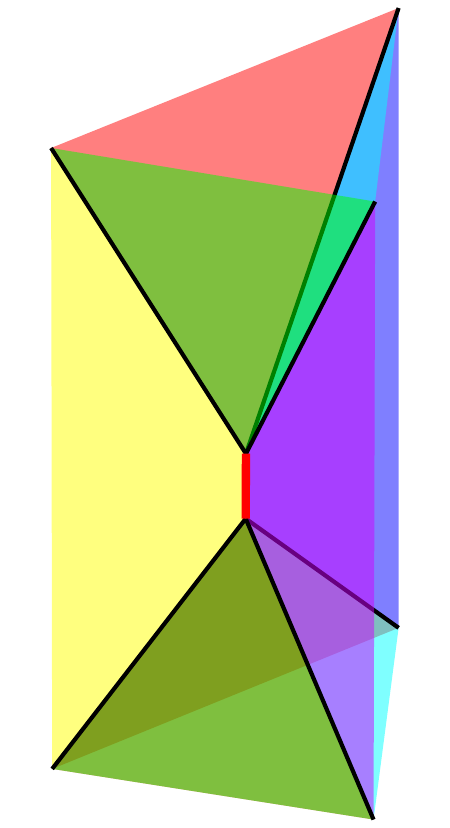}
		}%
		\subfloat[][]{		\label{subfig:circuit_g}
			\includegraphics[height=135.3pt]{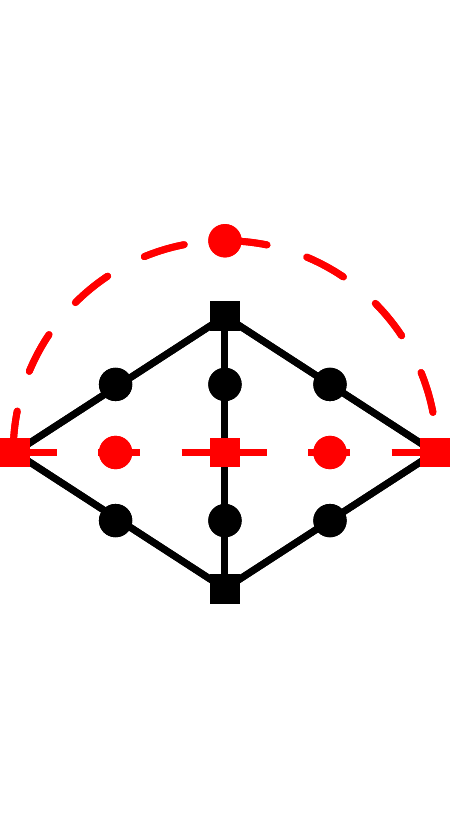}
		}%
		\qquad
		\caption{A line insertion corresponds to a circuit on the adjacency graph. (a) A five grain configuration and a circuit going around the point. (b) Every surface punctured by the circuit is extended by adding the inserted line to their adjacency lists. (c) The adjacency graph around the point. Edges along the circuit are dashed.}
		\label{fig:1stratum_ins}
	\end{figure}
	
	\subsubsection{Line insertions}
	Every possible line insertion corresponds to a circuit on the associated adjacency graph with one example shown in Figure \ref{fig:1stratum_ins}. This configuration frequently occurs for isotropic grain boundary energies, e.g., when a triple line collapses and two quadruple points are merged. The circuit shown in Figure \ref{subfig:circuit_stratum} passes through the front, left and right volumes, and every surface that is punctured by the circuit is adjacent to the inserted line. The circuit in Figure \ref{subfig:circuit_stratum} precisely corresponds to the circuit in Figure \ref{subfig:circuit_g}, and enumerating all possible line insertions is equivalent to enumerating all circuits on the adjacency graph. Algorithms to identify the circuits on a graph are available in the literature \cite{1969CommunACMPaton, 1969JourACMGibbs}. Not all possible circuits need to be considered though; if removing the nodes and edges of the circuit from the adjacency graph leaves only a single connected component, then the line insertion would create a spurious line and point that would subsequently be removed. The resulting algorithm is  described in detail in Section \ref{app:c_ins} of the SM.

	\begin{figure}
		\centering

		\subfloat[][]{		\label{subfig:2stratum_path}
			\includegraphics[height=135.3pt]{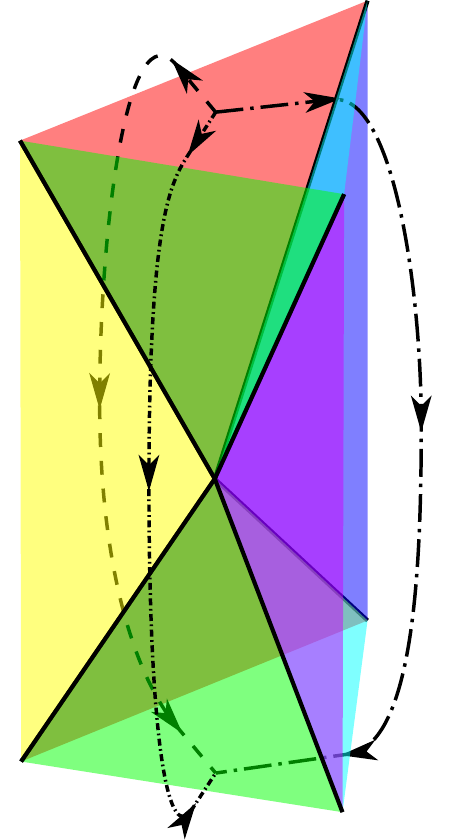}
		}%
		\subfloat[][]{		\label{subfig:2stratum_trigon}
			\includegraphics[height=135.3pt]{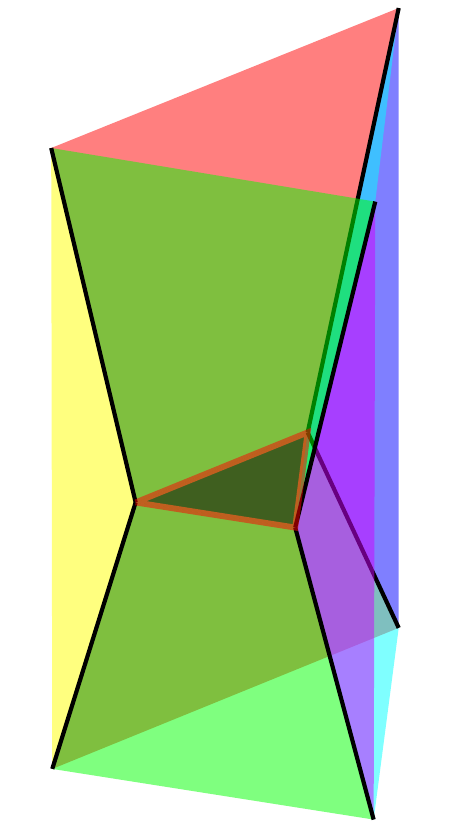}
		}%
		\subfloat[][]{		\label{subfig:2stratum_g}
			\includegraphics[height=135.3pt]{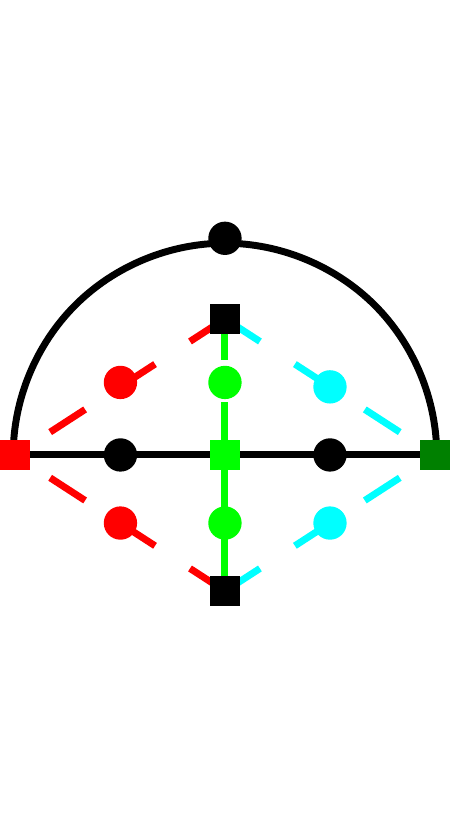}
		}%
		\caption{A surface insertion corresponds to a set of paths on the adjacency graph. (a) A five grain configuration, showing a set of three non-intersecting paths connecting the disconnected (top and bottom) volumes. (b) A surface is inserted between the disconnected volumes with one bounding line for each path. Each line is added to the adjacency lists of the surfaces punctured by the corresponding path.
			(c) The adjacency graph around the point. The color of punctured surfaces and edges on the graph match on (a) and (c).
		}
		\label{fig:2stratum_ins}
	\end{figure}
	
	\subsubsection{Surface insertions}
	Around a point a surface can only be inserted between two disconnected volumes. Given a pair of such volumes, the inserted surface is connected to the surrounding surfaces by some set of inserted lines. Each line corresponds to a path that starts on one of the disconnected volumes and ends on the other, as in Figure \ref{subfig:2stratum_path}. A set of such paths completely specifies the topology around the inserted surface. Every surface punctured by a path is adjacent to the corresponding inserted line, as in Figure \ref{subfig:2stratum_trigon}. The set of all possible surface insertions can be found by constructing all possible sets of non-intersecting paths between the nodes of the adjacency graph corresponding to the disconnected volumes. These paths can be found using a standard depth first search algorithm on the adjacency graph. Unlike line insertions, paths along surfaces that share a common edge are still acceptable, as the newly inserted line will bound the inserted surface and will be topologically different from any preexisting line. The resulting algorithm is described in detail in Section \ref{sec:c2_ins} of the SM.
	
	\subsection{Other considerations}
	
	\begin{figure}%
		\centering

		\subfloat[][]{		\label{subfig:non1c}
			\includegraphics[width=61.5pt]{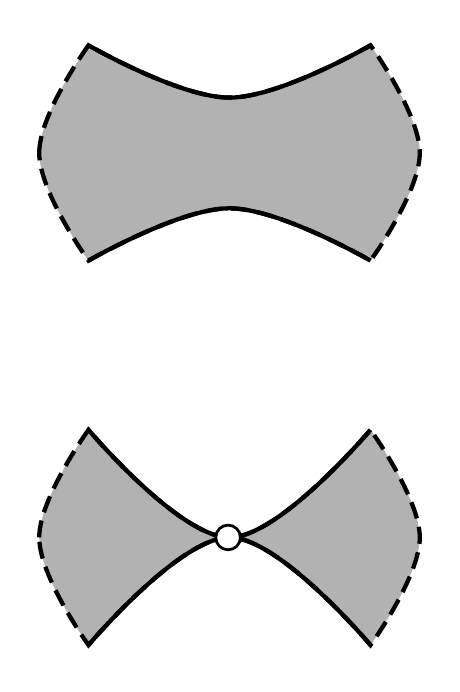}
		}%
		\subfloat[][]{		\label{subfig:non2c}
			\includegraphics[width=61.5pt]{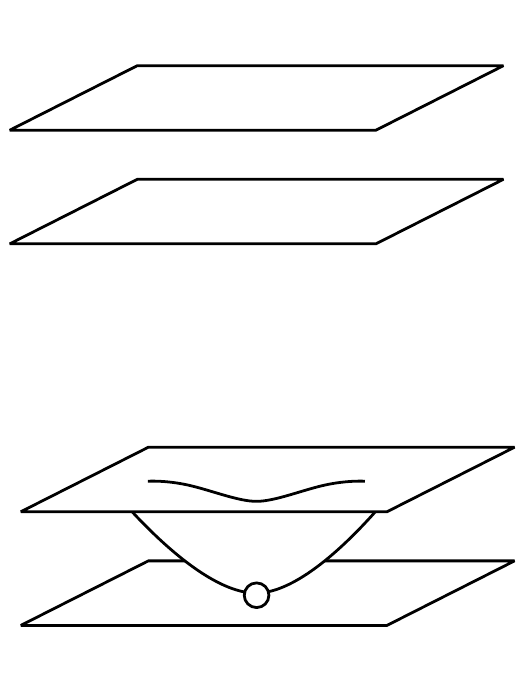}
		}%
		\subfloat[][]{		\label{subfig:non3c}
			\includegraphics[width=61.5pt]{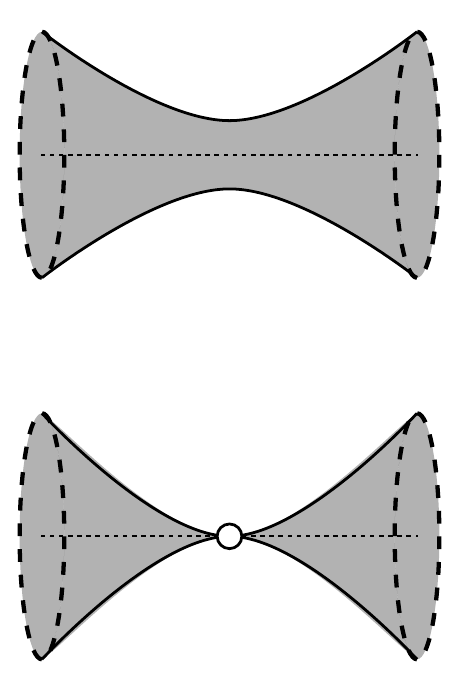}
		}%
		\caption{Topological transitions not considered here. (a) Two lines bounding the same surface meet to form a new point. (b) Two bounding surfaces of a volume meet to form a new point. (c) The cross section of a cylindrical volume is reduced to a point. }
		\label{fig:singular}
	\end{figure}
	
	The algorithms described in this section are conjectured to result in sets of topological transitions that include all those that occur during grain growth for a generic initial condition, even with anisotropic grain boundary energies. A generic initial condition is one for which the type of topological transition shown in Figure \ref{fig:singular} does not occur. That is, the only allowed topological transitions are those for which the length of a line, the area of a surface, or the volume of a grain passes through zero. This is not believed to be a serious constraint though, since the topological transitions in Figure \ref{fig:singular} are not expected to occur during grain growth in a generic physical system.

	\begin{figure}
		\centering
		\includegraphics[width=130pt]{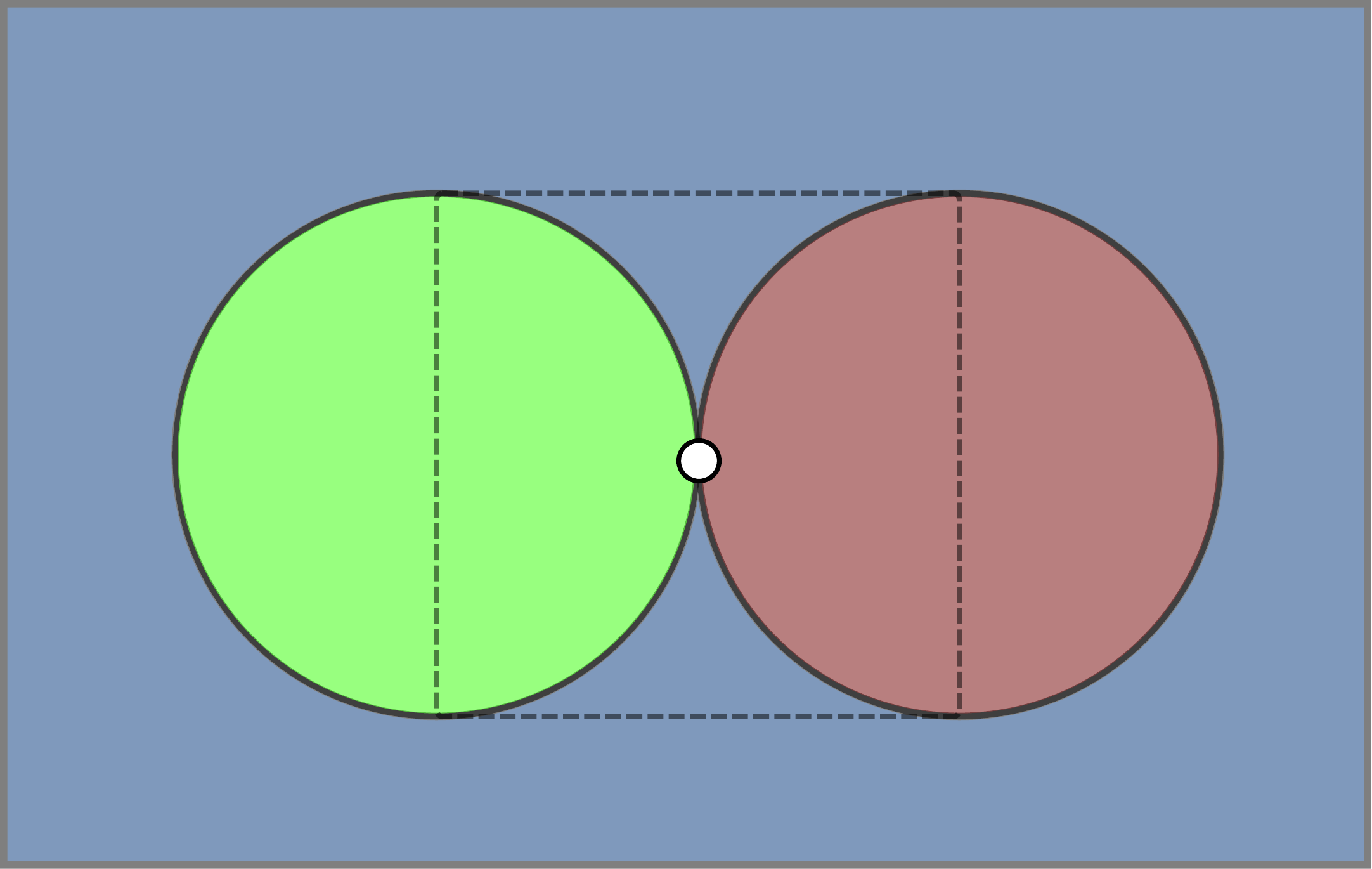}
		\caption{A point connected to two spherical grains, and two grains above and below the page. The neighborhood of the point is outlined by a dashed line. The surface in the page is represented twice in the neighborhood of the point. }
		\label{fig:spurious}
	\end{figure}
	
	There are some situations where the adjacency graph of the strata does not accurately reflect the topology around a point. For example, a single point could appear on the boundary of a surface more than once, as in Figure \ref{fig:spurious}. This is the reason that the adjacency graph is constructed from the microstructure components in a small neighborhood of the point. This can allow spurious insertions (in the sense of Section \ref{sec:MSrepresentation}) that are nevertheless required by the physical system, and any spurious strata can easily be removed after the topological transition is complete. The detection algorithm for spurious strata is provided in Section \ref{sec:spurious} of the SM. The construction of a small neighborhood necessarily involves the mesh, and will be considered further in Section \ref{subsec:tmesh}.

	\section{Operations on the mesh}
	
	Since the SCOREC library does not natively support changes to the topology of the finite element mesh, a set of fundamental and localized operations are proposed and implemented. Given that the microstructure is represented by means of a finite element mesh, the individual microstructure components are comprised of sets of simplicial mesh elements. These mesh elements will be referred to as tetrahedra, triangles, edges, and vertices, or occasionally as $ d $-simplices when that is simpler. The distinction between the topological and geometric components of the microstructure is reinforced in Figure \ref{fig:FEM}.
	
	Applying the stratum collapse and insertion operations described in Section \ref{sec:microstructure} on a simplicial finite element mesh requires some mesh modifications, both to prepare the mesh for these changes and to execute them. The two basic operations are lens collapse and lens expansion, associated with stratum collapse and insertion, respectively.  The lens split is an additional operation used to prepare the mesh around a stratum before stratum collapse or in the neighborhood of a point before stratum insertion. While the actual collapse and insertion operations are more complex than those described below, the underlying approach is the same.
	
	Remembering that the set of volumes, faces, lines and points and their connections compromise a topological structure called a stratified space, microstructural components will be called strata in this section, i.e., a volume will be called a $ 3 $-stratum, a surface will be called a $ 2 $-stratum, a line will be called a $ 1 $-stratum, and a point will be called a $ 0 $-stratum. For brevity, $ S^d $ will denote a $ d $-stratum and $ S^{d}_i $ more specifically the $ i $th $ d $-stratum. 
	
	\subsection{Stratum collapse}\label{subsec:gencolmesh}
	
	An $ S^d $ with $ d > 0 $ is represented by a collection of $ e $-dimensional mesh entities with $ e = 0,1,\dots,d $. Collapsing an $ S^d $ is equivalent to collapsing its constituents onto a single vertex. This can be further simplified to collapsing all edges within the $ S^d $ and its bounding strata, giving the central idea of stratum collapse. For simplicity, this section describes the procedure for a single collapsing edge. This is extended in Section \ref{app:generalized_collapse} of the SM to stratum collapses involving multiple collapsing edges.
	
	\begin{figure}
		\centering
		\includegraphics[width=196.8pt]{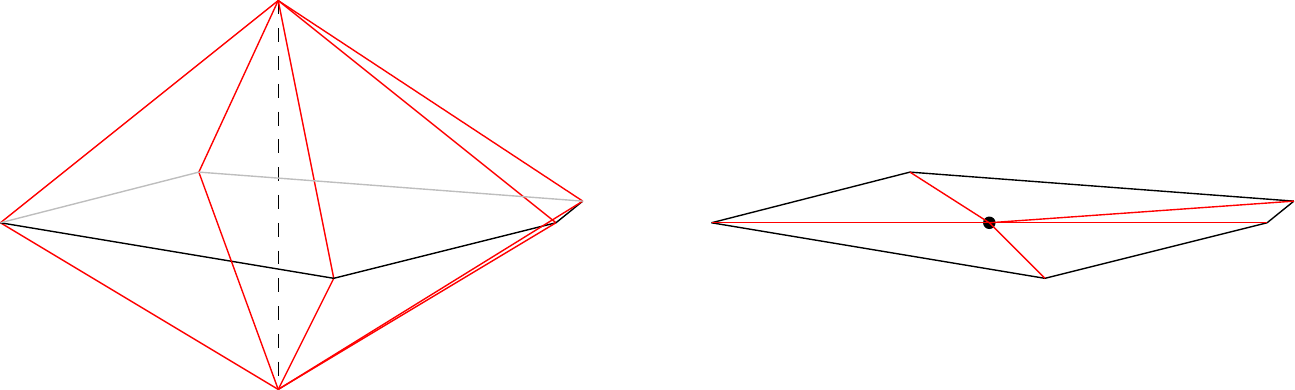}
		\caption{Lens collapse operation. Left, the lens composed of tetrahedra and triangles bounded by the collapsing dashed edge. Right, the disc obtained by collapsing the lens.}
		\label{fig:lenscol}
	\end{figure}
	
	On a simplicial mesh, an edge bounds a collection of tetrahedra and triangles forming a lens around that edge. As shown in Figure \ref{fig:lenscol}, the entities that are bounded by the collapsing edge will also collapse and need to be removed. For each collapsing triangle, the other two bounding edges form a merging couple. For each collapsing tetrahedron, the two triangles that are not collapsing form a merging couple. After the collapse, a new entity is generated for each merging couple. Such an entity belongs to the lower dimensional stratum of the merging couple, assuming the merging entities belong to the same or adjacent strata. 
	
	\begin{figure}
		\centering
		\includegraphics[width=196.8pt]{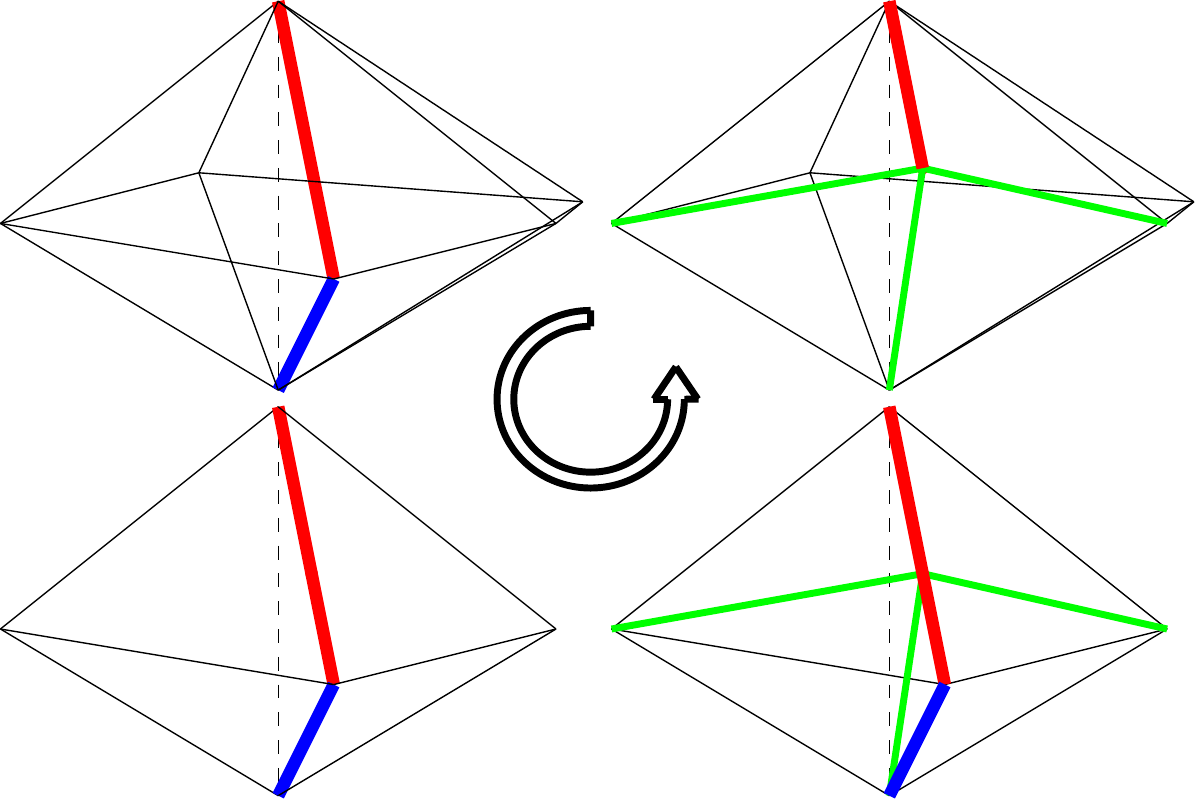}
		\caption{Edge split operation during preconditioning. The thicker edges in red and blue belong to strata $ S^{d}_{i} $ and $ S^{e}_{j} $, respectively. If $ S^{d}_{i} $ and $ S^{e}_{j} $ are not the same and one doesn't bound the other, collapse of the dashed vertical edge is not allowed. Splitting the red edge and all entities that are bounded by that edge into two creates new entities which by construction either belong to $ S^{d}_{i} $ or strata bounded by $ S^{d}_{i} $. }
		\label{fig:edgesplit}
	\end{figure}
	
\begin{figure}
		\centering

		\subfloat[][]{		\label{subfig:pc_merg}
			\includegraphics[width=76pt]{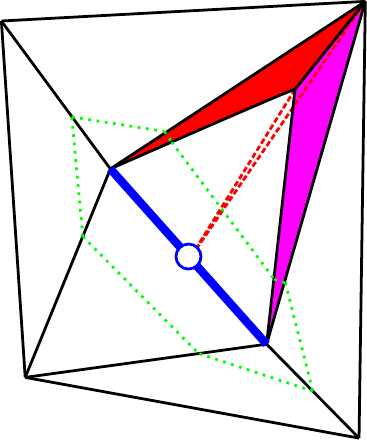}
		}%
		\subfloat[][]{		\label{subfig:pc_split}
			\includegraphics[width=76pt]{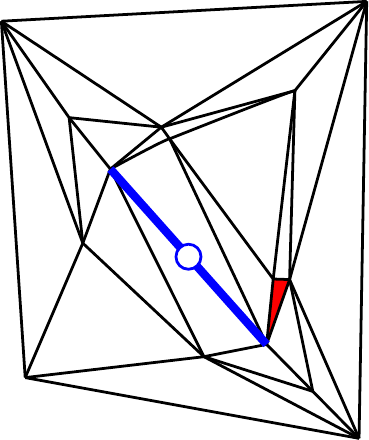}
		}%
		\subfloat[][]{		\label{subfig:pc_relax}
			\includegraphics[width=76pt]{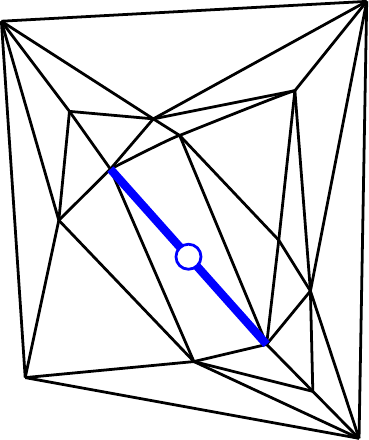}
		}%
		\caption{The effect of preconditioning for an $ S^1 $ collapse on a two-dimensional mesh. (a) Collapsing the blue $ S^1 $ and moving the vertices to the blue node would invert the red triangle and merge it with the purple triangle. The resulting triangle is shown in dashed lines. (b) The splitting procedure resolves this problem, but yields the red triangle that could invert during collapse. (c) Relaxation allows the $ S^1 $ to be collapsed without inverting any elements. }
		\label{fig:preconditioning}
	\end{figure}

	During the stratum collapse, three issues could arise that would invalidate the mesh. First, stratum collapse could cause an additional topological transition if any of the merging entities do not belong to the same or adjacent strata. Applying the edge split operation shown in Figure \ref{fig:edgesplit} to one of the edges of the problematic couple resolves this situation. Second, it is possible that two $ d $-dimensional entities could unintentionally merge. This could occur even if they do not belong to the the collapsing lens, but requires that they share $ d-1 $ vertices and that the remaining vertex of each be a distinct merging vertex as in Figure \ref{subfig:pc_merg}. The edge split procedure can also resolve this by isolating the collapsing entity, as shown in Figure \ref{fig:preconditioning}. A third issue that would invalidate the mesh is inversion of one of the surrounding entities during a collapse. This could occur if the initial and final positions of a merging vertex lie on distinct sides of the plane containing the opposite triangle of an adjacent tetrahedron.
	
	The three-dimensional equivalent of the preconditioning operation in Figure \ref{fig:preconditioning} is applied to edges that are adjacent to a single merging vertex to avoid all three situations. First, the midpoints of all edges emanating from the merging vertices are collected to compute their convex hull, and the emanating edges are split where they intersect the convex hull. This resolves the first two issues and yields a hull of triangles surrounding the collapsing stratum. While it is still possible for a surrounding tetrahedron to invert during the collapse, a relaxation procedure analogous to that in Figure \ref{subfig:pc_relax} and described in Section \ref{app:relaxation} of the SM is applied to vertices on the hull to avoid such an event. After preconditioning, the stratum memberships of the new entities associated with the merging entities are found. A new entity belongs to the lowest dimensional stratum that owns one of the merging entities; the preconditioning certifies that there is a single stratum of the lowest dimension.
	
	During the course of microstructure evolution, the criterion for collapsing a stratum is decided at the mesh level with a two step algorithm. First, the diameter of a stratum is approximated as that of an edge, square or cube with the same length, area or volume, respectively. If the diameter of a $ S^d $ is smaller than a threshold, then the time rate of change of the total length, area, or volume of the collapsing stratum is calculated using the velocities associated with the bounding vertices. If this is negative, then the stratum is collapsed. 

	\subsection{Stratum insertion}\label{subsec:insmesh}
	
	As described in Section \ref{subsec:stratum_ins}, the insertion of a $ S^1 $ or $ S^2 $ around a central $ S^0 $ initially involves finding circuits or paths in the adjacency graph of surfaces and volumes. For this to work on the mesh level, there should be at least one internal edge in each of the surrounding $ S^2 $ and $ S^3 $. This is ensured by two operations. First, a lens expansion is applied to each connected set of tetrahedra belonging to the same $ S^3 $. The $ S^2 $ triangles bounding such a set and adjacent to the $ S^0 $ form a disc that can be expanded. The expansion forms a new vertex, a new edge and a set of new triangles and tetrahedra corresponding to the disc triangles, all belonging to the specified $ S^3 $. Second, if there are any sets of connected triangles belonging to a $ S^2 $ that consist of a single triangle, the edge opposite the $ S^0 $ is split. Next, the split operation is applied to the edges bounded by the central vertex belonging to the $ S^0 $. The vertices created by these split operations are positioned on a sphere centered at the $ S^0 $ vertex location. The radius $ \rho $ of the sphere is smaller than the distance to the closest triangle opposite the central vertex in any surrounding tetrahedron. 
	
	Preconditioning achieves three things. First, it ensures that corresponding sets of triangles and edges can be found for each circuit associated with a $ S^1 $ insertion and each path associated with a $ S^2 $ insertion. These sets of triangles and edges form disc- or fin-like structures. Second, it forms a convex cavity of triangles, preventing element inversion after the insertion. Third, it reduces the size disparity of the surrounding triangles and the associated bias in the numerical scheme for vertex velocities.

	Stratum insertion requires expansion of a disc/fin, creation of triangles and tetrahedra with the same stratum membership as the edges and triangles on the disc/fin, and creation of edges and triangles belonging to the new strata. In the case of a $ S^1 $ insertion, a new $ S^0 $ vertex and a new $ S^1 $ vertex to be positioned at the interior of the new line are created. The disc associated with the circuit is used to create three discs, one for the old $ S^0 $ vertex, one for the new $ S^1 $ vertex, and one for the new $ S^0 $ vertex such that the disc entities belong to the same strata as in the initial disc. Two new $ S^1 $ edges are created to connect the $ S^1 $ vertex to the bounding $ S^0 $ vertices. The volume between the discs and around the new $ S^1 $ edges is filled by triangles and tetrahedra corresponding to edges and triangles on the discs.	In the case of a $ S^2 $ insertion, the entities bounded by the new $ S^2 $ entities need to be generated. A triangle belonging to the new $ S^2 $ is generated for each new $ S^1 $ edge, and a new tetrahedron belonging to the adjoining $ S^3 $ is generated for each new $ S^2 $. When inserting strata on a $ S^0 $ on the boundary of the simulation, the algorithm skips the creation of entities for the exterior $ S^3 $. The final step of the insertion is the relaxation described in Section \ref{sec:energy}.

	\subsection{Spurious stratum detection and insertion}\label{subsec:tmesh}
	
	\begin{figure}
		\centering
		
		\subfloat[][]{		\label{subfig:spur_i}
			\includegraphics[width=76pt]{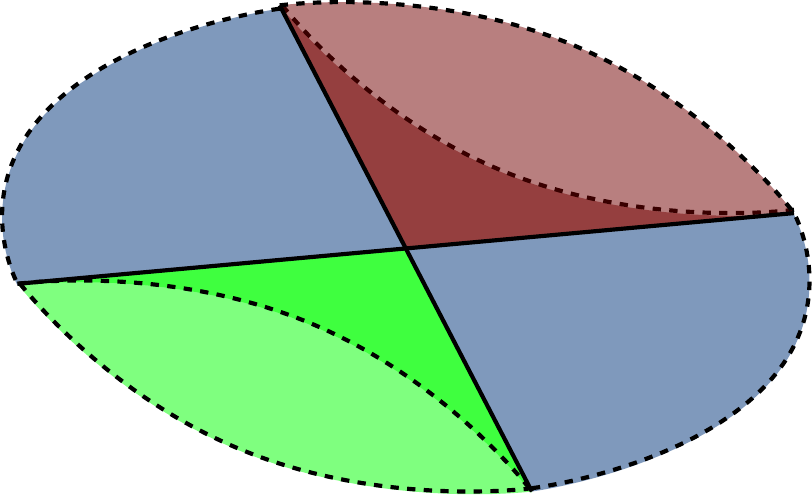}
		}%
		\subfloat[][]{		\label{subfig:spur_m}
			\includegraphics[width=76pt]{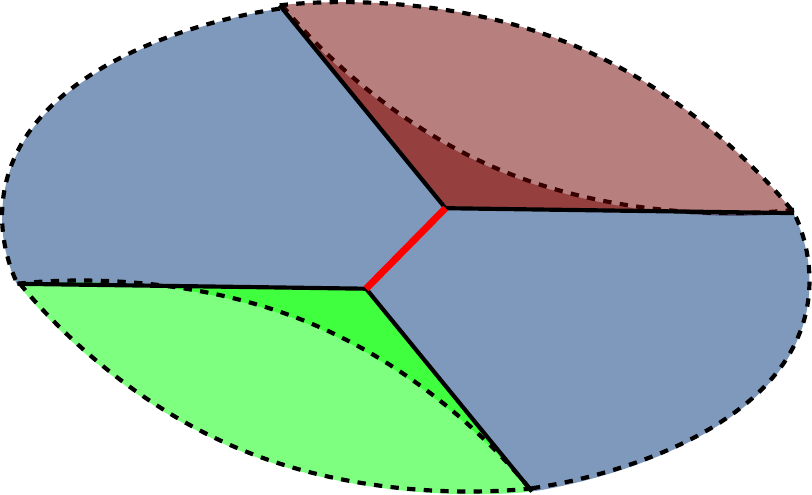}
		}%
		\subfloat[][]{		\label{subfig:spur_f}
			\includegraphics[width=76pt]{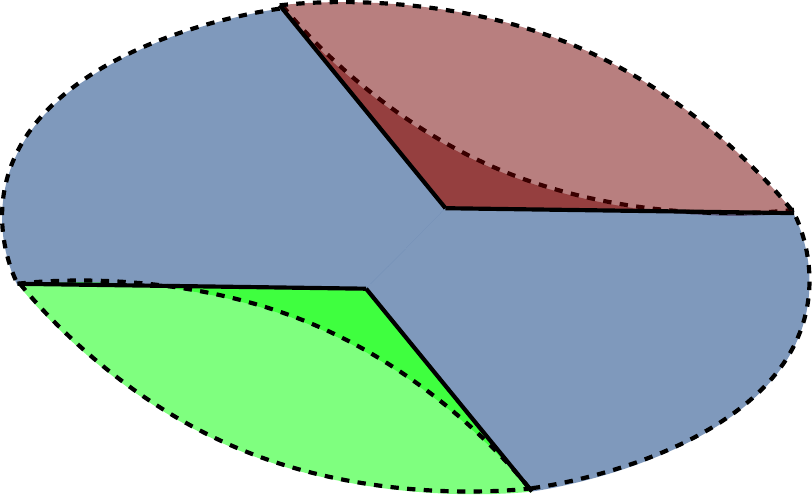}
		}%
		\caption{Steps of spurious line insertion. (a) A point, connected to two volumes top and bottom, and two grains above and below the page. (b) Insertion of the spurious line, adjacent to two surfaces both separating the same volumes. (c) The spurious line is removed and the two surfaces are merged. }
		\label{fig:spur}
	\end{figure}
	If an inserted stratum has fewer than the minimum number of higher-dimensional adjacencies, it is spurious and is removed by merging the higher-dimensional adjacencies. An example is given in Figure \ref{fig:spur}. This operation is sometimes necessary, e.g., when a $ S^0 $ is connected to multiple disjoint sets of triangles belonging to the same $ S^2 $ or disjoint sets of tetrahedra belonging to the same $ S^3 $. In this situation, the global connectivity of the stratification is not representative of the possible local insertions around the vertex. A local stratification of disjoint sets of entities belonging to the same stratum is generated, and the set of all possible insertions is found with the same circuit and path detection method as in the generic case. 

	\section{Boundary evolution and energy criteria} \label{sec:energy}
	
	\begin{figure}
		\centering
		
		\subfloat[][]{		\label{subfig:2dins11}
			\includegraphics[width=51.1pt]{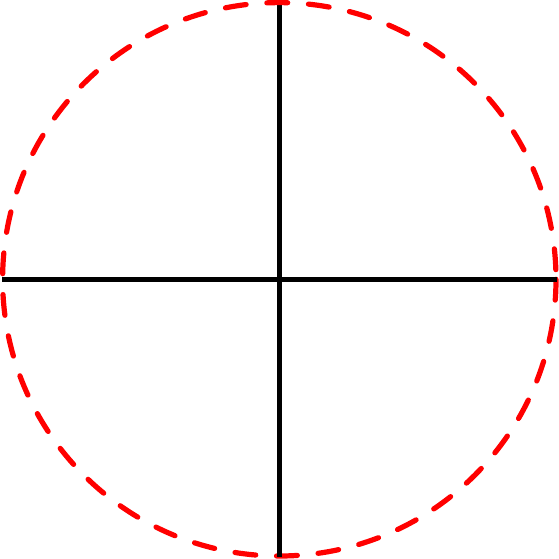}
		}%
		\subfloat[][]{		\label{subfig:2dins12}
			\includegraphics[width=51.1pt]{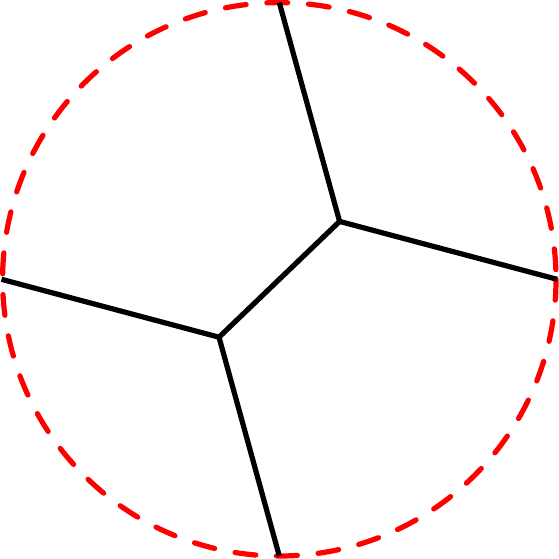}
		}%
		\subfloat[][]{		\label{subfig:2dins13}
			\includegraphics[width=51.1pt]{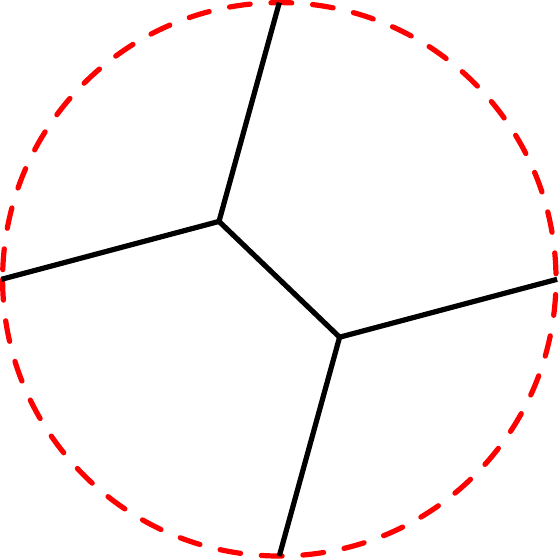}
		}%
		\\
		\subfloat[][]{		\label{subfig:2dins21}
			\includegraphics[width=51.1pt]{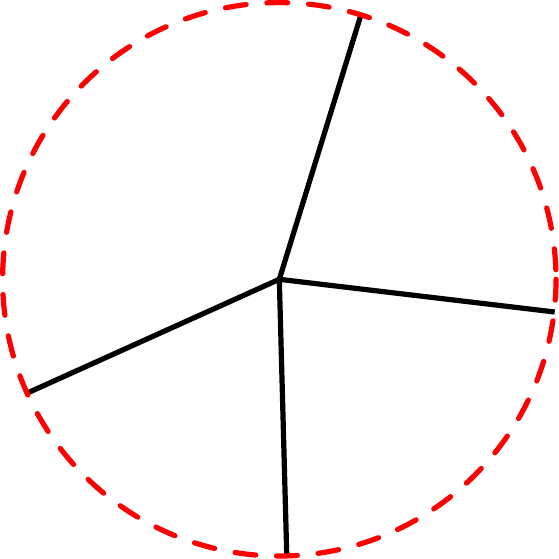}
		}%
		\subfloat[][]{		\label{subfig:2dins22}
			\includegraphics[width=51.1pt]{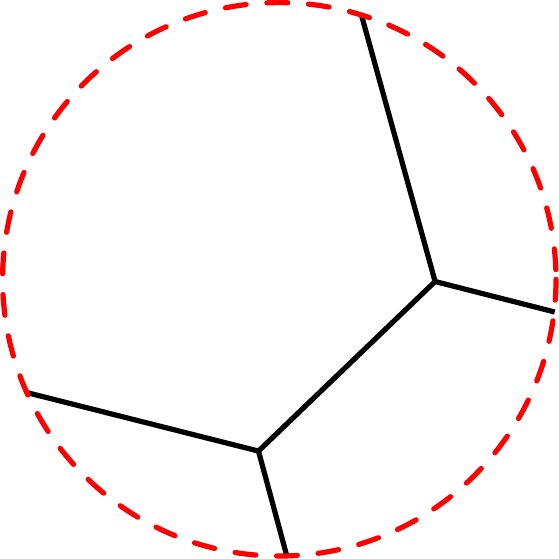}
		}%
		\subfloat[][]{		\label{subfig:2dins23}
			\includegraphics[width=51.1pt]{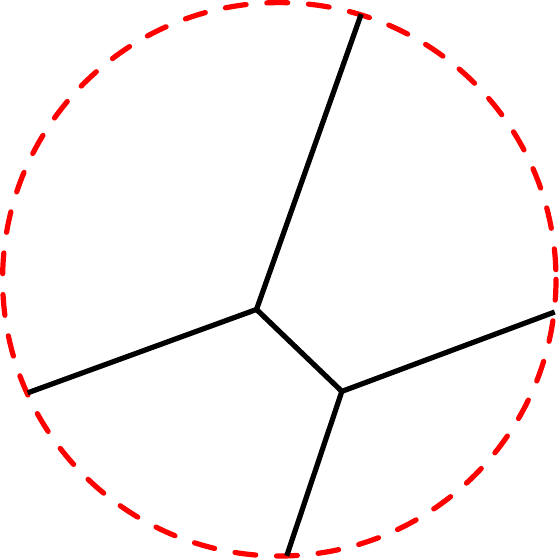}
		}%

		\caption{The choice of insertion can change the overall trajectory of the system. (a) A two-dimensional degenerate configuration with four grains could transition to either (b) or (c) since they are energetically equivalent. For (d), (e) and (f) both lower the energy, but (e) more so. }
		\label{fig:2dins}
	\end{figure}
	
	When inserting a new stratum, it is important that the geometry of the stratum maximizes the energy dissipation rate as the stratum expands. This is especially important when there is more than one possible stable insertion, as shown in Figure \ref{fig:2dins}. Even for a constant grain boundary energy, inaccurate calculations of the geometry could change the selected insertion and drastically alter the evolution of the system. 

	\begin{figure*}
		\centering
		
		\subfloat[][]{		\label{subfig:reorient_i}
			\includegraphics[width=103.3pt]{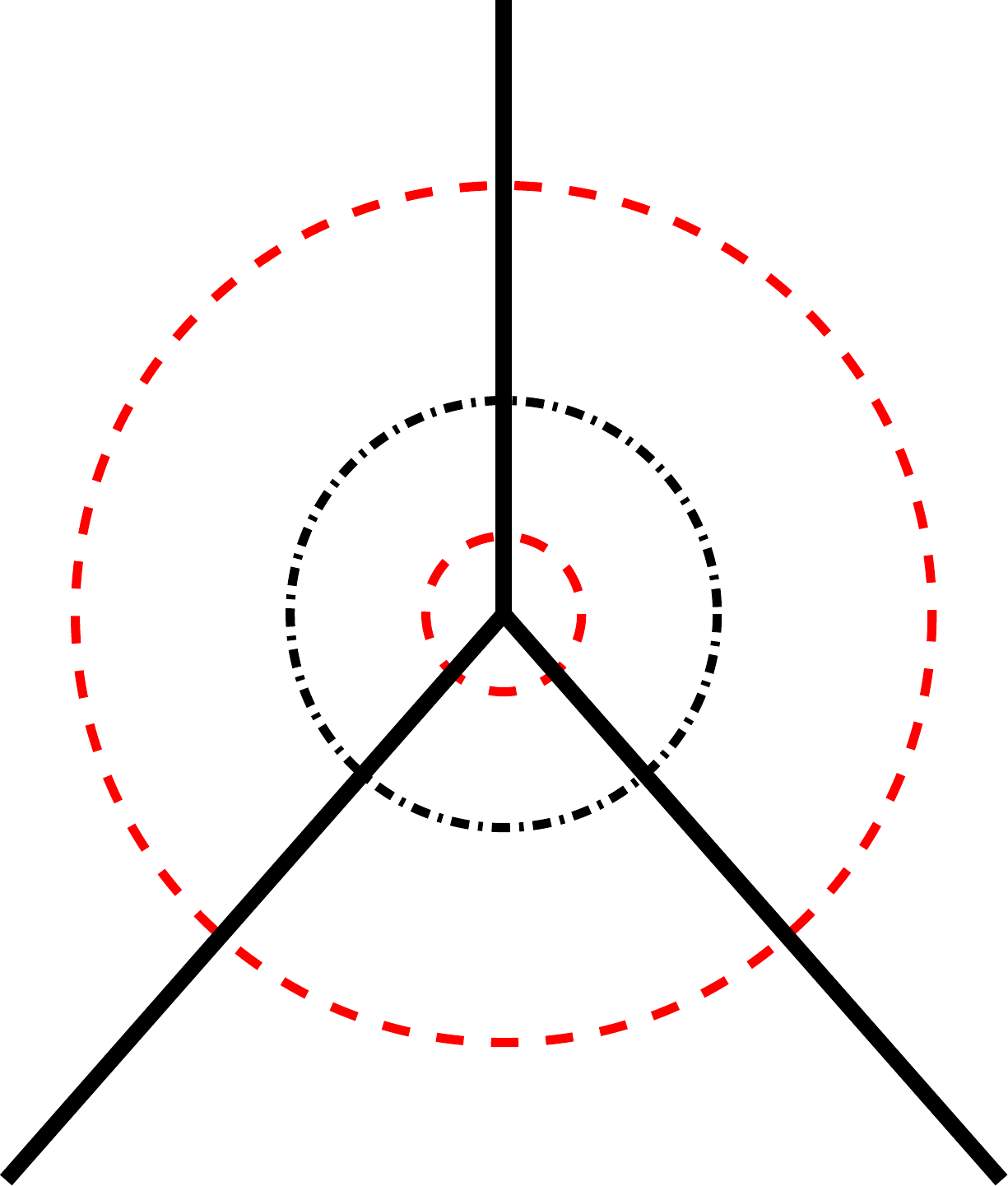}
		}%
		\subfloat[][]{		\label{subfig:reorient_ins}
			\includegraphics[width=103.3pt]{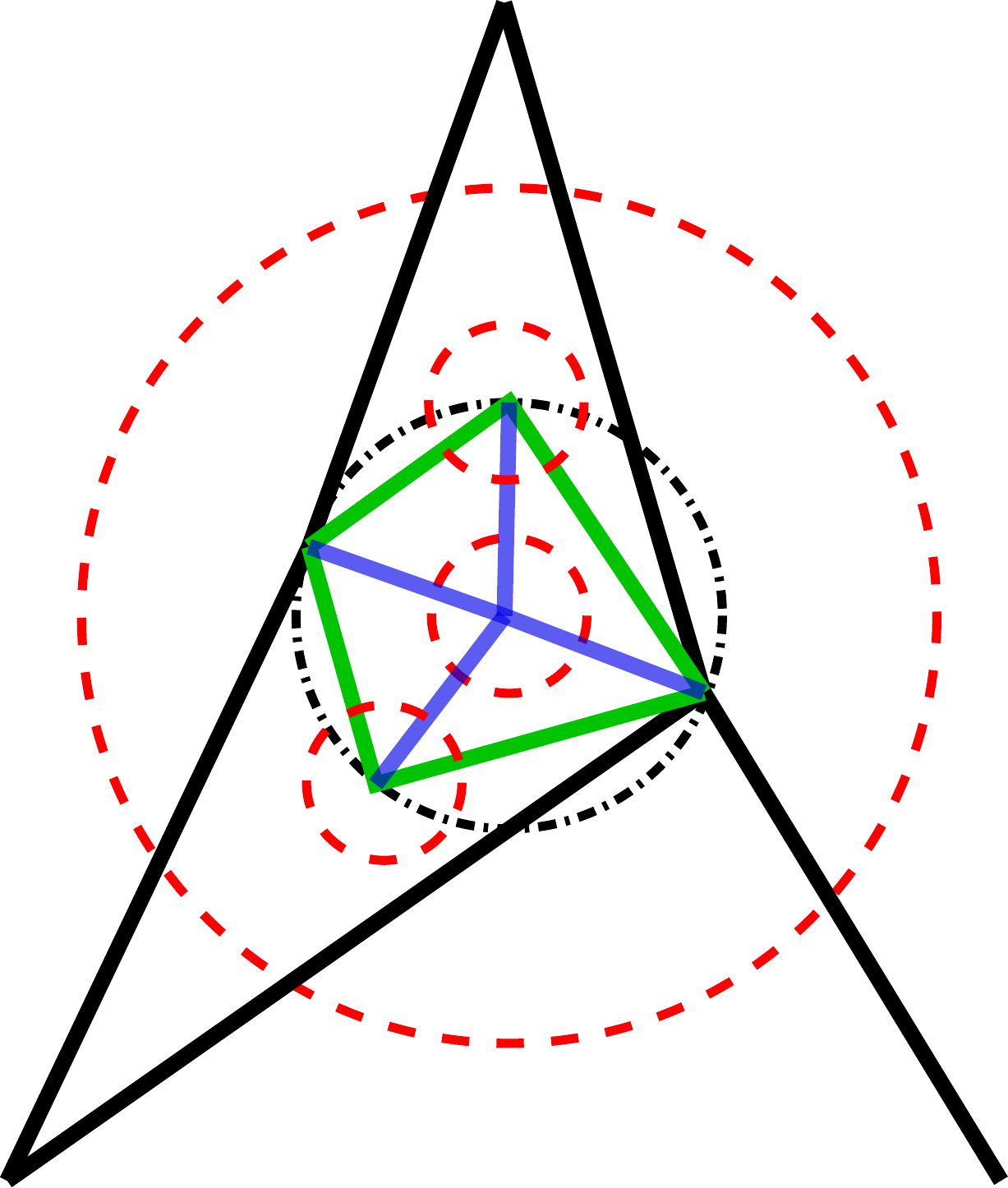}
		}%
		\subfloat[][]{		\label{subfig:reorient_exp}
			\includegraphics[width=103.3pt]{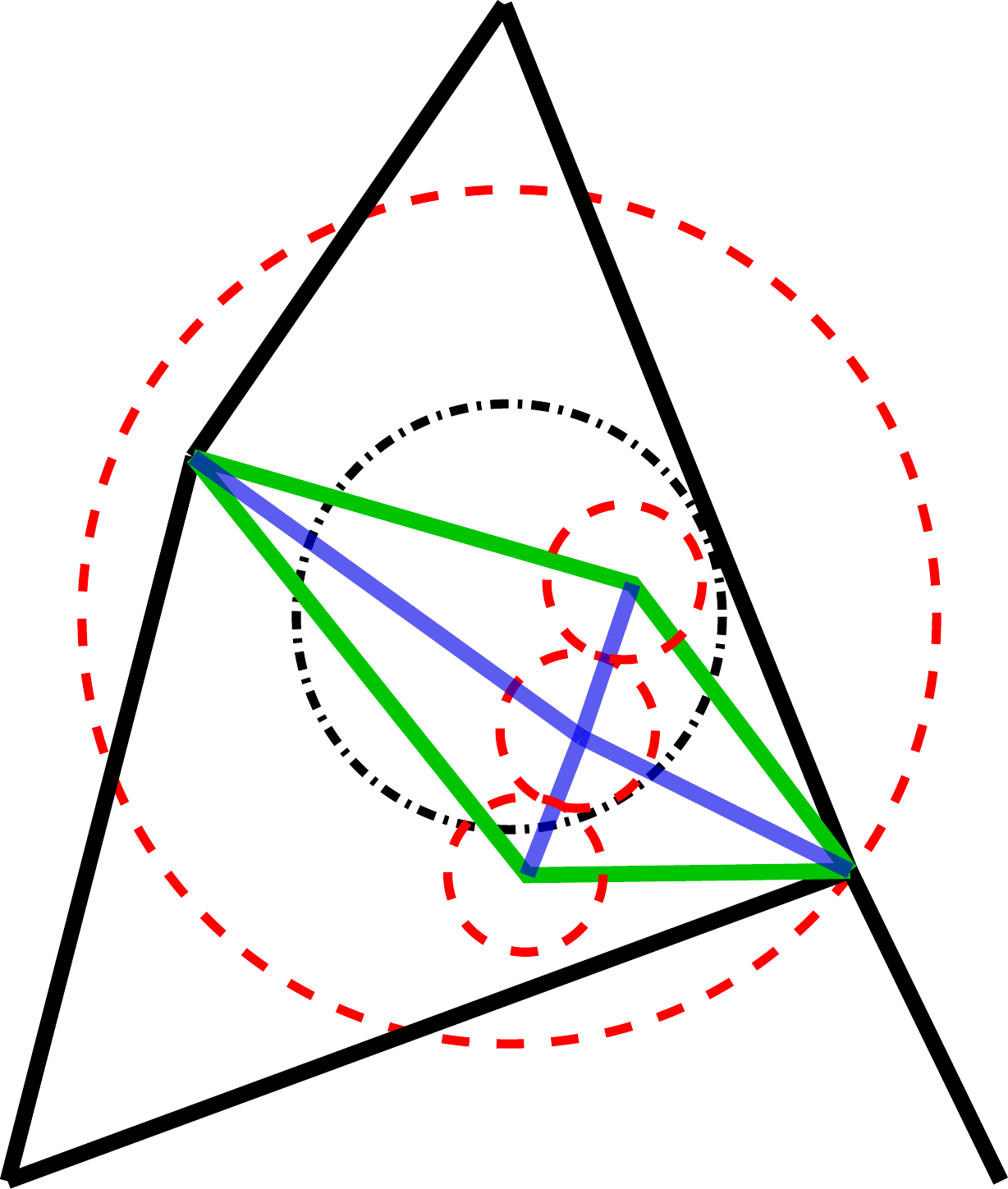}
		}%
		\subfloat[][]{		\label{subfig:reorient_sc}
			\includegraphics[width=103.3pt]{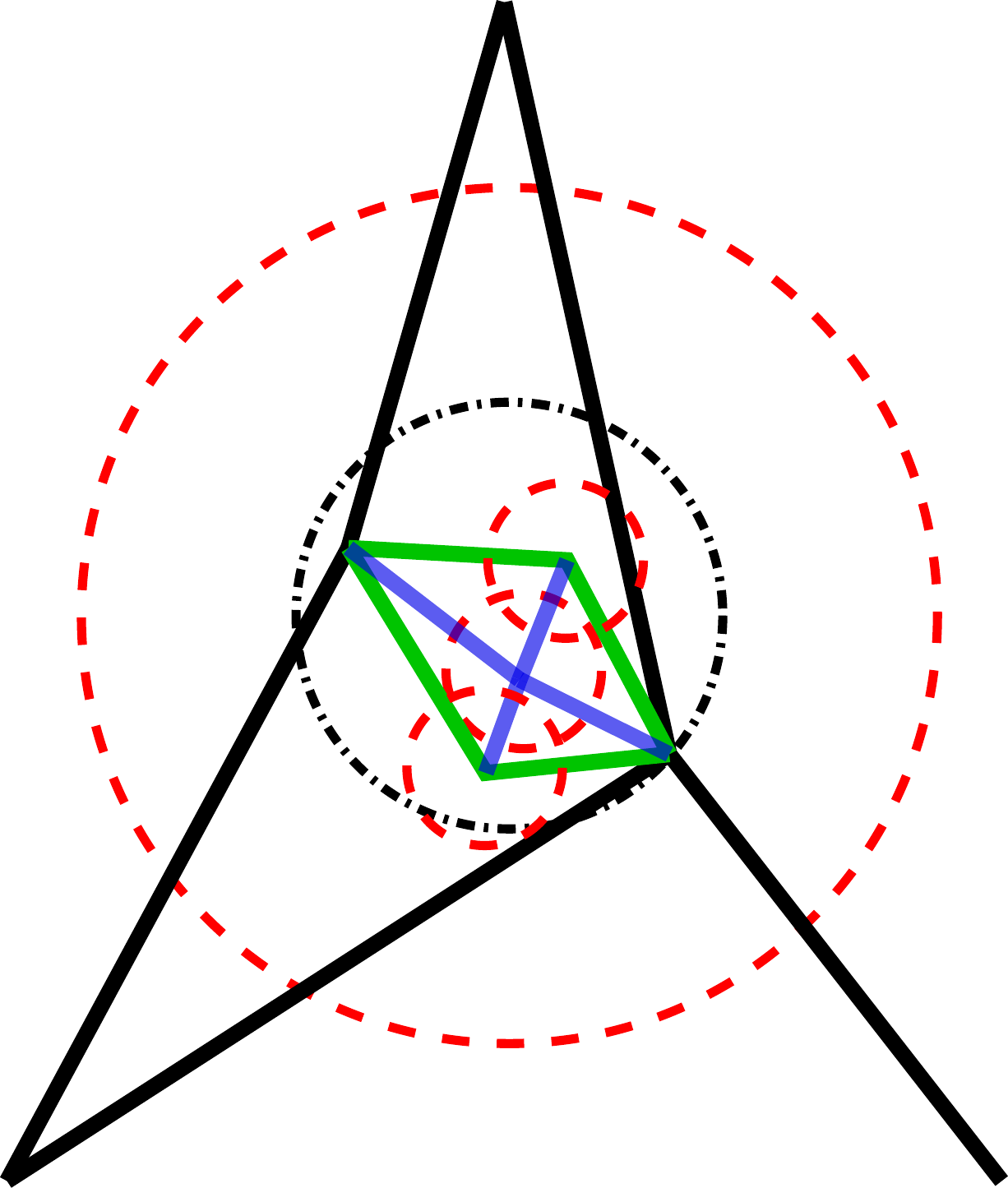}
		}%

		\caption{The steps of mesh level insertion and reorientation for a digon insertion. (a) Fins of triangles along paths, shown in bold black. (b) Insertion of the new digon, where $ S^1 $ edges are shown as green lines and $ S^2 $ edges are shown as blue lines. (c) The vertices are allowed to move until one of the ending criteria is reached. (d) The digon is scaled to be within the projection sphere, and relaxation continues until the energies converge.}
		\label{fig:reorientation}
	\end{figure*}
	The calculation of the geometry of an inserted stratum begins by isolating the mesh around the old $ S^0 $ vertex and applying the relaxation algorithm shown in Figure \ref{fig:reorientation} for a digon insertion. The bold black lines in Figure \ref{subfig:reorient_i} represent the fins of triangles on the paths. A new digon is inserted by expanding the two selected fins, changing the topology as shown in Figure \ref{subfig:reorient_ins}. The projection sphere of radius $ r $ is represented by the black dot-dashed circle and the inner (one for each $ S^1 $ and $ S^2 $ vertex) and outer bounding spheres are represented by red dashed circles. The vertices are then allowed to move according to the equations of motion (Figure \ref{subfig:reorient_exp}) until a minimum energy is reached or one of the moving vertices intersects an inner or outer bounding sphere. If one of the inner spheres is intersected, the insertion is discarded. If the outer sphere is intersected, the inserted stratum is scaled to be contained within the projection sphere. The steps in Figure \ref{subfig:reorient_exp} and \ref{subfig:reorient_sc} are repeated until both the energy at the intersection and the energy after the scaling converge to the final and initial energies $ E_f $ and $ E_i $. 
	
	Since the thermodynamically-driven system follows a gradient flow of the energy, the physical system will transition to the state with the the highest energy dissipation rate. After the process converges, the energy dissipation rate is calculated for the expanding insertions at the singular configuration where all the new vertices are positioned at the old vertex position. Assuming the contributions of the newly generated strata to the forces acting on the vertices are vanishingly small in this configuration, the dissipation rate of initial expansion is given by
	
	\begin{equation*}
	W = - \sum_{i} \vec{F}_i \cdot \vec{v}_i
	\end{equation*}
	where $ F_i $ and $ v_i $ are the force acting on and the velocity of vertex $ i $ and the sum is over all newly inserted bounding vertices. 

	Our energy dissipation rate criterion is similar to the depinning force which Shya and Weygand use to repeatedly split a node by edge insertions \cite{2010ModelSimuMaterSciEngSyha}. The difference is that our approach instead compares all possible single stratum insertions at once using the energy dissipation rate criterion, presumably more closely following the evolution of the physical system. Moreover, the relaxation algorithm discards insertions that do not expand, allowing for stable high valency junctions that could form, e.g., at intersecting deformation twins in TWIP steels.	
	
	\section{Modified MacPherson-Srolovitz relation} \label{sec:Lazar}
	
	All numerical approaches should be benchmarked against experimental or analytical results. One benchmark for polycrystalline microstructures evolving under constant grain boundary energy is the MacPherson-Srolovitz relation \cite{2007NatureMacPhersonSrolovitz}, the three-dimensional extension of the von Neumann-Mullins relation \cite{vonNeumann1952,Mullins1956}. For a constant grain boundary energy, this relation should be satisfied by each grain at every moment in time except for when a topological transition occurs. 
	
	The MacPherson-Srolovitz \cite{2007NatureMacPhersonSrolovitz} relation governing the rates of change of volumes is given by:
	\begin{equation} \label{eqn:sromacp}
	\frac{dV(D)}{d t} = -2\pi \mu \gamma \left[\mathcal{L}(D)
	- \frac{1}{6} \mathcal{M}(D)
	\right],
	\end{equation}
	where $ \mu $ is the constant grain boundary mobility, $ \gamma $ is the constant grain boundary energy, $ \mathcal{L}(D) $ is the mean width which measures the the total mean curvature of grain $ D $, and $ \mathcal{M}(D) $ is the total length of the triple lines of grain $ D $. Lazar et al.\ describe a discretized form of the MacPherson-Srolovitz relation that can be used to calculate the rate of volume change for grains composed of discretized linear elements \cite{2011ActaMateLazar}. For this case, $ \mathcal{L}(D) $ and $ \mathcal{M}(D) $ reduce to
	
	\begin{align*} 
	\mathcal{L}(D) &= \frac{1}{2\pi} \sum_{i}e_i \alpha_i, \\
	\mathcal{M}(D) &= \sum_{j}l_j,
	\end{align*}
	where $ e_i $ is the length of the $ i $th boundary edge, $ \alpha_i $ is the exterior angle around the $ i $th boundary edge with respect to the grain $ D $, and $ l_j $ is the length of the $ j $th triple line edge. 
	
	The coefficient of $ \mathcal{M}(D) $ is related to the equilibrium exterior angle of $ \pi/3 $. For periodic boundary conditions and when all junctions are composed of triple junctions and quadruple points, this is the expected exterior angle everywhere. As will be further discussed in Section \ref{sec:RnD} though, when using an exterior boundary or allowing higher valency junctions due to the discretized mesh, the MS relation needs to be modified to include more general exterior angle conditions. Eq.\ (\ref{eqn:sromacp}) is reformulated as
	
	\begin{align} 
	\frac{dV(D)}{d t} &= - \mu \gamma \left[
	2\pi\mathcal{L}(D) - \mathcal{N}(D)
	\right], \label{eqn:sromacp_mod1}\\
	\mathcal{N}(D) &= \sum_{j}\beta_jl_j,
	\end{align}
	where $ \beta_j $ is the equilibrium exterior angle around the $ j $th junction line edge. This is determined from the geometric relation
	\begin{equation*} \label{eqn:alphaext}
	(\pi - \beta_j)n = \xi_j
	\end{equation*}
	where $ n $ is the number of grains and $ \xi_j $ is the total interior angle available for all grains around the $ j $th junction line edge. For a stable interior $ S^1 $, $ \xi_j = 2\pi$, $ n = 3 $, $ \beta_j = \pi/3 $ and Eq.\ (\ref{eqn:sromacp_mod1}) reduces to Eq.\ (\ref{eqn:sromacp}). Assuming a cubic simulation cell, the stable configuration for a $ S^1 $ on a simulation cell edge has $ n = 1$, $ \xi_j = \pi/2 $ and $ \beta_j = \pi/2 $, the stable configuration for a $ S^1 $ on a simulation cell face has $ n = 2$, $ \xi_j = \pi $ and $ \beta_j = \pi/2 $. It is possible to have unstable junctions with $ n $ larger than that for the stable configurations.
	
	\section{Results and Discussion} \label{sec:RnD}
	
	We consider some example configurations to enumerate the possible insertions, and show the effect of local geometry on the selection criterion and the inserted stratum shape. Then we demonstrate the importance of enumerating all possible insertions with a relatively simple microstructure that could lead to transitions not often considered in previous FEM-based methods. Finally, we show the evolution of a trial microstructure as a demonstration of the capabilities of our implementation.
	
	\begin{figure*}
		\centering
		\includegraphics[width=510pt]{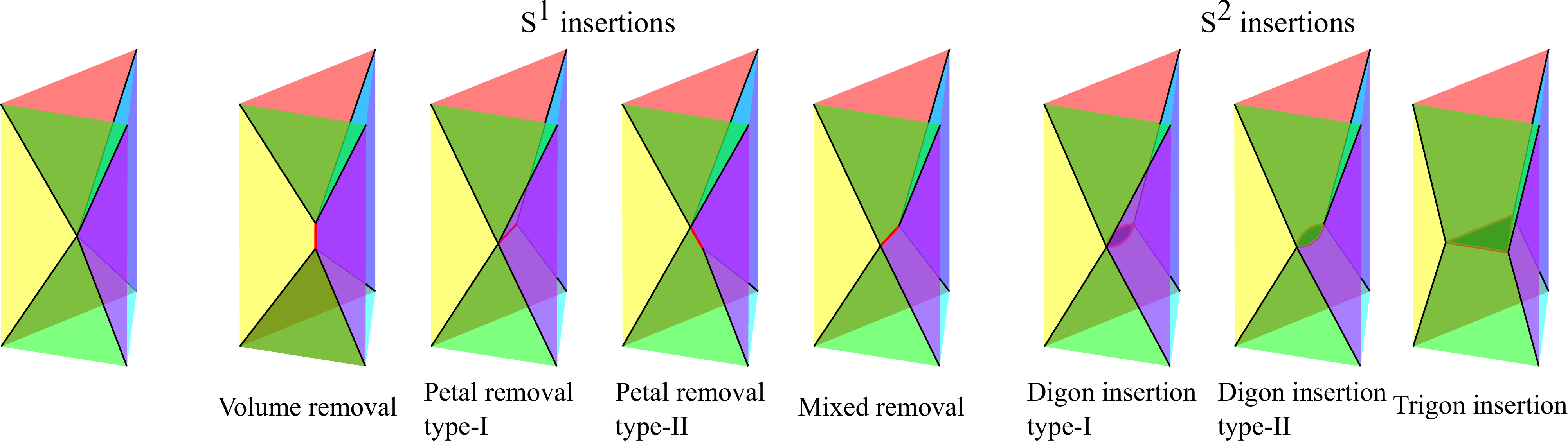}
		\caption{All possible insertions for the canonical configuration, classified by symmetry groups. Observe that digon insertions are obtained by decomposing circuits containing disconnected $ 3 $-stratum couples into two paths connecting the couples and using these to insert a $ 2 $-stratum. Digon insertion type-I is related to petal removal type-I and digon insertion type-II is related to mixed removal. }
		\label{fig:split}
	\end{figure*}
	
	To verify that all insertions are considered, consider the five grain configuration previously described in Figure \ref{subfig:circuit_stratum}. All possible insertions can be found by applying the circuit and path detection algorithms, and these are shown in Figure \ref{fig:split} (grouped by their symmetries). There are four classes of $ S^1 $ insertions and three classes of $ S^2 $ insertions. The volume removal and trigon insertion are generally handled by all grain growth codes, but the other insertions are usually not since a $ S^1 $ collapse is always followed by a trigon insertion for a uniform boundary energy. Digons can also be inserted, with the two types shown in Figure \ref{fig:split}.
	
	To be specific, there is one volume removal, three petal removal type-Is, six petal removal type-IIs, and six mixed removals possible, all of which are found by circuit analysis. There are three type-I, six type-II digon, and one trigon insertions possible, as well. Note that digon insertion type-I and type-II use paths that can be constructed by decomposing the circuits of petal removal type-I or mixed removal, respectively. When discussing the energy dissipation rates, it will be shown that these additional operations could be relevant depending on the grain boundary energy function. 
	
	Depending on the geometry of the boundaries, each insertion has a different energy dissipation rate associated with the subsequent evolution. The energy dissipation rate criterion states that the insertion with the highest positive dissipation rate is the one that will be realized. To test this criterion, a mesh was generated for the configuration in Figure \ref{fig:split}. If the geometry is such that the three $ S^1 $s on top and three $ S^1 $s on the bottom are separated by the tetrahedral angle, a degenerate configuration is created where any insertion results in an unstable configuration with increased energy. If the angles between the $ S^1 $s are instead larger than the tetrahedral angle, a trigon insertion is favored. Conversely, if the angles between the $ S^1 $s are smaller than the tetrahedral angle, a volume removal is favored.
	
	\begin{figure*}
		\centering
		
		\subfloat[][]{		\label{subfig:split_E}
			\includegraphics[height=130pt]{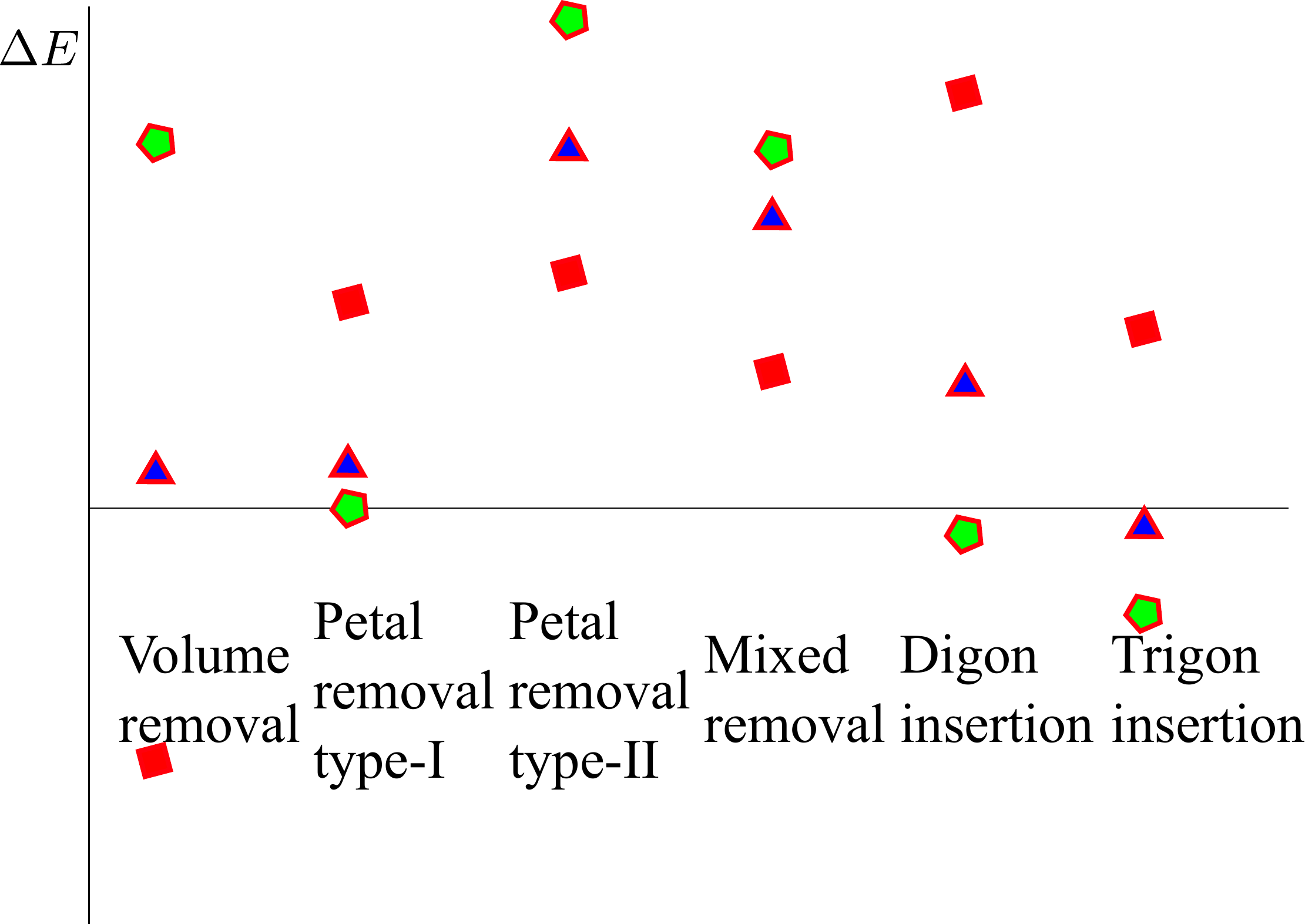}
		}%
		\subfloat[][]{		\label{subfig:split_W}
			\includegraphics[height=130pt]{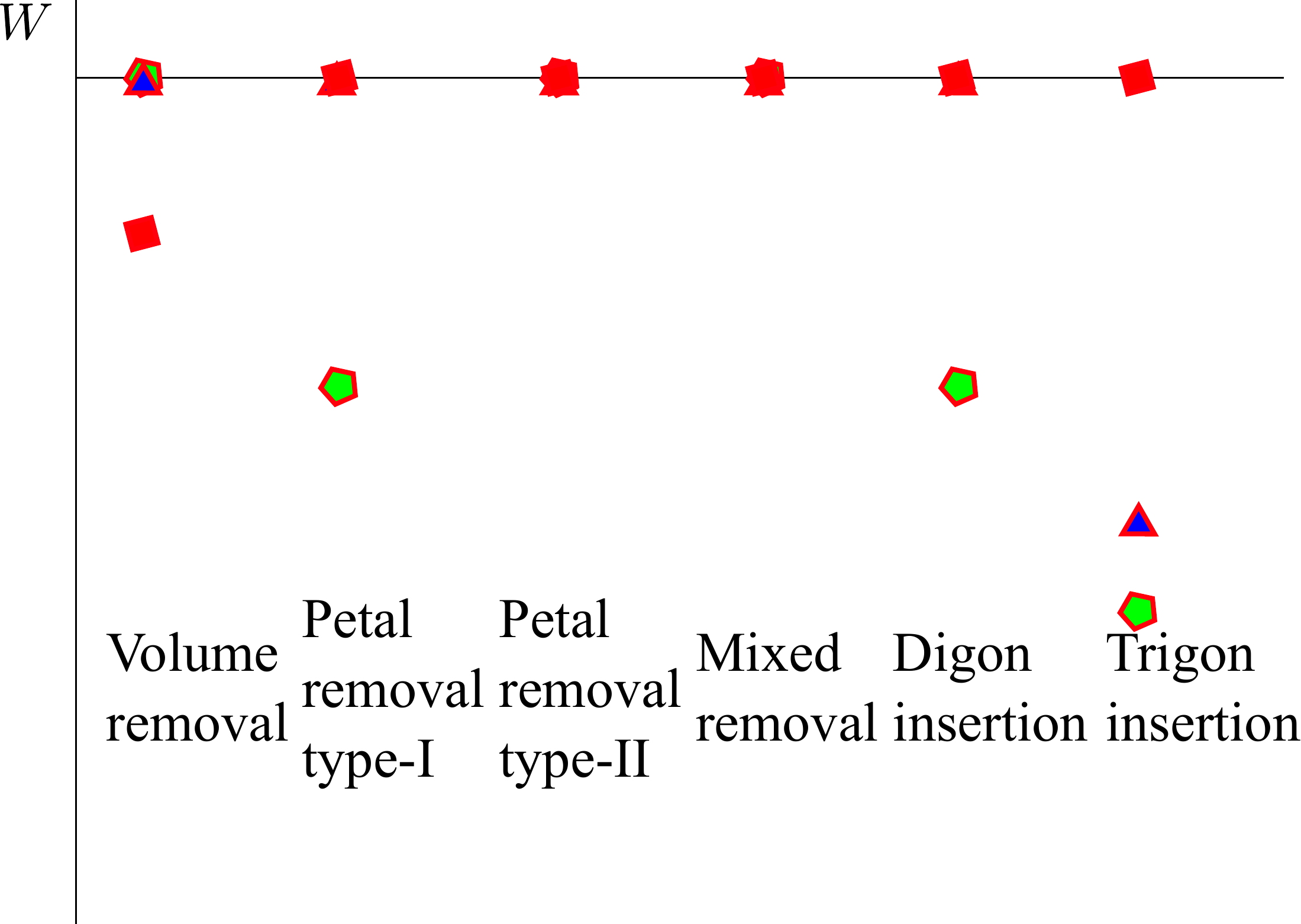}
		}%
		\subfloat[][]{		\label{subfig:split_degen}
			\includegraphics[height=130pt]{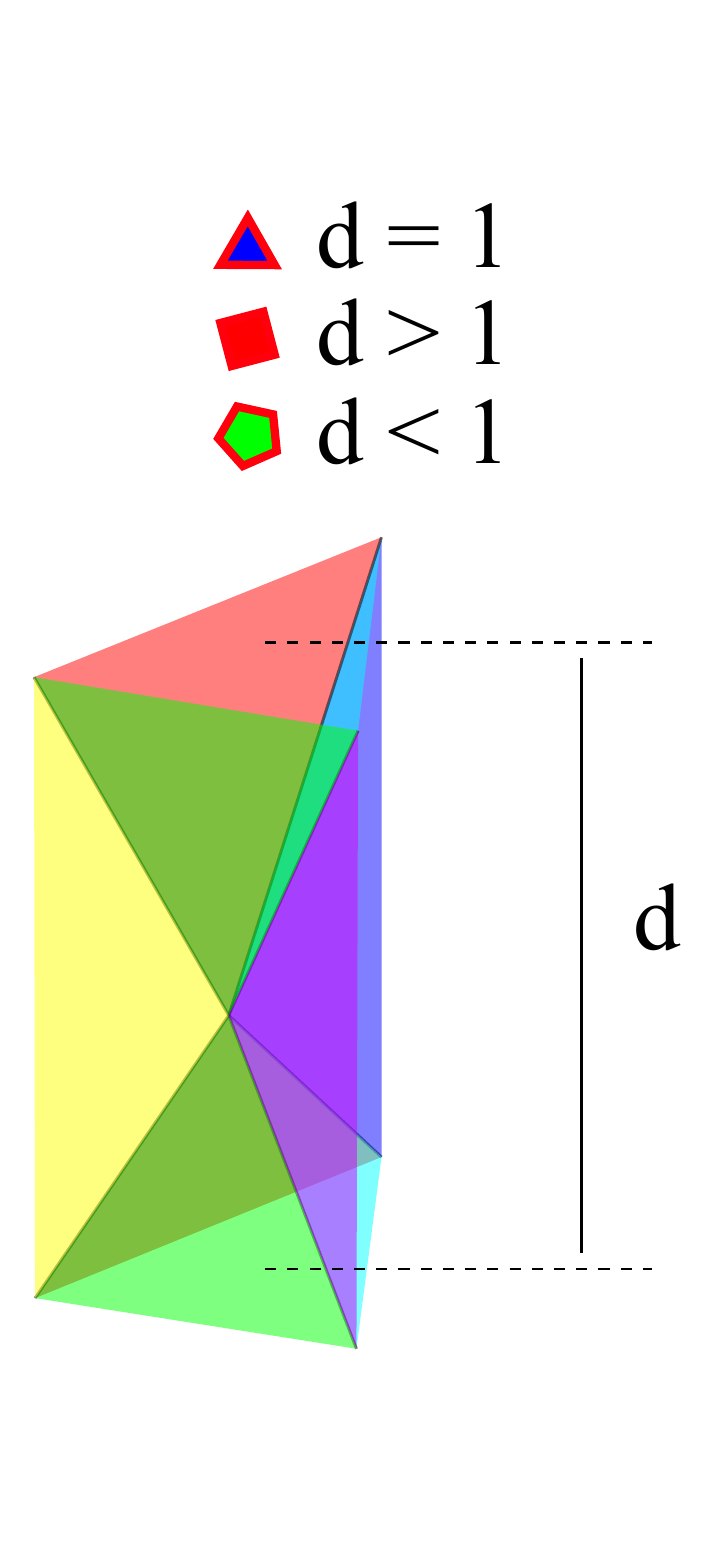}
		}%
		\caption{(a) The variation in energy change of insertion with changing configuration. Blue triangles show the energies for the configuration when the $ S^1 $ angles in (c) are tetrahedral angles. Red squares denote the energies for the stretched case, and the green pentagons show the compressed case. (b) The dissipation rates for the expanding insertions at the singular configuration, where the volume removal and the trigon insertion are energetically favorable for the stretched and compressed cases, respectively. }
		\label{fig:5split_rate}
	\end{figure*}

	The changes in energy for each insertion are shown in Figure \ref{fig:5split_rate}. The energies in Figure \ref{subfig:split_E} are calculated with the new vertices on the outer projection sphere. For the compressed case where trigon insertion is favored, it is significant that the digon insertion is also energy decreasing and the petal removal type-I is nearly energy neutral. The dissipation rates associated with the expanding insertions are compared in Figure \ref{subfig:split_W} to select the most energetically favorable insertion. 
	
	In the current scheme the inserted triangles apply lower forces than the surrounding triangles due to the discretized equations of motion, and there is a small bias towards trigon insertions in the degenerate configuration as is visible in Figure \ref{fig:5split_rate}. The bias depends on the selection of the ratio of the radii of the inner and outer spheres in Figure \ref{fig:reorientation}. By increasing the ratio, smaller radius insertions are discarded, effectively creating a range of $ d $ around the value corresponding to the degenerate case where no insertion is valid. However that can also make high aspect ratio $ S^2 $ insertions hit the inner sphere and be discarded until their aspect ratio lowers on the consecutive time steps.

	\begin{figure}
		\centering
		
		\subfloat[][]{		\label{subfig:ort_0}
			\includegraphics[width=76pt]{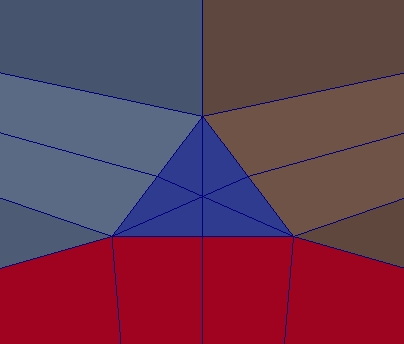}
		}%
		\subfloat[][]{		\label{subfig:ort_1}
			\includegraphics[width=76pt]{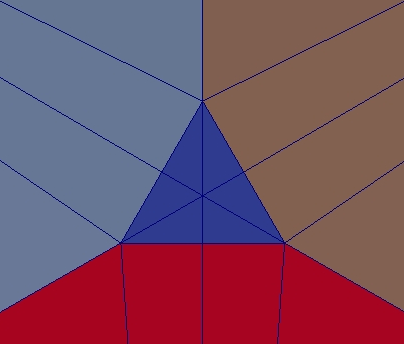}
		}%
		\subfloat[][]{		\label{subfig:ort_2}
			\includegraphics[width=76pt]{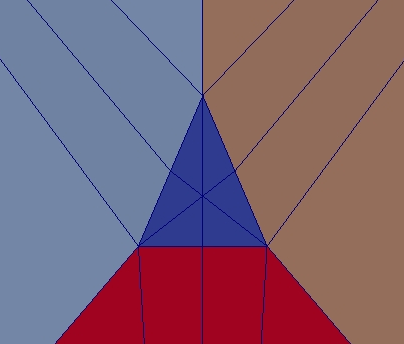}
		}%
		\caption{The effect of orthogonal stretching on the trigon shape. (b) Starting configuration, where dihedral angles between surfaces separating the surrounding $ S^3 $ are equal. (a)-(c) After stretching (compressing) the configuration in the lateral direction, running the relaxation yields a laterally stretched (compressed) $ S^2 $. }
		\label{fig:orthogonal}
	\end{figure}
	
	Whereas the vertical stretch changes which insertion is energetically favored, lateral stretches change the energy-minimizing shape of the inserted stratum and are reflected in the relaxation scheme. Without this, insertions of equilateral $ S^2 $ could increase the energy artificially and cause a physical insertion to be overlooked. Relaxation mitigates the problem, and as shown in Figure \ref{fig:orthogonal}, the shape of the inserted $ S^2 $ changes depending on the geometry.
	
	\begin{figure}
		\centering
		
		\subfloat[][]{		\label{subfig:sim_i}
			\includegraphics[width=83pt]{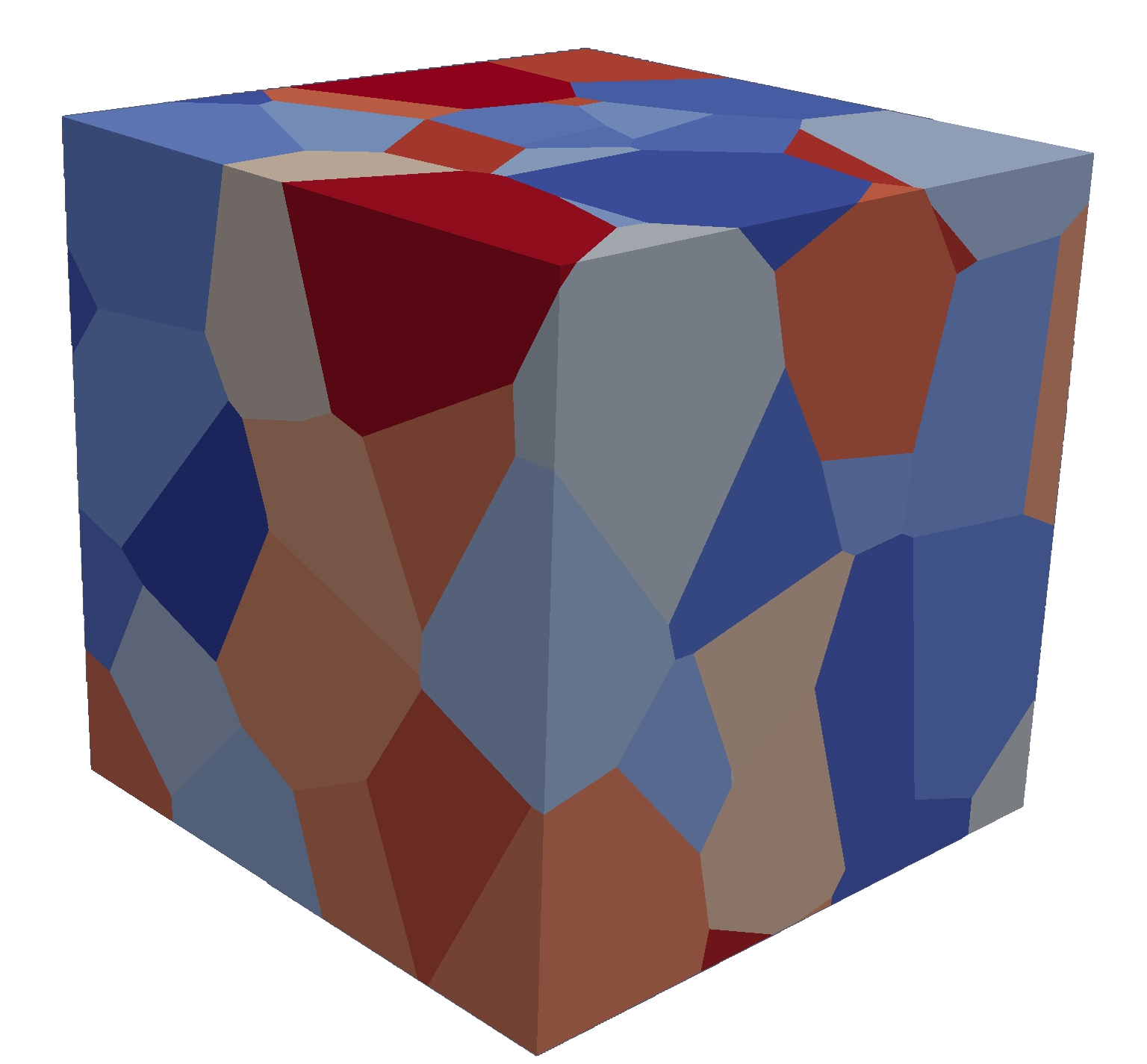}
		}%
		\subfloat[][]{		\label{subfig:sim_m}
			\includegraphics[width=83pt]{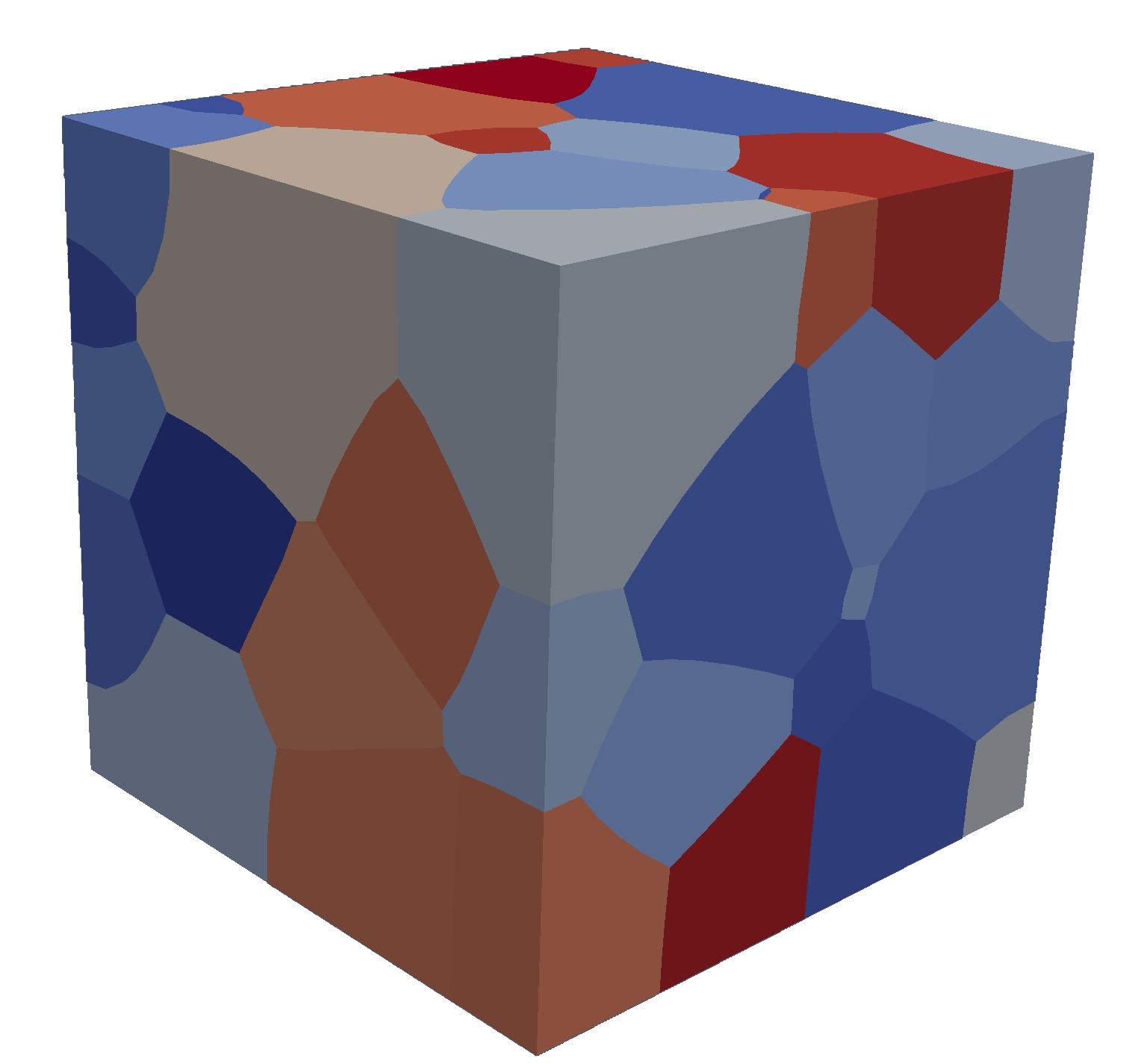}
		}%
		\caption{Simulation of a microstructure composed of $ 100 $ grains under isotropic grain boundary energy. (a) Initial configuration. (b) The number of grains is about one half of the starting number. }
		\label{fig:sim}
	\end{figure}
	
	Finally, we simulate the evolution of some artificial microstructures generated using Neper \cite{2011CompMethAppMechQuey}. These microstructures are not periodic, and their evolution requires imposing a local volume preservation constraint on the exterior vertices. This relaxes the connectivity constraint on grain surfaces on the exterior, and requires some additional operations described in Section \ref{app:shell} of the SM. 
	
	\begin{figure}
		\centering
		\includegraphics[width=236pt]{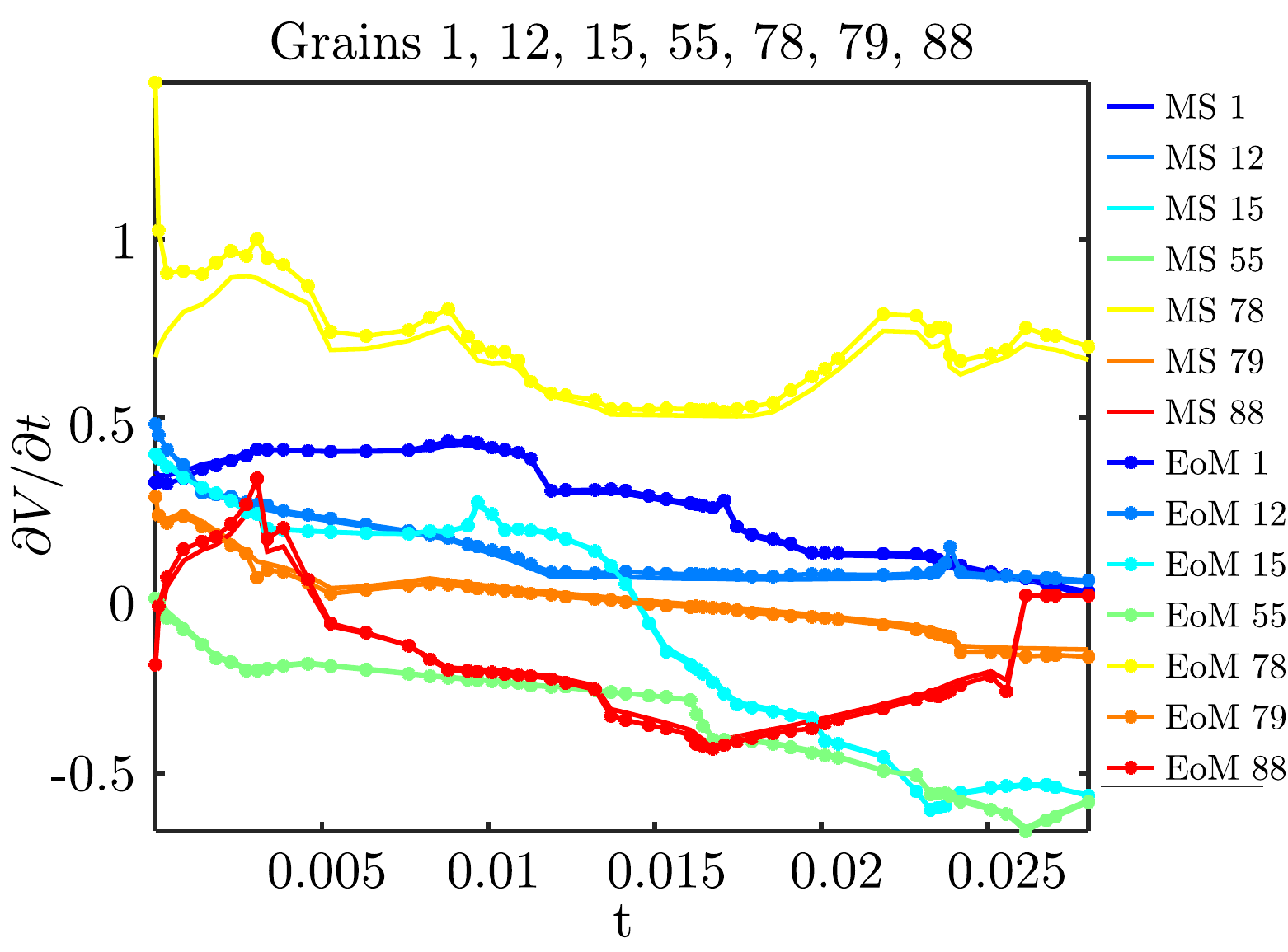}
		\caption{The rates of volume change for example grains as calculated by the modified MacPherson-Srolovitz (MS) relation, and first-order approximation using the equations of motion (EoM).  }
		\label{fig:MSROC}
	\end{figure}
	
	To demonstrate the capabilities of VDLIB, a trial microstructure composed of $ 100 $ grains is generated as a Voronoi tesselation using Neper \cite{2011CompMethAppMechQuey}. The simulation cell is a cube with unit edge length. The mesh is adaptively refined, with a target edge length set to a fraction of the median edge length of cubes with equivalent volumes to the grains. In addition, the $ S^1 $ are required to contain at least two edges to provide sufficient degrees of freedom. The microstructure is evolved using equations of motion by Mason \cite{2017ActMateMason} with unit surface drag coefficient and grain boundary energy. The volume constraint is implemented by the method described in Section \ref{app:shell} of the SM. The time iteration is implemented by a second order Runge-Kutta scheme with the time step at each iteration given by 
	$ \text{min}(t_{\text{inv}}/20, t_{\text{fixed}}) $, where $ t_{\text{inv}} $ is the shortest time step to invert any element and $ t_{\text{fixed}} $ is the maximum fixed time step of $ 5.0 \times 10^{-5}$. One iteration loop involves nine sub-iterations of the equations of motion, checking for and implementing collapses, followed by checking for and implementing insertions. Some snapshots from the resulting system evolution are shown in Figure \ref{fig:sim}. 
	
	The modified MacPherson-Srolovitz relation in Section \ref{sec:Lazar} can be used to calculate the rate of volume change for grains composed of discretized linear elements. The resulting actual rates of volume change for a select number of grains and the predictions of the modified MacPherson-Srolovitz relation are given in Figure \ref{fig:MSROC}. The initial discrepancy is mainly due to the deviation from the equilibrium angle conditions in the initial condition. The discrepancy falls as the initial microstructure evolves and the angles around the junction lines approach the equilibrium values. Topological transitions can also cause temporary deviations (e.g.,\ grain 78 around $ t = 0.003 $ in Figure \ref{fig:MSROC}) which decrease in time. Despite using linear elements and an explicit time integration scheme, there is overall good agreement with the MacPherson-Srolovitz relation.
	
	\section{Conclusion}
	
	A computational framework with an explicit grain boundary representation is proposed to predict grain growth for anisotropic grain boundary energies and mobilities. This establishes the foundations of a massively parallelizable general-purpose framework to model microstructure evolution during, e.g., high-temperature and finite-strain processes. There does not appear to be any other software with these capabilities, that uses an explicit boundary representation, and that supports general changes to the grain boundary network.
	
	Predictive simulations of microstructure evolution during thermomechanical processing require the ability to represent features such as stable quadruple junction lines in low stacking-fault energy metals. This in turn requires the ability to handle anisotropic properties and more general topologies than usually assumed in the literature. Moreover, the mesh should be partitioned across multiple processing units to reach physically relevant scales, and the equations of motion should be local to keep the computational cost linearly proportional to the number of grains. The discrete equations of motion proposed by Mason \cite{2017ActMateMason} can accommodate anisotropic grain boundary energies and drag coefficients. They are local and scalable, and have been implemented to describe the boundary motion. 
	
	A generic method to enumerate the singular transitions is proposed and implemented. An energy-based insertion selection criterion is proposed and implemented. The method can utilize models for anisotropic energies, and once experimental grain boundary energy functions are available, the framework will be used to simulate grain growth under these conditions. Finally, the work is done in the context of a massively parallelizable finite element based library that can support volumetric physics. 
	
	\section{Acknowledgments}
	E.E.\ was partially supported by the Takamura and Erhardt Family Fellowship.
	

	\newpage

	\makeatletter
	\onecolumngrid
	\newpage
	\pagebreak
	
	\setcounter{equation}{0}
	\setcounter{section}{0}
	\setcounter{figure}{0}
	\setcounter{table}{0}
	\setcounter{page}{1}
	
	\makeatletter
	\renewcommand{\theequation}{S\arabic{equation}}
	\renewcommand{\thefigure}{S\arabic{figure}}
	\renewcommand{\bibnumfmt}[1]{[S#1]}
	\renewcommand{\citenumfont}[1]{S#1}

	\begin{center}
		\textbf{\large Supplemental Materials: Topological transitions during grain growth on a finite element mesh}
	\end{center}
	
	\section{Notation}
	
	For brevity, let $ \mathbf{S}^d $ be the set of $ d $-dimensional strata and $ S^{d}_i $ the $ i $th $ d $-stratum. $ \left|A\right| $ is the number of elements in the set $ A $, $ A^e(S^d_i) $ is the collection of $ S^e $  adjacent to $ S^d_{i} $, and $ A^e_j(S^d_i) $ the $ j $th $ S^e $ adjacent to $ S^d_{i} $. $ A^{f,e}(S^d_i) $ are the $ f $-dimensional strata adjacent to the $ e $-dimensional adjacencies of $ S^d_i $. $ \tilde{S}^{d}_i $ is a newly inserted stratum.
	
	The adjacency graph of $ S^2 $ and $ S^3 $ in the neighborhood of a 0-stratum $ S^0_i $ will be used to enumerate the possible changes to the local microstructure. Let the adjacency graph around $ S^0_i $ be $ G_i $, and have nodes corresponding to the set $ A^2(S^0_i) \cup  A^3(S^0_i) $ and edges for each incidence of an $ S^2 $ and $ S^3$. Similarly, $ G'_i $ is the adjacency graph with nodes corresponding to the set $ A^1(S^0_i) \cup  A^2(S^0_i) $ and edges for each incidence of an $ S^1 $ and $ S^2$. Paths and circuits on $ G_i $ will be used to find possible $ S^2 $ and $ S^1 $ insertions around $ S^0_i $, where a path is a sequence of non-repeating nodes connected by edges and a circuit is a path that begins and ends at the same node. 
	
	\begin{figure*}
		\centering
		\subfloat[][]{		\label{subfig:hesse_i}
			\includegraphics[width=91.3pt]{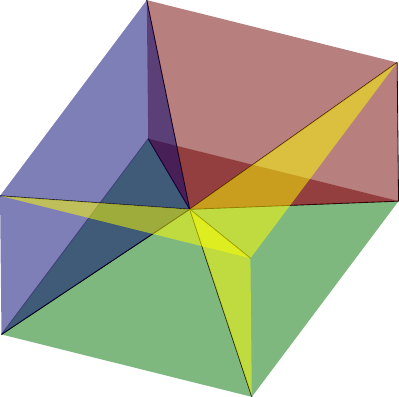}	
		}%
		\subfloat[][]{		\label{subfig:hesse_1}
			\includegraphics[width=91.3pt]{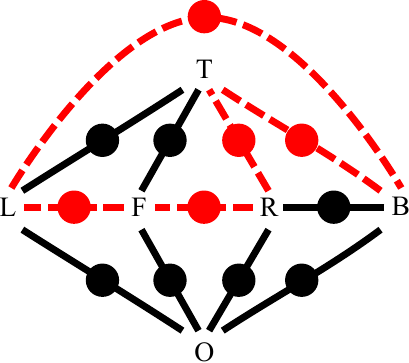}	
		}%
		\subfloat[][]{		\label{subfig:hesse_2}
			\includegraphics[width=91.3pt]{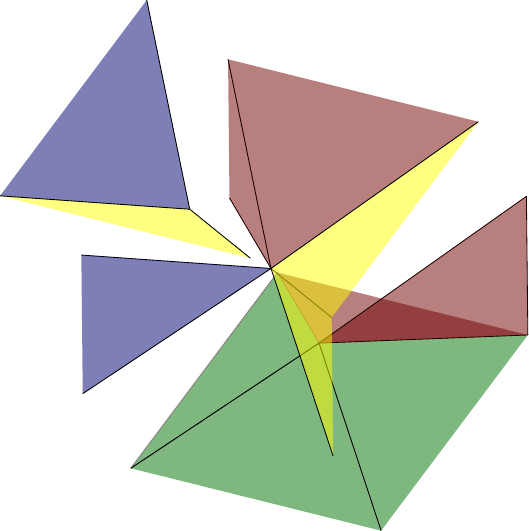}	
		}%
		\subfloat[][]{		\label{subfig:hesse_3}
			\includegraphics[width=91.3pt]{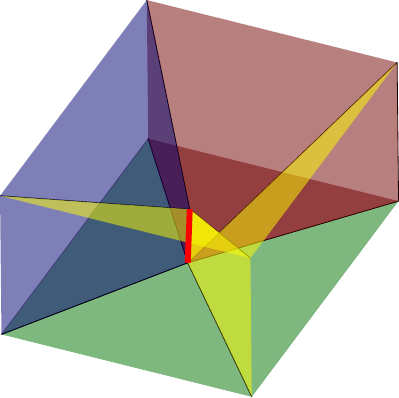}
		}%
		\subfloat[][]{		\label{subfig:hesse_4}
			\includegraphics[width=91.3pt]{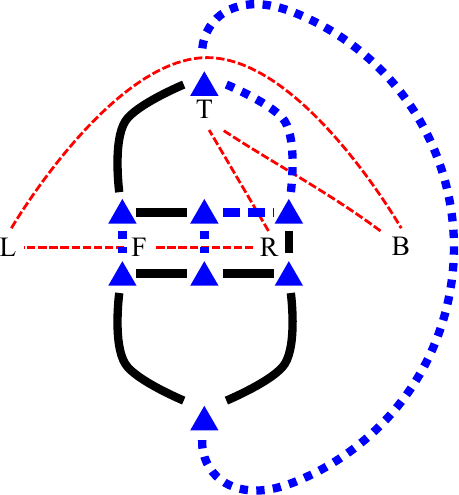}	
		}%
		\caption{A representation of how the circuit $ Q^i_j $ divides nodes into disjoint graphs $ H_1^{i;j} $ and $ H_2^{i;j} $, which are connected over $ S^1 $. (a) The initial microstructure consisting of six grains. (b) Removal of the red dashed circuit $ Q^i_j $ going through (T)op, (B)ack, (L)eft, (F)ront and (R)ight grains leaves two disjoint graphs. $ H_1^{i;j} $ consists of the $ S^2 $ connecting the grains T-F and T-L. $ H_2^{i;j} $ consists of the b(O)ttom grain, the $ S^2 $ bounding grain O in the neighborhood of $ S^0_i $, and the $ S^2 $ connecting the grains R-B.
			(c) At the microstructure level, it is easy to see how the components of $ H_1^{i;j} $ are connected by $ S^1 $s. 
			(d) The final configuration after the insertion with the associated $ S^1 $ colored red.
			(e) $ G'_i $, where $ Q^i_j $ is shown superposed and the dotted blue edges correspond to $ S^3-S^2-S^3 $ components of $ Q^i_j $. The nodes of the two subgraphs of $ G'_i $ can be seen to be connected by solid edges.}
		\label{fig:subgraph}
	\end{figure*}
	
	Let $ Q^i $ be the set of circuits on $ G_i $ and $ Q^i_j $ be the $ j $th such circuit. Removing the circuit $ Q^i_j $ from the graph leaves two disjoint graphs of nodes which will be denoted as $ H_1^{i;j} $ and $ H_2^{i;j} $. The $ H_k^{i;j} $ could be disconnected over $ G_i $, but the corresponding strata around $ S^0_i $ can always be connected through shared $ S^1 $, as shown in Figure \ref{fig:subgraph}. If one of the $ H_k^{i;j} $ is empty, that implies that the circuit $ Q^i_j $ is associated with an existing $ S^1 $ and should be discarded.
	
	Let $ P^{i;j,k}$ be the set of paths between $ S^3_j $ and $ S^3_k $ in the vicinity of $ S^0_i $, and $ P^{i;j,k}_l $ be the $ l $th such path. Let $ \powerset (P^{i;j,k})$ be the set of sets of paths between $ S^3_j $ and $ S^3_k $, where every set contains at least two paths and none of the paths in the same set intersect. If $ l $ is the index for this set, then subgraph $ H_m^{i;j,k;l}$ is the $ m $th connected component remaining when the $ l $th set of paths with the end points $ S^3_j $ and $ S^3_k $ is subtracted from $ G_i $. 
	
	The mesh is composed of simplicial finite elements, including the $ 0$-dimensional vertices, $ 1$-dimensional edges, $ 2 $-dimensional triangles and $ 3 $-dimensional tetrahedra. An $ n $-dimensional simplicial element belongs to the stratum of lowest dimension in which it is contained, i.e.,\ a vertex may belong to an $ S^0 $, $ S^1 $, $ S^2 $, or $ S^3 $, an edge may belong to an $ S^1 $, $ S^2 $, or $ S^3 $, etc. Similar to the notation for strata, we denote the $ i $th member of the set of $ d $-dimensional simplicial entities as $ \Delta_i^d $ and use the adjacency operator $ A^e(\cdot) $ in the same way to obtain the set of adjacent mesh entities of dimension $ e $. Additionally, the stratum membership of a simplicial entity is indicated as $ \Delta_i^d \in S^e_j$, or $\Delta_i^d$ belongs to $S^e_j$. The set of $ e $-dimensional simplicial entities belonging to $ S^d_i $ is obtained by the membership operator $ M^e(S^d_i) $. $ S(\Delta^d_i) $ is the stratum that owns the simplicial entity $ \Delta^d_i $. A sample microstructure showing the simplicial entities outlined in red is provided in Figure \ref{subfig:FEMFEM}.

	\section{Stratum collapse} \label{app:gen_col}

	Given a stratum $ S^d_i $ to collapse and a final point $ \hat{S}^0 $, recursively collapse the bounding lower dimensional strata and then remove $ S^d_i $. If $ \hat{S}^0 $ is not specified, it is always possible to pick the first bounding $ S^0 $ (otherwise there are no $ S^0 $ remaining after collapse). Update the adjacency lists of the surrounding strata. 

	\begin{algorithm} [H]\label{algo:gen_col}
		\begin{algorithmic}\\
			\textit{Implement changes in the stratification when collapsing $S^d_i$.}
			\If{$ d = 0 $}
			\State return
			\EndIf
			\If{$ \hat{S}^0 = \emptyset $}\Comment{Assign $ \hat{S}^0 $ if not specified.}
			\If {$ A^0(S^d_i) \neq \emptyset $}
			\State$ \hat{S}^0 \coloneqq A^0_{1}(S^d_i) $
			\EndIf
			\EndIf
			
			\If{$ d = 1 $} \Comment{Replace the merging $ S^0 $ with $ \hat{S}^0 $.}
			\For{$ S^1_j \in A^{1,0}(S^1_i)$}
			\For{$ S^0_k \in A^{0}(S^1_j)$}
			\If {$ S^0_k \in A^{0}(S^1_i)$}
			\State $ A_k^0(A^{1,0}_j(S^d_i)) \coloneqq \hat{S}^0 $
			\EndIf
			\EndFor
			\If {$ | A^0(S^1_j)| = 2$ and $ A_1^0(S^1_j) = A_2^0(S^1_j) = \hat{S}^0 $} \Comment{If $ \hat{S}^0 $ is repeated remove one. }
			\State $ A^0(S^1_j) \coloneqq \{\hat{S}^0\} $
			\EndIf
			
			\EndFor
			\Else \Comment{Collapse the bounding strata.}
			\For{$ S^d_j \in A^{d-1} (S^d_i) $}
			\State Collapse $( S^d_j, \hat{S}^0)$
			\EndFor
			\EndIf
			
			\If{$ d < 3 $} \Comment{Remove the collapsing strata.}
			\For{$ S^{d+1}_j \in A^{d+1} (S^d_i) $}
			\State $ A^{d} (S^{d+1}_j) \coloneqq A^{d} (S^{d+1}_j) \ \backslash \ \{S^{d}_{i}\} $
			\EndFor
			\EndIf
			\caption{Collapse $ (S^d_i, \hat{S}^0  \coloneqq \emptyset ) $}
			
		\end{algorithmic}
	\end{algorithm}

	\section{1-stratum insertion} \label{app:c_ins}

	Given a candidate $ S^0_i $ and a circuit $Q^i_j$ on $ G_i $ insert the new stratum $ \tilde{S}^1 $ corresponding to $Q^i_j$. Add the new strata $ \tilde{S}^0 $ and $ \tilde{S}^1 $ to the stratification and set the $ S^0 $ adjacencies of $ \tilde{S}^1 $ as $\{S^0_i, \tilde{S}^0\}  $. Update the adjacency lists of the surrounding strata. 

	\begin{algorithm} [H]
		\begin{algorithmic}\\
			\textit{Implement changes in the stratification when inserting $ \tilde{S}^1$ using $Q^i_j$.}
			\State Create new strata $ \tilde{S}^1 $, $ \tilde{S}^0 $.
			\State $ A^0(\tilde{S}^1) \coloneqq \{S^0_i, \tilde{S}^0\} $ \Comment{Adjacency of $ \tilde{S}^1 $.}
			\For{$ S^2_k \in Q^i_j$}  \Comment{Add $ \tilde{S}^1 $ to $ A^1(S^2_k)  $, for $ S^2 $ on $ Q^i_j $.}
			\State $ A^1(S^2_k) \coloneqq A^1(S^2_k) \cup \{\tilde{S}^1\}$
			\EndFor
			\For{$ S^2_k \in  H_2^{i;j} $}    \Comment{Replace $ S^0_i $ with $ \tilde{S}^0 $.}
			
			\For{$ S^1_l \in A^1(S^2_k)$}
			\If{$ S^0_i \in A^0(S^1_l) $}
			\State $ A^0(S^1_l) \coloneqq A^0(S^1_l) \cup \{\tilde{S}^0\} \ \backslash \ \{S^{0}_{i}\}$
			\EndIf
			
			\EndFor
			
			\EndFor
			
			\caption{$ S^1 $ insertion $(S^0_i, Q^i_j)$.}
			
		\end{algorithmic}
	\end{algorithm}

	\section{2-stratum insertion}\label{sec:c2_ins}
	
	Given a candidate $ S^0_i $ and a set of paths $\powerset_l (P^{i;j,k})$ on $ G_i $, insert the corresponding new stratum $ \tilde{S}^2 $. Add the new strata $ \tilde{S}^2 $, $ \tilde{S}^0_m $ for $ m = 1:|\powerset_l (P^{i;j,k})| - 1 $, and $ \tilde{S}^1_m$ for $ m = 1:|\powerset_l (P^{i;j,k})|$ to the stratification. Set the $ (d-1) $-dimensional adjacency lists of the new strata $ \tilde{S}^2 $ and $ \tilde{S}^1_m$ for $ m = 1:|\powerset_l (P^{i;j,k})|$. Update the adjacency lists of the surrounding strata. 

	\begin{algorithm} [H]
		\begin{algorithmic}\\
			\textit{Implement changes in the stratification when inserting a $ \tilde{S}^2$ using $\powerset_l (P^{i;j,k})$.}
			\State Create new stratum $ \tilde{S}^2$.
			\State Create new strata $ \tilde{S}^0_m$ for $ m \coloneqq 1:|\powerset_l (P^{i;j,k})|-1$. \Comment{In addition to $ S^0_i $.}
			\State Create new strata $ \tilde{S}^1_m$ for $ m \coloneqq 1:|\powerset_l (P^{i;j,k})|$.
			
			\State $ A^2(S^3_j) \coloneqq A^2(S^3_j) \cup \{\tilde{S}^2\}$ \Comment{Add $ \tilde{S}^2 $ to $ A^2(S^3_j) $.}
			\State $ A^2(S^3_k) \coloneqq A^2(S^3_k) \cup \{\tilde{S}^2\}$ \Comment{Add $ \tilde{S}^2 $ to $ A^2(S^3_k) $.}
			
			\State $ A^1(\tilde{S}^2) \coloneqq \{\tilde{S}^1_1, \tilde{S}^1_2, \dots, \tilde{S}^1_m\} $ with $ m \coloneqq |\powerset_l (P^{i;j,k})|$ \Comment{Set the adjacency of $ \tilde{S}^2 $.}
			\State $ A^0(\tilde{S}^1_1) \coloneqq \{S^0_{i}, \tilde{S}^0_{1}\} $ \Comment{Set the adjacencies of $ \tilde{S}^1 $.}
			\For{$ m \coloneqq 2:|\powerset_l (P^{i;j,k})|-1$} 
			\State $ A^0(\tilde{S}^1_m) \coloneqq \{\tilde{S}^0_{m-1}, \tilde{S}^1_{m}\} $ 
			\EndFor
			\State $ A^0(\tilde{S}^1_m) \coloneqq \{\tilde{S}^0_{m-1}, S^0_{i}\}$ with $ m \coloneqq |\powerset_l (P^{i;j,k})|$ 
			
			\For{$ P_m^{i;j,k} \in \powerset_l (P^{i;j,k})$} \Comment{Add $ \tilde{S}^1 $ to the adjacency lists of the $ S^2 $ on path $ P_m^{i;j,k} $.}
			\For{$ S^2_o \in P_m^{i;j,k} $}
			\State $ A^1(S^2_o) \coloneqq A^1(S^2_o) \cup \{\tilde{S}^1_m\}$
			\EndFor
			\EndFor
			
			\For{$ m \coloneqq 2:|\powerset_l (P^{i;j,k})|$} 
			\Comment{For $ S^1 $ adjacent to $ S^2 \in H_m^{i;j,k;l} $, replace the $ S^{0}_{i} $ with $ \tilde{S}^0_{m}$.}
			\For{$ S^2_o \in H_m^{i;j,k;l}$}
			\For{$ S^1_p \in A^1(S^2_o)$}
			\If{$ S^0_i \in A^0(S^1_p) $}
			\State $ A^0(S^1_p) \coloneqq A^0(S^1_p) \cup \{\tilde{S}^0_{m-1}\} \backslash \{S^0_i\} $
			\EndIf
			\EndFor
			\EndFor
			\EndFor
			
			\caption{$ S^2 $ insertion $(\powerset_l (P^{i;j,k}))$.}
			
		\end{algorithmic}
	\end{algorithm}
	
	\section{Check spurious strata}\label{sec:spurious}
	
	Spurious stratum insertions can occur for a $ S^1 $ insertion if there are two $ S^2 $ on the circuit that bound the same $ S^3 $, or for a $ S^2 $ insertion between two components of the same $ S^3 $. An example configuration leading to such an event is shown in Figure \ref{fig:spurious} of the main text. 

	\begin{algorithm} [H]
		\begin{algorithmic}\\
			
			\textit{Compare the upper adjacencies of $(S^d_i)$ to check if it is spurious.}
			\If{$ d = 0 \ or \ d = 1 $} \Comment{Higher adjacency rule for valid $ S^0 $ and $ S^1 $.}
			\State return $ |A^{d+1}(S^d_i)| < 3 $
			\Else 
			\If {d = 2 \ and \  $ |A^{d+1}(S^d_i)| = 2 $} \Comment{If $ S^2 $ bounds the same $ S^3 $, it is spurious.}
			\State return $ A^{d+1}_1(S^d_i) = A^{d+1}_2(S^d_i) $ 
			\EndIf
			\EndIf
			\State return \textbf{FALSE} 
			
			\caption{Check spurious $(S^d_i)$.}
			
		\end{algorithmic}
	\end{algorithm}

	\section{Relaxation during collapse} \label{app:relaxation}
	
	\begin{figure}
		\centering
		\includegraphics[width=260pt]{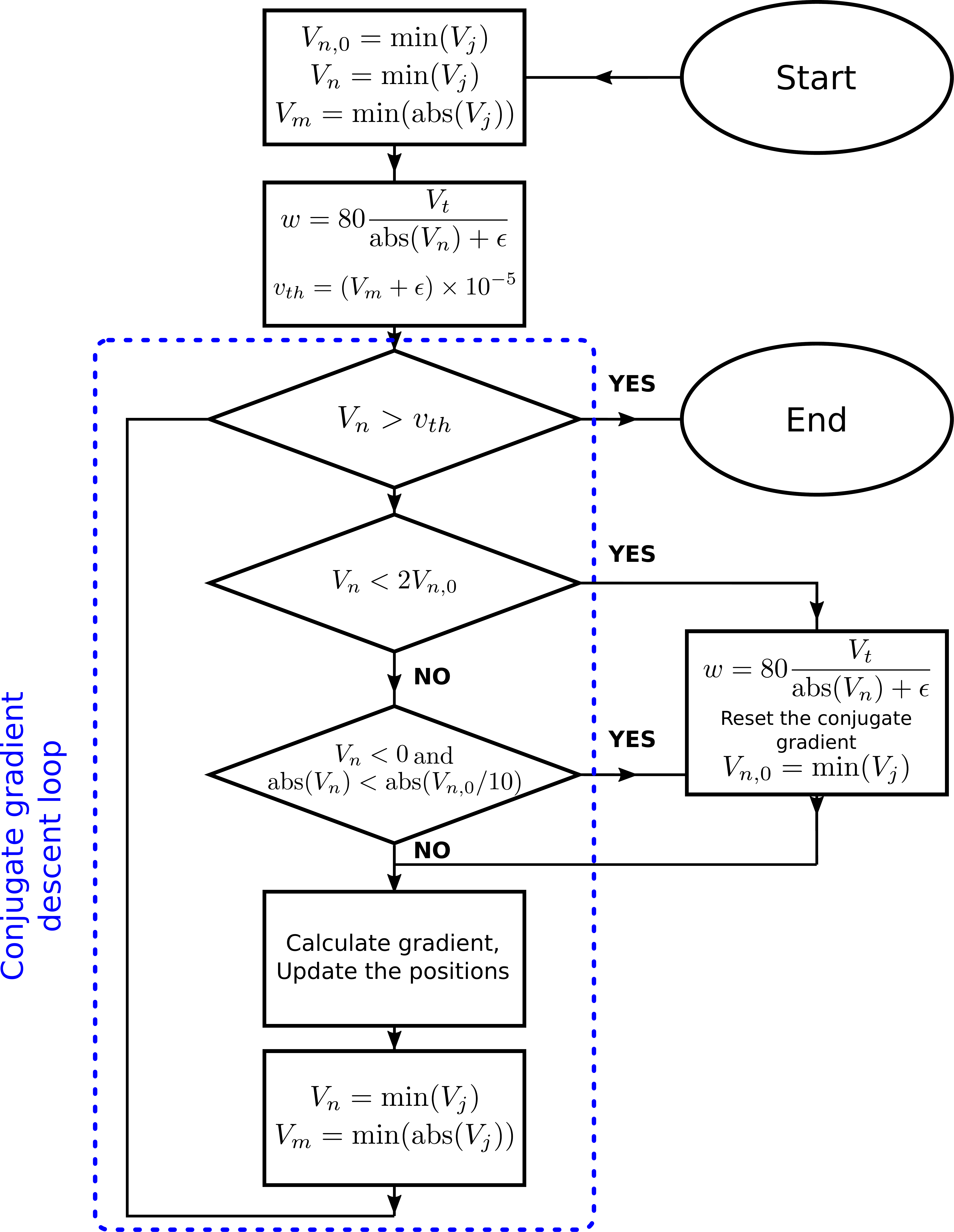}
		\caption{The flow chart for the calculation of the volumes and the update of the parameters $ w $, $ v_{th} $ required for the conjugate gradient descent calculations. }
		\label{fig:conj_grad}
	\end{figure}
	
	The preconditioning operation relaxes the positions of the surrounding vertices to prevent inversions of the surrounding tetrahedra during a collapse. More specifically, the vertices connected to the collapsing strata by an edge form a hull. Positions of the vertices on the hull are found such that the surrounding tetrahedra do not invert during collapse of the stratum by a conjugate gradient search to minimize the positive definite potential $ \phi $. This is defined as
	\begin{align*} 
	\phi &= \sum_{i \in \Delta^{0}} \phi_i, \\ 
	\phi_i &= \frac{1}{w}\ln \left\{\sum_{\Delta^3_j \in A^{3}(\Delta_i^0)} \exp\left[-w V_j(\bar{v})/V_t \right]  + 1\right\}, 
	\end{align*}
	
	where $ \Delta^{0} $ is the set of vertices on the hull, $ \bar{v} $ is the position vector of all vertices, $ V_j $ is the volume of $ j $th tetrahedron, and $ w $ is a weight for scaling the exponent. $ w $ is defined as $ 80 \frac{V_t}{\mbox{abs}(V_n) + \epsilon}$, where $ V_t $ is the total starting volume of all tetrahedra surrounding the hull vertices, and $V_n = \mathrm{min}(V_j)$. In practice $ \epsilon $ is $ 2.22507 \times 10^{-298} $, $ 10^{10} $ times the smallest representable double. Using the algorithm shown in Figure \ref{fig:conj_grad}, the volumes are updated until $ V_n > v_{th} $, where $ v_{th} $ is the desired volume ratio of the surrounding tetrahedra at the collapsed configuration defined as $ v_{th} = (V_m + \epsilon) \times 10^{-5}$ where $ V_m = \mathrm{min}(\mathrm{abs}(V_j))$. After each time the positions of the hull vertices are updated, $ V_n $ and $ V_m $ (but not $ w $ or $ v_{th} $) are updated. If the smallest volume $ V_n $ is smaller than twice the starting most negative volume $ V_{n,0} $, or $ V_n < 0 $ and $ \mathrm{abs}(V_n) < \mathrm{abs}(V_{n,0}/10) $, $ w $ is updated and the conjugate gradient is reinitialized to increase the convergence rate. 
	
	The negative of the gradient of the potential is given by
	\begin{align*} 
	- \overline{\nabla}_i\phi &= - \overline{\nabla}_i\phi_i - \sum_{\Delta^0_j \in A^{0,1}(\Delta^0_i)} \overline{\nabla}_i\phi_{j}, \\ 
	- \overline{\nabla}_i \phi_i &= \sum_{k} \left[\frac{\exp(-w A_k)}{\sum_{l} \exp(-w A_l) + 1} \overline{\nabla}_i V_k \right], 
	\end{align*}
	where $ k,l \in A^{3}(\Delta^0_i) $ and $ A_k = V_k(\bar{v})/V_t $. The form of $ - \overline{\nabla}_i \phi_j $ is the same but $ k,l \in A^{3}(\Delta^0_i) \cap A^{3}(\Delta^0_j) $. It is possible that due to the starting geometry a non-inverting configuration with a minimum volume of $ v_{th} $ cannot be found. The relaxation continues until either a non-inverting configuration is found or the limit for number of iterations is reached. 

	\section{Generalized collapse of multiple lenses} \label{app:generalized_collapse}
	
	\begin{figure}
		\centering
		\subfloat[][]{		\label{subfig:genlens1}
			\includegraphics[height=81.3pt]{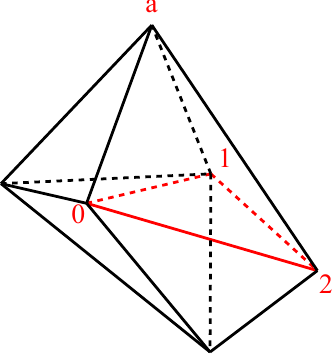}	
		}%
		\subfloat[][]{		\label{subfig:genlens2}
			\includegraphics[height=81.3pt]{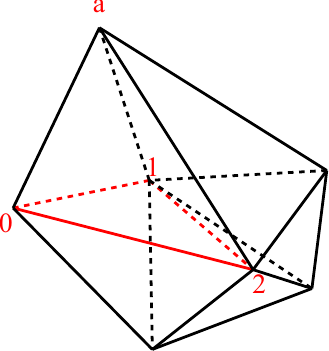}	
		}%
		\qquad
		\subfloat[][]{		\label{subfig:genlens3}
			\includegraphics[height=81.3pt]{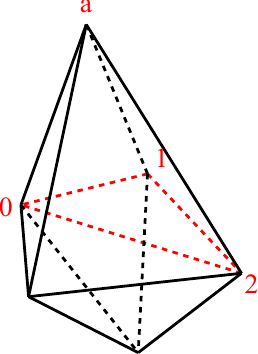}	
		}%
		\subfloat[][]{		\label{subfig:genlens4}
			\includegraphics[height=81.3pt]{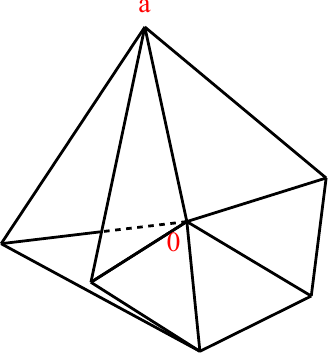}
		}%
		\caption{The generalized lens collapse corresponding to the triangle $ (0,1,2) $. Merging entities form sets rather than couples and an entity might be merging in one lens and collapsing in another, and will collapse during the stratum collapse. The lens corresponding to edge (a) $ (0,1) $, (b) $ (1,2) $, (c) $ (0,2) $. In the lenses corresponding to edges $ (0,2) $ and $ (1,2) $ the triangle $ (0,1,a) $ is merging, but since the triangle is collapsing in the lens of edge $ (0,1) $, it is collapsing. Edges $ (0,a) $, $ (1,a) $, and $ (2,a) $ form a merging set. (d) The final configuration after the collapse. }
		\label{fig:genlensmesh}
	\end{figure}

	The generalized stratum collapse follows the same procedure described in Section \ref{subsec:gencolmesh} for lens collapse, i.e., the main steps are preconditioning the mesh, finding the stratum memberships of the remaining entities, destroying old entities, and regenerating the entities using the last remaining vertex. When there is more than one edge in the collapsing stratum, it is possible that some merging entities form sets rather than couples to form a new entity. Furthermore, an entity can be a merging or a collapsing entity in different lenses, in which case all merging entities in the associated set will collapse as shown in Figure \ref{fig:genlensmesh}. Similar to lens collapse, for each set of merging entities a new entity will be regenerated by replacing the merging vertex with the final vertex and using the new stratum membership. 

	\section{Using an exterior shell for volume preservation} \label{app:shell}
	
	Evolving down the gradient of surface energy, a non-periodic mesh will not preserve volume without additional constraints. Volume preservation is achieved by creating a stratification composed of the simulation cell corners, edges and surfaces. These strata are called \mbox{0-}, \mbox{1-}, and \mbox{2-}shells, respectively. The shells are determined at the start of the simulation. In this section, surface strata will indicate strata on the simulation cell boundary. Each $ S^0 $ is first tested to identify those on the simulation cell corners, edges or surfaces, and ones on the corners are attached to $ 0 $-shells. Next, the $ S^1 $ on the simulation cell boundaries are tested to identify those on the simulation cell edges by a depth first search, and ones on the edges are attached to 1-shells. The 2-shells are constructed similarly. During the simulation the motions of the vertices on these shells are projected onto the corresponding shell to preserve the total volume. 
	 
	\begin{figure*}
		\centering
		\subfloat[][]{		\label{subfig:spurext1}
			\includegraphics[width=81.3pt]{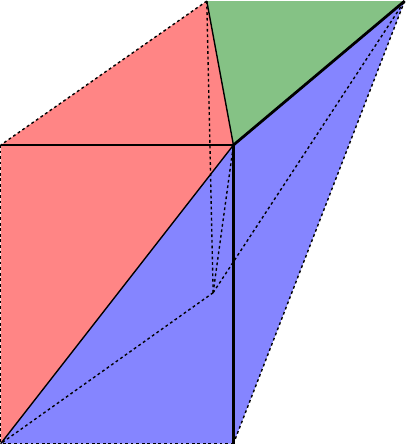}	
		}%
		\subfloat[][]{		\label{subfig:spurext2}
			\includegraphics[width=81.3pt]{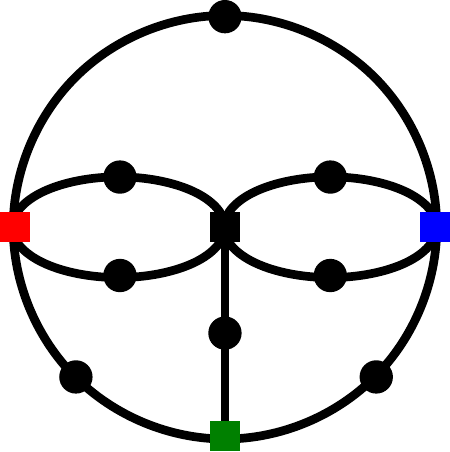}	
		}%
		\subfloat[][]{		\label{subfig:spurext3}
			\includegraphics[width=81.3pt]{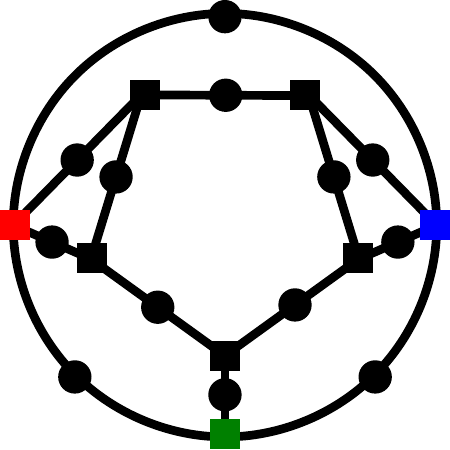}	
		}%
		\subfloat[][]{		\label{subfig:spurext4}
			\includegraphics[width=81.3pt]{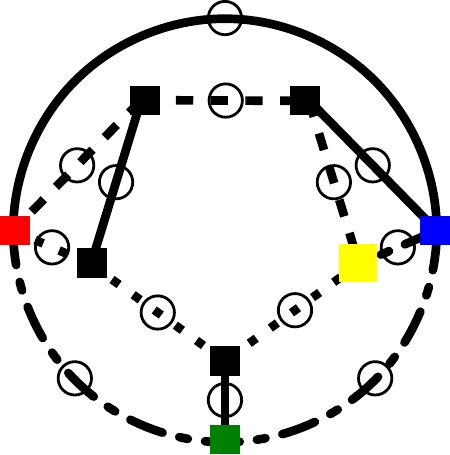}	
		}%
		\subfloat[][]{		\label{subfig:spurext5}
			\includegraphics[width=81.3pt]{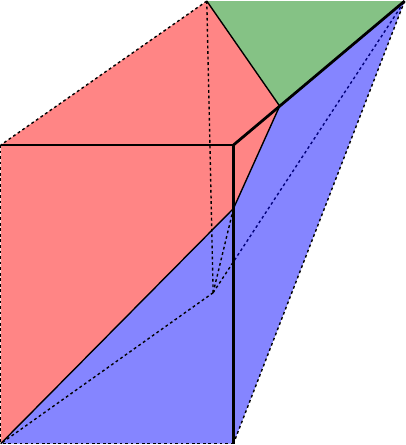}	
		}%
		\caption{Exterior $ S^0 $ insertions require additional exterior $ S^2 $ and $ S^3 $ in the adjacency graph. (a) The red, blue and green $ S^3 $ at the corner of the simulation cell, having two, two, and one surface $ S^2 $, respectively. (b) The adjacency graph obtained by including a node for the exterior $ S^3 $ which doesn't allow any new surface $ S^2 $ insertion. (c) Instead of a single exterior $ S^3 $, $ S^2 $ and $ S^3 $ are included in the adjacency graph for each surface $ S^1 $ and $ S^2 $, respectively. (d) The augmented adjacency graph in (c) can be used to detect surface $ S^2 $ insertions, e.g., a trigon insertion using the dashed, dotted and dash-dotted paths between the exterior and the red grain. (e) The corresponding change in the microstructure.}
		\label{fig:spuriousext}
	\end{figure*}
	
	Any newly inserted strata during stratum insertions around exterior $ S^0 $ are associated with the appropriate shells. Since the exterior can be multiply connected to the volumes touching the exterior surface, one artificial exterior $ S^3 $ is created for each disconnected mesh component of the surface $ S^2 $, as shown in Figure \ref{fig:spuriousext}. The paths and circuits detected on this augmented adjacency graph contain multiplicities as there is actually a single exterior $ S^3 $. These are removed by replacing all artificial strata on the paths and circuits with the only exterior $ S^3 $ and only allowing uninterrupted segments of the artificial strata on a single circuit or path.
	
	Finally, computing the convex hull for collapses of strata touching the exterior shell requires that the positions of vertices belonging to the collapsing stratum be added to the set of points.
	
	\section{Six grain configuration}
\begin{figure*}
		\centering
		\includegraphics[width=480pt]{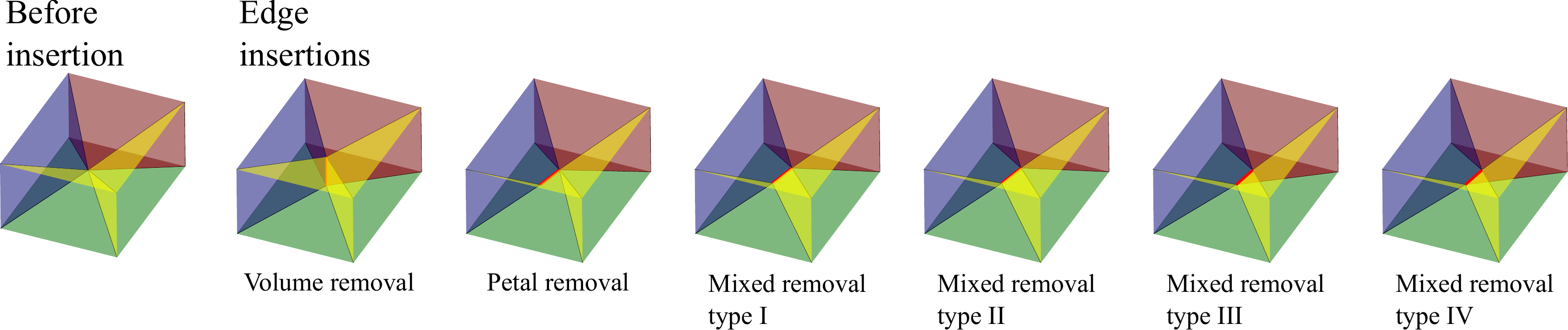}
		\caption{To demonstrate the capabilities of the method, a configuration formed by six grains meeting at the center of a cube is generated. The possible classes of insertions are more numerous than in Figure \ref{fig:split}. Some examples are shown denoted by the number of $ 2 $-strata on the $ H_1^{i;j}/Q_j^i/H_2^{i;j} $, though this is not a complete descriptor (e.g.,\ there are three $ 3/6/3 $ type insertions). In addition to these, digon, trigon and tetragon insertions between each three disconnected $ S^3 $ couples are possible. }
		\label{fig:6grain_1stratum}
	\end{figure*}

\begin{figure*}
		\centering
		\includegraphics[width=350pt]{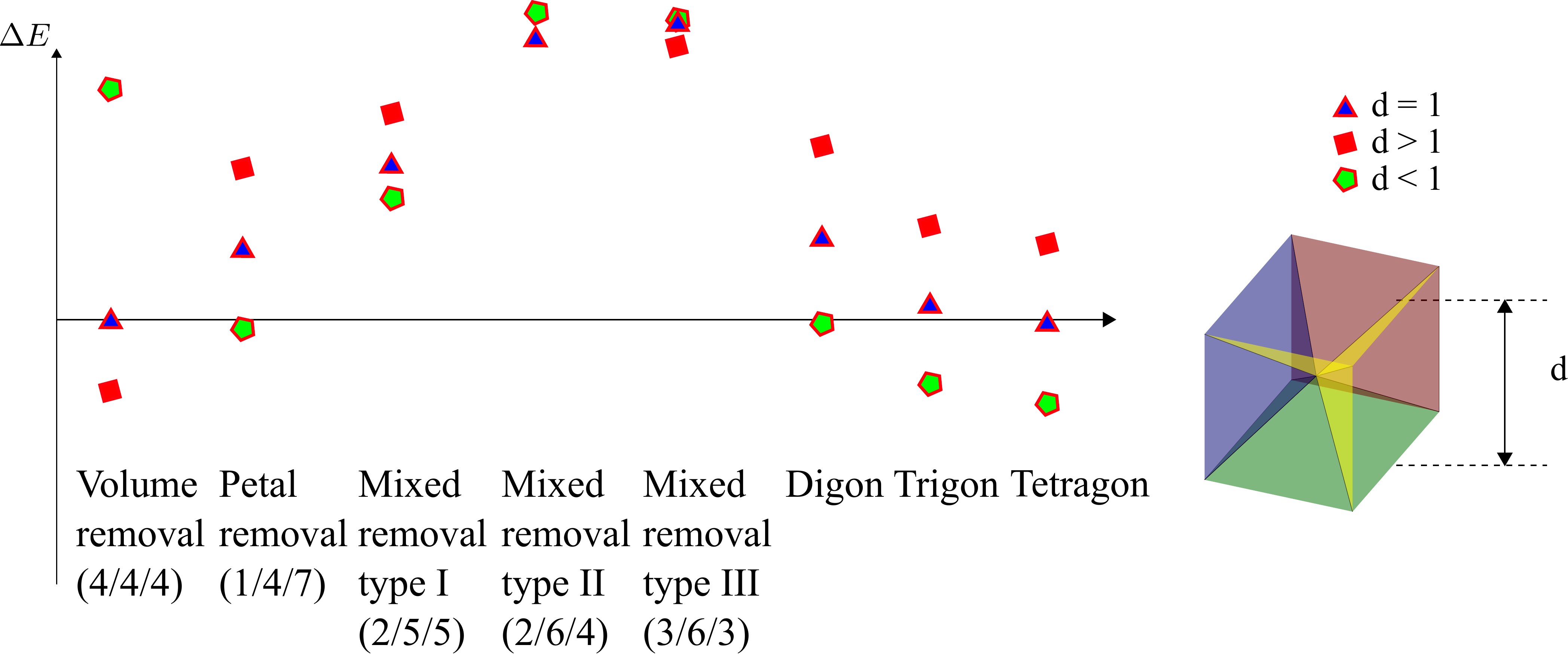}
		\caption{The variation of energy with changing dihedral angle configuration. The degenerate configuration is when the outer dimensions correspond to a cube. Red squares denote the case of the cube stretched in one direction and green pentagons denote the compressed case.}
		\label{fig:6split_rate}
	\end{figure*}

	As a further demonstration of the insertion detection, a more complicated configuration with six grains is generated and some of the possible $ S^1 $ insertions are shown in Figure \ref{fig:6grain_1stratum}. This list is not exhaustive, but demonstrates the capability of the detection algorithm. In addition to these, digon, trigon, and rectangle $ S^2 $ insertions are possible. Similar to Figure \ref{fig:5split_rate}, the dependence of the energy change for different insertions on the dihedral angle is shown in Figure \ref{fig:6split_rate}.
	
	
\begin{thebibliography}{45}
	\expandafter\ifx\csname natexlab\endcsname\relax\def\natexlab#1{#1}\fi
	\expandafter\ifx\csname bibnamefont\endcsname\relax
	\def\bibnamefont#1{#1}\fi
	\expandafter\ifx\csname bibfnamefont\endcsname\relax
	\def\bibfnamefont#1{#1}\fi
	\expandafter\ifx\csname citenamefont\endcsname\relax
	\def\citenamefont#1{#1}\fi
	\expandafter\ifx\csname url\endcsname\relax
	\def\url#1{\texttt{#1}}\fi
	\expandafter\ifx\csname urlprefix\endcsname\relax\def\urlprefix{URL }\fi
	\providecommand{\bibinfo}[2]{#2}
	\providecommand{\eprint}[2][]{\url{#2}}
	
	\bibitem[{\citenamefont{Council}(2008)}]{2008ICME}
	\bibinfo{author}{\bibfnamefont{N.~R.} \bibnamefont{Council}},
	\emph{\bibinfo{title}{Integrated Computational Materials Engineering: A
			Transformational Discipline for Improved Competitiveness and National
			Security}} (\bibinfo{publisher}{The National Academies Press},
	\bibinfo{address}{Washington, DC}, \bibinfo{year}{2008}), ISBN
	\bibinfo{isbn}{978-0-309-11999-3}.
	
	\bibitem[{\citenamefont{Zaefferer}(2005)}]{2005MatSciFormZaefferer}
	\bibinfo{author}{\bibfnamefont{S.}~\bibnamefont{Zaefferer}}, in
	\emph{\bibinfo{booktitle}{Textures of Materials - ICOTOM 14}}
	(\bibinfo{publisher}{Trans Tech Publications}, \bibinfo{year}{2005}), vol.
	\bibinfo{volume}{495} of \emph{\bibinfo{series}{Materials Science Forum}},
	pp. \bibinfo{pages}{3--12}.
	
	\bibitem[{\citenamefont{Li and Suter}(2013)}]{2013ApplCrysLi}
	\bibinfo{author}{\bibfnamefont{S.~F.} \bibnamefont{Li}} \bibnamefont{and}
	\bibinfo{author}{\bibfnamefont{R.~M.} \bibnamefont{Suter}},
	\bibinfo{journal}{Journal of Applied Crystallography}
	\textbf{\bibinfo{volume}{46}}, \bibinfo{pages}{512} (\bibinfo{year}{2013}).
	
	\bibitem[{\citenamefont{Li et~al.}(2014)\citenamefont{Li, Mason, Lind, and
			Kumar}}]{2014ActaMateLi}
	\bibinfo{author}{\bibfnamefont{S.}~\bibnamefont{Li}},
	\bibinfo{author}{\bibfnamefont{J.}~\bibnamefont{Mason}},
	\bibinfo{author}{\bibfnamefont{J.}~\bibnamefont{Lind}}, \bibnamefont{and}
	\bibinfo{author}{\bibfnamefont{M.}~\bibnamefont{Kumar}},
	\bibinfo{journal}{Acta Materialia} \textbf{\bibinfo{volume}{64}},
	\bibinfo{pages}{220 } (\bibinfo{year}{2014}), ISSN \bibinfo{issn}{1359-6454}.
	
	\bibitem[{\citenamefont{Morawiec}(2000)}]{2000ActaMateMorawiec}
	\bibinfo{author}{\bibfnamefont{A.}~\bibnamefont{Morawiec}},
	\bibinfo{journal}{Acta Materialia} \textbf{\bibinfo{volume}{48}},
	\bibinfo{pages}{3525 } (\bibinfo{year}{2000}), ISSN
	\bibinfo{issn}{1359-6454}.
	
	\bibitem[{\citenamefont{Saylor et~al.}(2003{\natexlab{a}})\citenamefont{Saylor,
			Morawiec, and Rohrer}}]{2003Saylorexp}
	\bibinfo{author}{\bibfnamefont{D.~M.} \bibnamefont{Saylor}},
	\bibinfo{author}{\bibfnamefont{A.}~\bibnamefont{Morawiec}}, \bibnamefont{and}
	\bibinfo{author}{\bibfnamefont{G.~S.} \bibnamefont{Rohrer}},
	\bibinfo{journal}{Acta Materialia} \textbf{\bibinfo{volume}{51}},
	\bibinfo{pages}{3663 } (\bibinfo{year}{2003}{\natexlab{a}}), ISSN
	\bibinfo{issn}{1359-6454}.
	
	\bibitem[{\citenamefont{Saylor et~al.}(2003{\natexlab{b}})\citenamefont{Saylor,
			Morawiec, and Rohrer}}]{2003Saylortheo}
	\bibinfo{author}{\bibfnamefont{D.~M.} \bibnamefont{Saylor}},
	\bibinfo{author}{\bibfnamefont{A.}~\bibnamefont{Morawiec}}, \bibnamefont{and}
	\bibinfo{author}{\bibfnamefont{G.~S.} \bibnamefont{Rohrer}},
	\bibinfo{journal}{Acta Materialia} \textbf{\bibinfo{volume}{51}},
	\bibinfo{pages}{3675 } (\bibinfo{year}{2003}{\natexlab{b}}), ISSN
	\bibinfo{issn}{1359-6454}.
	
	\bibitem[{\citenamefont{Bulatov et~al.}(2014)\citenamefont{Bulatov, Reed, and
			Kumar}}]{2014ActMateBulatov}
	\bibinfo{author}{\bibfnamefont{V.}~\bibnamefont{Bulatov}},
	\bibinfo{author}{\bibfnamefont{B.}~\bibnamefont{Reed}}, \bibnamefont{and}
	\bibinfo{author}{\bibfnamefont{M.}~\bibnamefont{Kumar}},
	\bibinfo{journal}{Acta Materialia} \textbf{\bibinfo{volume}{65}},
	\bibinfo{pages}{161} (\bibinfo{year}{2014}).
	
	\bibitem[{\citenamefont{Runnels et~al.}(2016)\citenamefont{Runnels, Beyerlein,
			Conti, and Ortiz}}]{2016JourMechPhySolidRunnels}
	\bibinfo{author}{\bibfnamefont{B.}~\bibnamefont{Runnels}},
	\bibinfo{author}{\bibfnamefont{I.~J.} \bibnamefont{Beyerlein}},
	\bibinfo{author}{\bibfnamefont{S.}~\bibnamefont{Conti}}, \bibnamefont{and}
	\bibinfo{author}{\bibfnamefont{M.}~\bibnamefont{Ortiz}},
	\bibinfo{journal}{Journal of the Mechanics and Physics of Solids}
	\textbf{\bibinfo{volume}{94}}, \bibinfo{pages}{388 } (\bibinfo{year}{2016}),
	ISSN \bibinfo{issn}{0022-5096}.
	
	\bibitem[{\citenamefont{Srolovitz et~al.}(1986)\citenamefont{Srolovitz, Grest,
			and Anderson}}]{1986ActaMetalMateSrolovitzGrestAndersonvol1}
	\bibinfo{author}{\bibfnamefont{D.}~\bibnamefont{Srolovitz}},
	\bibinfo{author}{\bibfnamefont{G.}~\bibnamefont{Grest}}, \bibnamefont{and}
	\bibinfo{author}{\bibfnamefont{M.}~\bibnamefont{Anderson}},
	\bibinfo{journal}{Acta Metallurgica} \textbf{\bibinfo{volume}{34}},
	\bibinfo{pages}{1833 } (\bibinfo{year}{1986}), ISSN
	\bibinfo{issn}{0001-6160}.
	
	\bibitem[{\citenamefont{Holm et~al.}(1991)\citenamefont{Holm, Glazier,
			Srolovitz, and Grest}}]{1991PhysRevAHolm}
	\bibinfo{author}{\bibfnamefont{E.~A.} \bibnamefont{Holm}},
	\bibinfo{author}{\bibfnamefont{J.~A.} \bibnamefont{Glazier}},
	\bibinfo{author}{\bibfnamefont{D.~J.} \bibnamefont{Srolovitz}},
	\bibnamefont{and} \bibinfo{author}{\bibfnamefont{G.~S.} \bibnamefont{Grest}},
	\bibinfo{journal}{Phys. Rev. A} \textbf{\bibinfo{volume}{43}},
	\bibinfo{pages}{2662} (\bibinfo{year}{1991}).
	
	\bibitem[{\citenamefont{Raabe}(2002)}]{2002AnnuRevMatRsRaabe}
	\bibinfo{author}{\bibfnamefont{D.}~\bibnamefont{Raabe}},
	\bibinfo{journal}{Annual Review of Materials Research}
	\textbf{\bibinfo{volume}{32}}, \bibinfo{pages}{53} (\bibinfo{year}{2002}).
	
	\bibitem[{\citenamefont{Ding et~al.}(2006)\citenamefont{Ding, He, Liu, and
			Ding}}]{2006JourCrysGrowDing}
	\bibinfo{author}{\bibfnamefont{H.}~\bibnamefont{Ding}},
	\bibinfo{author}{\bibfnamefont{Y.}~\bibnamefont{He}},
	\bibinfo{author}{\bibfnamefont{L.}~\bibnamefont{Liu}}, \bibnamefont{and}
	\bibinfo{author}{\bibfnamefont{W.}~\bibnamefont{Ding}},
	\bibinfo{journal}{Journal of Crystal Growth} \textbf{\bibinfo{volume}{293}},
	\bibinfo{pages}{489 } (\bibinfo{year}{2006}), ISSN \bibinfo{issn}{0022-0248}.
	
	\bibitem[{\citenamefont{Janssens}(2010)}]{2010MathemCompSimJanssens}
	\bibinfo{author}{\bibfnamefont{K.}~\bibnamefont{Janssens}},
	\bibinfo{journal}{Mathematics and Computers in Simulation}
	\textbf{\bibinfo{volume}{80}}, \bibinfo{pages}{1361 } (\bibinfo{year}{2010}),
	ISSN \bibinfo{issn}{0378-4754}, \bibinfo{note}{{M}ultiscale modeling of
		moving interfaces in materials}.
	
	\bibitem[{\citenamefont{Mason et~al.}(2015)\citenamefont{Mason, Lind, Li, Reed,
			and Kumar}}]{2015ActaMateMason82}
	\bibinfo{author}{\bibfnamefont{J.}~\bibnamefont{Mason}},
	\bibinfo{author}{\bibfnamefont{J.}~\bibnamefont{Lind}},
	\bibinfo{author}{\bibfnamefont{S.}~\bibnamefont{Li}},
	\bibinfo{author}{\bibfnamefont{B.}~\bibnamefont{Reed}}, \bibnamefont{and}
	\bibinfo{author}{\bibfnamefont{M.}~\bibnamefont{Kumar}},
	\bibinfo{journal}{Acta Materialia} \textbf{\bibinfo{volume}{82}},
	\bibinfo{pages}{155 } (\bibinfo{year}{2015}), ISSN \bibinfo{issn}{1359-6454}.
	
	\bibitem[{\citenamefont{Mason}(2015)}]{2015ActaMateMason94}
	\bibinfo{author}{\bibfnamefont{J.}~\bibnamefont{Mason}}, \bibinfo{journal}{Acta
		Materialia} \textbf{\bibinfo{volume}{94}}, \bibinfo{pages}{162 }
	(\bibinfo{year}{2015}), ISSN \bibinfo{issn}{1359-6454}.
	
	\bibitem[{\citenamefont{Zhang et~al.}(2012)\citenamefont{Zhang, Rollett,
			Bartel, Wu, and Lusk}}]{2012ActaMateZhang}
	\bibinfo{author}{\bibfnamefont{L.}~\bibnamefont{Zhang}},
	\bibinfo{author}{\bibfnamefont{A.~D.} \bibnamefont{Rollett}},
	\bibinfo{author}{\bibfnamefont{T.}~\bibnamefont{Bartel}},
	\bibinfo{author}{\bibfnamefont{D.}~\bibnamefont{Wu}}, \bibnamefont{and}
	\bibinfo{author}{\bibfnamefont{M.~T.} \bibnamefont{Lusk}},
	\bibinfo{journal}{Acta Materialia} \textbf{\bibinfo{volume}{60}},
	\bibinfo{pages}{1201 } (\bibinfo{year}{2012}), ISSN
	\bibinfo{issn}{1359-6454}.
	
	\bibitem[{\citenamefont{Steinbach and Pezzolla}(1999)}]{1999PhysicaDSteinbach}
	\bibinfo{author}{\bibfnamefont{I.}~\bibnamefont{Steinbach}} \bibnamefont{and}
	\bibinfo{author}{\bibfnamefont{F.}~\bibnamefont{Pezzolla}},
	\bibinfo{journal}{Physica D: Nonlinear Phenomena}
	\textbf{\bibinfo{volume}{134}}, \bibinfo{pages}{385 } (\bibinfo{year}{1999}),
	ISSN \bibinfo{issn}{0167-2789}.
	
	\bibitem[{\citenamefont{Moelans et~al.}(2008)\citenamefont{Moelans, Blanpain,
			and Wollants}}]{2008CalphadMoelans}
	\bibinfo{author}{\bibfnamefont{N.}~\bibnamefont{Moelans}},
	\bibinfo{author}{\bibfnamefont{B.}~\bibnamefont{Blanpain}}, \bibnamefont{and}
	\bibinfo{author}{\bibfnamefont{P.}~\bibnamefont{Wollants}},
	\bibinfo{journal}{Calphad} \textbf{\bibinfo{volume}{32}}, \bibinfo{pages}{268
	} (\bibinfo{year}{2008}), ISSN \bibinfo{issn}{0364-5916}.
	
	\bibitem[{\citenamefont{Gruber et~al.}(2006)\citenamefont{Gruber, Ma, Wang,
			Rollett, and Rohrer}}]{2006ModSimMatSciEngGruber}
	\bibinfo{author}{\bibfnamefont{J.}~\bibnamefont{Gruber}},
	\bibinfo{author}{\bibfnamefont{N.}~\bibnamefont{Ma}},
	\bibinfo{author}{\bibfnamefont{Y.}~\bibnamefont{Wang}},
	\bibinfo{author}{\bibfnamefont{A.~D.} \bibnamefont{Rollett}},
	\bibnamefont{and} \bibinfo{author}{\bibfnamefont{G.~S.}
		\bibnamefont{Rohrer}}, \bibinfo{journal}{Modelling and Simulation in
		Materials Science and Engineering} \textbf{\bibinfo{volume}{14}},
	\bibinfo{pages}{1189} (\bibinfo{year}{2006}).
	
	\bibitem[{\citenamefont{Vedantam and Patnaik}(2006)}]{2006PhysRevEVedantam}
	\bibinfo{author}{\bibfnamefont{S.}~\bibnamefont{Vedantam}} \bibnamefont{and}
	\bibinfo{author}{\bibfnamefont{B.~S.~V.} \bibnamefont{Patnaik}},
	\bibinfo{journal}{Phys. Rev. E} \textbf{\bibinfo{volume}{73}},
	\bibinfo{pages}{016703} (\bibinfo{year}{2006}).
	
	\bibitem[{\citenamefont{Miyoshi et~al.}(2017)\citenamefont{Miyoshi, Takaki,
			Ohno, Shibuta, Sakane, Shimokawabe, and Aoki}}]{2017NPJMiyoshi}
	\bibinfo{author}{\bibfnamefont{E.}~\bibnamefont{Miyoshi}},
	\bibinfo{author}{\bibfnamefont{T.}~\bibnamefont{Takaki}},
	\bibinfo{author}{\bibfnamefont{M.}~\bibnamefont{Ohno}},
	\bibinfo{author}{\bibfnamefont{Y.}~\bibnamefont{Shibuta}},
	\bibinfo{author}{\bibfnamefont{S.}~\bibnamefont{Sakane}},
	\bibinfo{author}{\bibfnamefont{T.}~\bibnamefont{Shimokawabe}},
	\bibnamefont{and} \bibinfo{author}{\bibfnamefont{T.}~\bibnamefont{Aoki}},
	\bibinfo{journal}{npj Computational Materials} \textbf{\bibinfo{volume}{3}},
	\bibinfo{pages}{25} (\bibinfo{year}{2017}), ISSN \bibinfo{issn}{2057-3960}.
	
	\bibitem[{\citenamefont{Dorr et~al.}(2010)\citenamefont{Dorr, Fattebert,
			Wickett, Belak, and Turchi}}]{2010DorrJourCompPhy}
	\bibinfo{author}{\bibfnamefont{M.}~\bibnamefont{Dorr}},
	\bibinfo{author}{\bibfnamefont{J.-L.} \bibnamefont{Fattebert}},
	\bibinfo{author}{\bibfnamefont{M.}~\bibnamefont{Wickett}},
	\bibinfo{author}{\bibfnamefont{J.}~\bibnamefont{Belak}}, \bibnamefont{and}
	\bibinfo{author}{\bibfnamefont{P.}~\bibnamefont{Turchi}},
	\bibinfo{journal}{Journal of Computational Physics}
	\textbf{\bibinfo{volume}{229}}, \bibinfo{pages}{626 } (\bibinfo{year}{2010}),
	ISSN \bibinfo{issn}{0021-9991}.
	
	\bibitem[{\citenamefont{Jin et~al.}(2015)\citenamefont{Jin, Bozzolo, Rollett,
			and Bernacki}}]{2015CompMatSciJin}
	\bibinfo{author}{\bibfnamefont{Y.}~\bibnamefont{Jin}},
	\bibinfo{author}{\bibfnamefont{N.}~\bibnamefont{Bozzolo}},
	\bibinfo{author}{\bibfnamefont{A.}~\bibnamefont{Rollett}}, \bibnamefont{and}
	\bibinfo{author}{\bibfnamefont{M.}~\bibnamefont{Bernacki}},
	\bibinfo{journal}{Computational Materials Science}
	\textbf{\bibinfo{volume}{104}}, \bibinfo{pages}{108 } (\bibinfo{year}{2015}),
	ISSN \bibinfo{issn}{0927-0256}.
	
	\bibitem[{\citenamefont{Ribot et~al.}(2019)\citenamefont{Ribot, Agrawal, and
			Runnels}}]{2019ModSimMatSciRibot}
	\bibinfo{author}{\bibfnamefont{J.~G.} \bibnamefont{Ribot}},
	\bibinfo{author}{\bibfnamefont{V.}~\bibnamefont{Agrawal}}, \bibnamefont{and}
	\bibinfo{author}{\bibfnamefont{B.}~\bibnamefont{Runnels}},
	\bibinfo{journal}{Modelling and Simulation in Materials Science and
		Engineering} \textbf{\bibinfo{volume}{27}}, \bibinfo{pages}{084007}
	(\bibinfo{year}{2019}).
	
	\bibitem[{\citenamefont{Kawasaki et~al.}(1989)\citenamefont{Kawasaki, Nagai,
			and Nakashima}}]{1989PhilMagBKawasaki}
	\bibinfo{author}{\bibfnamefont{K.}~\bibnamefont{Kawasaki}},
	\bibinfo{author}{\bibfnamefont{T.}~\bibnamefont{Nagai}}, \bibnamefont{and}
	\bibinfo{author}{\bibfnamefont{K.}~\bibnamefont{Nakashima}},
	\bibinfo{journal}{Philosophical Magazine B} \textbf{\bibinfo{volume}{60}},
	\bibinfo{pages}{399} (\bibinfo{year}{1989}).
	
	\bibitem[{\citenamefont{Nagai et~al.}(1990)\citenamefont{Nagai, Ohta, Kawasaki,
			and Okuzono}}]{1990PhaseTransNagai}
	\bibinfo{author}{\bibfnamefont{T.}~\bibnamefont{Nagai}},
	\bibinfo{author}{\bibfnamefont{S.}~\bibnamefont{Ohta}},
	\bibinfo{author}{\bibfnamefont{K.}~\bibnamefont{Kawasaki}}, \bibnamefont{and}
	\bibinfo{author}{\bibfnamefont{T.}~\bibnamefont{Okuzono}},
	\bibinfo{journal}{Phase Transitions} \textbf{\bibinfo{volume}{28}},
	\bibinfo{pages}{177} (\bibinfo{year}{1990}).
	
	\bibitem[{\citenamefont{Roters et~al.}(2010)\citenamefont{Roters, Eisenlohr,
			Hantcherli, Tjahjanto, Bieler, and Raabe}}]{2010ActaMateRoters}
	\bibinfo{author}{\bibfnamefont{F.}~\bibnamefont{Roters}},
	\bibinfo{author}{\bibfnamefont{P.}~\bibnamefont{Eisenlohr}},
	\bibinfo{author}{\bibfnamefont{L.}~\bibnamefont{Hantcherli}},
	\bibinfo{author}{\bibfnamefont{D.}~\bibnamefont{Tjahjanto}},
	\bibinfo{author}{\bibfnamefont{T.}~\bibnamefont{Bieler}}, \bibnamefont{and}
	\bibinfo{author}{\bibfnamefont{D.}~\bibnamefont{Raabe}},
	\bibinfo{journal}{Acta Materialia} \textbf{\bibinfo{volume}{58}},
	\bibinfo{pages}{1152 } (\bibinfo{year}{2010}), ISSN
	\bibinfo{issn}{1359-6454}.
	
	\bibitem[{\citenamefont{Log� et~al.}(2008)\citenamefont{Log�, Bernacki,
			Resk, Delannay, Digonnet, Chastel, and Coupez}}]{2008PhilMagLoge}
	\bibinfo{author}{\bibfnamefont{R.}~\bibnamefont{Log�}},
	\bibinfo{author}{\bibfnamefont{M.}~\bibnamefont{Bernacki}},
	\bibinfo{author}{\bibfnamefont{H.}~\bibnamefont{Resk}},
	\bibinfo{author}{\bibfnamefont{L.}~\bibnamefont{Delannay}},
	\bibinfo{author}{\bibfnamefont{H.}~\bibnamefont{Digonnet}},
	\bibinfo{author}{\bibfnamefont{Y.}~\bibnamefont{Chastel}}, \bibnamefont{and}
	\bibinfo{author}{\bibfnamefont{T.}~\bibnamefont{Coupez}},
	\bibinfo{journal}{Philosophical Magazine} \textbf{\bibinfo{volume}{88}},
	\bibinfo{pages}{3691} (\bibinfo{year}{2008}).
	
	\bibitem[{\citenamefont{Tonks et~al.}(2012)\citenamefont{Tonks, Gaston,
			Millett, Andrs, and Talbot}}]{2012CompMatSciTonks}
	\bibinfo{author}{\bibfnamefont{M.~R.} \bibnamefont{Tonks}},
	\bibinfo{author}{\bibfnamefont{D.}~\bibnamefont{Gaston}},
	\bibinfo{author}{\bibfnamefont{P.~C.} \bibnamefont{Millett}},
	\bibinfo{author}{\bibfnamefont{D.}~\bibnamefont{Andrs}}, \bibnamefont{and}
	\bibinfo{author}{\bibfnamefont{P.}~\bibnamefont{Talbot}},
	\bibinfo{journal}{Computational Materials Science}
	\textbf{\bibinfo{volume}{51}}, \bibinfo{pages}{20 } (\bibinfo{year}{2012}),
	ISSN \bibinfo{issn}{0927-0256}.
	
	\bibitem[{\citenamefont{Raabe and Becker}(2000)}]{2000ModellSimulMatSciRaabe}
	\bibinfo{author}{\bibfnamefont{D.}~\bibnamefont{Raabe}} \bibnamefont{and}
	\bibinfo{author}{\bibfnamefont{R.~C.} \bibnamefont{Becker}},
	\bibinfo{journal}{Modelling and Simulation in Materials Science and
		Engineering} \textbf{\bibinfo{volume}{8}}, \bibinfo{pages}{445}
	(\bibinfo{year}{2000}).
	
	\bibitem[{\citenamefont{Kuprat}(2000)}]{2000SIAMJourSciCompKuprat}
	\bibinfo{author}{\bibfnamefont{A.}~\bibnamefont{Kuprat}},
	\bibinfo{journal}{SIAM Journal on Scientific Computing}
	\textbf{\bibinfo{volume}{22}}, \bibinfo{pages}{535} (\bibinfo{year}{2000}).
	
	\bibitem[{\citenamefont{Gruber et~al.}(2005)\citenamefont{Gruber, George,
			Kuprat, Rohrer, and Rollett}}]{2005ScripMateGruber}
	\bibinfo{author}{\bibfnamefont{J.}~\bibnamefont{Gruber}},
	\bibinfo{author}{\bibfnamefont{D.~C.} \bibnamefont{George}},
	\bibinfo{author}{\bibfnamefont{A.~P.} \bibnamefont{Kuprat}},
	\bibinfo{author}{\bibfnamefont{G.~S.} \bibnamefont{Rohrer}},
	\bibnamefont{and} \bibinfo{author}{\bibfnamefont{A.~D.}
		\bibnamefont{Rollett}}, \bibinfo{journal}{Scripta Materialia}
	\textbf{\bibinfo{volume}{53}}, \bibinfo{pages}{351} (\bibinfo{year}{2005}),
	ISSN \bibinfo{issn}{1359-6462}.
	
	\bibitem[{\citenamefont{Syha and Weygand}(2010)}]{2010ModelSimuMaterSciEngSyha}
	\bibinfo{author}{\bibfnamefont{M.}~\bibnamefont{Syha}} \bibnamefont{and}
	\bibinfo{author}{\bibfnamefont{D.}~\bibnamefont{Weygand}},
	\bibinfo{journal}{Modelling and Simulation in Materials Science and
		Engineering} \textbf{\bibinfo{volume}{18}}, \bibinfo{pages}{015010}
	(\bibinfo{year}{2010}).
	
	\bibitem[{\citenamefont{Lazar et~al.}(2011)\citenamefont{Lazar, Mason,
			MacPherson, and Srolovitz}}]{2011ActaMateLazar}
	\bibinfo{author}{\bibfnamefont{E.~A.} \bibnamefont{Lazar}},
	\bibinfo{author}{\bibfnamefont{J.~K.} \bibnamefont{Mason}},
	\bibinfo{author}{\bibfnamefont{R.~D.} \bibnamefont{MacPherson}},
	\bibnamefont{and} \bibinfo{author}{\bibfnamefont{D.~J.}
		\bibnamefont{Srolovitz}}, \bibinfo{journal}{Acta Materialia}
	\textbf{\bibinfo{volume}{59}}, \bibinfo{pages}{6837 } (\bibinfo{year}{2011}),
	ISSN \bibinfo{issn}{1359-6454}.
	
	\bibitem[{\citenamefont{MacPherson and
			Srolovitz}(2007)}]{2007NatureMacPhersonSrolovitz}
	\bibinfo{author}{\bibfnamefont{R.~D.} \bibnamefont{MacPherson}}
	\bibnamefont{and} \bibinfo{author}{\bibfnamefont{D.~J.}
		\bibnamefont{Srolovitz}}, \bibinfo{journal}{Nature}
	\textbf{\bibinfo{volume}{466}}, \bibinfo{pages}{1053 }
	(\bibinfo{year}{2007}).
	
	\bibitem[{\citenamefont{Tucker et~al.}(2015)\citenamefont{Tucker, III,
			Ingraffea, and Rollett}}]{2015ModSimMatSciTucker}
	\bibinfo{author}{\bibfnamefont{J.~C.} \bibnamefont{Tucker}},
	\bibinfo{author}{\bibfnamefont{A.~R.~C.} \bibnamefont{III}},
	\bibinfo{author}{\bibfnamefont{A.~R.} \bibnamefont{Ingraffea}},
	\bibnamefont{and} \bibinfo{author}{\bibfnamefont{A.~D.}
		\bibnamefont{Rollett}}, \bibinfo{journal}{Modelling and Simulation in
		Materials Science and Engineering} \textbf{\bibinfo{volume}{23}},
	\bibinfo{pages}{035003} (\bibinfo{year}{2015}).
	
	\bibitem[{\citenamefont{Mason}(2017)}]{2017ActMateMason}
	\bibinfo{author}{\bibfnamefont{J.~K.} \bibnamefont{Mason}},
	\bibinfo{journal}{Acta Materialia} \textbf{\bibinfo{volume}{125}},
	\bibinfo{pages}{286 } (\bibinfo{year}{2017}), ISSN \bibinfo{issn}{1359-6454}.
	
	\bibitem[{sco()}]{scorecweb}
	\emph{\bibinfo{title}{{Scientific Computation Research Center at Rensselaer
				Polytechnic Institute. http://scorec.rpi.edu/ (accessed 1 November 2020)}}}.
	
	\bibitem[{\citenamefont{Seol et~al.}(2012)\citenamefont{Seol, Smith, Ibanez,
			and Shephard}}]{2012SCCompanionSeolShephardPUMI}
	\bibinfo{author}{\bibfnamefont{S.}~\bibnamefont{Seol}},
	\bibinfo{author}{\bibfnamefont{C.~W.} \bibnamefont{Smith}},
	\bibinfo{author}{\bibfnamefont{D.~A.} \bibnamefont{Ibanez}},
	\bibnamefont{and} \bibinfo{author}{\bibfnamefont{M.~S.}
		\bibnamefont{Shephard}}, in \emph{\bibinfo{booktitle}{2012 SC Companion: High
			Performance Computing, Networking Storage and Analysis}}
	(\bibinfo{year}{2012}), pp. \bibinfo{pages}{1124--1132}.
	
	\bibitem[{\citenamefont{Paton}(1969)}]{1969CommunACMPaton}
	\bibinfo{author}{\bibfnamefont{K.}~\bibnamefont{Paton}},
	\bibinfo{journal}{Commun. ACM} \textbf{\bibinfo{volume}{12}},
	\bibinfo{pages}{514} (\bibinfo{year}{1969}), ISSN \bibinfo{issn}{0001-0782}.
	
	\bibitem[{\citenamefont{Gibbs}(1969)}]{1969JourACMGibbs}
	\bibinfo{author}{\bibfnamefont{N.~E.} \bibnamefont{Gibbs}},
	\bibinfo{journal}{J. ACM} \textbf{\bibinfo{volume}{16}}, \bibinfo{pages}{564}
	(\bibinfo{year}{1969}), ISSN \bibinfo{issn}{0004-5411}.
	
	\bibitem[{\citenamefont{von Neumann}(1952)}]{vonNeumann1952}
	\bibinfo{author}{\bibfnamefont{J.}~\bibnamefont{von Neumann}}, in
	\emph{\bibinfo{booktitle}{Metal Interfaces}} (\bibinfo{publisher}{American
		Society for Metals}, \bibinfo{address}{Cleveland, Ohio},
	\bibinfo{year}{1952}), pp. \bibinfo{pages}{108--110}.
	
	\bibitem[{\citenamefont{Mullins}(1956)}]{Mullins1956}
	\bibinfo{author}{\bibfnamefont{W.~W.} \bibnamefont{Mullins}},
	\bibinfo{journal}{Journal of Applied Physics} \textbf{\bibinfo{volume}{27}},
	\bibinfo{pages}{900} (\bibinfo{year}{1956}).
	
	\bibitem[{\citenamefont{Quey et~al.}(2011)\citenamefont{Quey, Dawson, and
			Barbe}}]{2011CompMethAppMechQuey}
	\bibinfo{author}{\bibfnamefont{R.}~\bibnamefont{Quey}},
	\bibinfo{author}{\bibfnamefont{P.}~\bibnamefont{Dawson}}, \bibnamefont{and}
	\bibinfo{author}{\bibfnamefont{F.}~\bibnamefont{Barbe}},
	\bibinfo{journal}{Computer Methods in Applied Mechanics and Engineering}
	\textbf{\bibinfo{volume}{200}}, \bibinfo{pages}{1729 }
	(\bibinfo{year}{2011}), ISSN \bibinfo{issn}{0045-7825}.
	
\end{thebibliography}
\end{document}